\documentclass[journal]{IEEEtran}
\usepackage{setspace}
 \usepackage{cite}
 \usepackage{calligra}
 \usepackage{latexsym}
 \usepackage{times}
 \usepackage{graphicx}
 \usepackage{color}
\usepackage{algorithm}
\usepackage{algorithmic}
 \usepackage{amssymb,amsmath}
 \usepackage{multirow}
 \usepackage{colortbl} 
\usepackage{enumerate}
\usepackage{graphicx}
\usepackage{silence}
\usepackage[font=footnotesize]{caption}
\usepackage{subcaption}
\usepackage{mathrsfs}
\usepackage{amsbsy}
\usepackage{xcolor, colortbl}
\usepackage{mathtools}
\usepackage[utf8]{inputenc}
\usepackage{setspace}
 \usepackage{calligra}
 \usepackage{latexsym}
 \usepackage{times}
 \usepackage{graphicx}
\usepackage{algorithm}
\usepackage{algorithmic}
 \usepackage{amssymb,amsmath}
 \usepackage{multirow}
 \usepackage{mathtools}
\usepackage{commath}
\usepackage{tikz}
\usepackage{tabularx}
\usepackage{enumitem, kantlipsum}
\usepackage[switch]{lineno}
\usepackage{booktabs, makecell}
\usepackage[hidelinks]{hyperref}
\usepackage[automake]{glossaries}
\newacronym{3GPP}{3GPP}{3rd Generation Partnership Project}
\newacronym{4G}{4G}{4th generation}
\newacronym{5G}{5G}{5th generation}
\newacronym{5GS}{5GS}{5G system}
\newacronym{6G}{6G}{6th generation}
\newacronym{5GAA}{5GAA}{5G automotive association}
\newacronym{ADE}{ADE}{average displacement error}
\newacronym{AF}{AF}{application function}
\newacronym{AI}{AI}{artificial intelligence}
\newacronym{AL}{AL}{alert limit}
\newacronym{AoA}{AoA}{angle of arrival}
\newacronym{Apps}{Apps}{applications}
\newacronym{A-AoA}{A-AoA}{azimuth angle of arrival}
\newacronym{acm}{ACM}{Association for Computing Machinery}
\newacronym{AMF}{AMF}{access and mobility function}
\newacronym{AoD}{AoD}{angle of departure}
\newacronym{AP}{AP}{access point}
\newacronym{API}{API}{application programming interface}
\newacronym{ASIL}{ASIL}{automotive safety integrity level}
\newacronym{AWGN}{AWGN}{additive white Gaussian noise}
\newacronym{B5G}{B5G}{Beyond 5G}
\newacronym{BS}{BS}{base station}
\newacronym{C-ITS}{C-ITS}{cooperative intelligent transportation systems}
\newacronym{CAM}{CAM}{cooperative awareness messages}
\newacronym{CPM}{CPM}{collective perception messages}
\newacronym{CAPIF}{CAPIF}{Common API Framework}
\newacronym{CCO}{CCO}{capacity and coverage optimization}
\newacronym{CCRB}{CCRB}{constrained CRB}
\newacronym{CDF}{CDF}{cumulative density function}
\newacronym{CEP}{CEP}{circular error probability}
\newacronym{CID}{CID}{cell identity}
\newacronym{CIR}{CIR}{channel impulse response}
\newacronym{CIS}{CIS}{continuous intelligent surface}
\newacronym{CFAR}{CFAR}{constant false alarm rate}
\newacronym{CoO}{CoO}{cell of origin}
\newacronym{CP}{CP}{cooperative positioning}
\newacronym{CRB}{CRB}{Cram\'er-Rao bound}
\newacronym{CNN}{CNN}{convolutional neural network}
\newacronym{CSI}{CSI}{channel state information}
\newacronym{CSI-RS}{CSI-RS}{channel state information reference signal}
\newacronym{CSI-RSRP}{CSI-RSRP}{channel state information RSRP}
\newacronym{CSI-RSRQ}{CSI-RSRQ}{channel state information RSRQ}
\newacronym{CX}{CX}{customer experience}
\newacronym{C-V2X}{C-V2X}{cellular V2X}
\newacronym{DBSCAN}{DBSCAN}{density-based spatial clustering of applications with noise}
\newacronym{DEN}{DEN}{decentralized environmental notification}
\newacronym{DFL}{DFL}{device-free localization}
\newacronym{DL}{DL}{downlink}
\newacronym{DL-AoD}{DL-AoD}{downlink angle-of-departure}
\newacronym{DL-PRS}{DL-PRS}{downlink positioning reference signal}
\newacronym{DL-TDoA}{DL-TDoA}{downlink time difference of arrival}
\newacronym{DMRS}{DMRS}{demodulation reference signal}
\newacronym{DNN}{DNN}{deep neural network}
\newacronym{NN}{NN}{neural network}
\newacronym{DOA}{DOA}{direction-of-arrival}
\newacronym{ECDF}{ECDF}{empirical cumulative density function}
\newacronym{eCID}{eCID}{enhanced CID}
\newacronym{eLCS}{eLCS}{enhanced LCS}
\newacronym{EFIM}{EFIM}{equivalent Fisher information matrix}
\newacronym{EGNOS}{EGNOS}{european geostationary navigation overlay service}
\newacronym{EM}{EM}{electro-magnetic}
\newacronym{EMF}{EMF}{electro-magnetic field}
\newacronym{eNB}{eNB}{eNodeB}
\newacronym{ESMLC}{ESMLC}{enhanced mobile location server}
\newacronym{ESPRIT}{ESPRIT}{estimation of signal parameters via rotational invariant techniques}
\newacronym{ETSI}{ETSI}{European Telecommunication Standards Institute}
\newacronym{FCC}{FCC}{Federal Communication Commission}
\newacronym{FDE}{FDE}{final displacement error}
\newacronym{FE}{FE}{feature engineering}
\newacronym{FIM}{FIM}{Fisher information matrix}
\newacronym{FTM}{FTM}{fine time measurement}
\newacronym{FLC}{FLC}{Fuzzy Logic Controller}
\newacronym{gNB}{gNB}{gNodeB}
\newacronym{GAN}{GAN}{generative adversarial networks }
\newacronym{GDOP}{GDOP}{geometric dilution of precision}
\newacronym{GBTs}{GBTs}{gradient boosted trees}
\newacronym{GMLC}{GMLC}{gateway mobile location center}
\newacronym{GNSS}{GNSS}{global navigation satellite system}
\newacronym{GoB}{GoB}{Grid of Beams}
\newacronym{GSM}{GSM}{global system for mobile communication}
\newacronym{GPS}{GPS}{global positioning system}
\newacronym{GRU}{GRU}{gated recurrent unit}
\newacronym{HARA}{HARA}{hazard assessment and risk analysis}
\newacronym{HD}{HD}{high definition}
\newacronym{HDBSCAN}{HDBSCAN}{hierarchical DBSCAN}
\newacronym{ICRB}{ICRB}{intrinsic CRB}
\newacronym{IIoT}{IIoT}{industrial internet of things}
\newacronym{IMU}{IMU}{inertial measurement unit}
\newacronym{INS}{INS}{inertial navigation system}
\newacronym{IOO}{IOO}{indoor open office}
\newacronym{IoT}{IoT}{internet of things}
\newacronym{IPv6}{IPv6}{internet protocol version 6}
\newacronym{IS}{IS}{intelligent surface}
\newacronym{ISAC}{ISAC}{integrated sensing and communication}
\newacronym{ISD}{ISD}{inter site distance}
\newacronym{ITS}{ITS}{intelligent transport systems}
\newacronym{ITS-S}{ITS-S}{ITS station}
\newacronym{JRC}{JRC}{joint radar and communication}
\newacronym{KPI}{KPI}{key performance indicator}
\newacronym{LCS}{LCS}{localization service}
\newacronym{LDM}{LDM}{local dynamic map}
\newacronym{LEO}{LEO}{Low-Earth Orbit}
\newacronym{LM}{LM}{localized measurement}
\newacronym{LMF}{LMF}{location management function}
\newacronym{LoS}{LoS}{line-of-sight}
\newacronym{LPP}{LPP}{LTE positioning protocol}
\newacronym{LPI}{LPI}{location privacy indication}
\newacronym{LSTM}{LSTM}{long short-term memory }
\newacronym{LTE}{LTE}{long term evolution}
\newacronym{LRF}{LRF}{location retrieval function}
\newacronym{LS}{LS}{least squares}
\newacronym{MANO}{MANO}{management and orchestration}
\newacronym{MBS}{MBS}{metropolitan beacon system}
\newacronym{MCRB}{MCRB}{misspecified CRB}
\newacronym{MDE}{MDE}{mean distance error}
\newacronym{MDT}{MDT}{Minimization of Drive Tests}
\newacronym{MLE}{MLE}{maximum likelihood estimation}
\newacronym{MLP}{MLP}{multi-layer perception}
\newacronym{MPC}{MPC}{multi-path component}
\newacronym{MIMO}{MIMO}{multiple-input-multiple-output}
\newacronym{M-MIMO}{M-MIMO}{massive multiple-input-multiple-output}
\newacronym{MMSE}{MMSE}{minimum mean squared error}
\newacronym{NSI}{NSI}{Network Synthetic Image}
\newacronym{NCAP}{NCAP}{New Car Assessment Program}
\newacronym{MEO}{MEO}{Medium-Earth Orbit}
\newacronym{mm-Wave}{mm-Wave}{millimeter wave}
\newacronym{MO-LR}{MO-LR}{mobile originated location request}
\newacronym{MPCs}{MPCs}{multiple signal components}
\newacronym{MR-DC}{MR-DC}{multi radio - dual connectivity}
\newacronym{MSE}{MSE}{mean squared error}
\newacronym{MT-LR}{MT-LR}{mobile terminated location request}
\newacronym{MU}{MU}{multi-user}
\newacronym{MUSIC}{MUSIC}{MUltiple SIgnal Classification}
\newacronym{MUT}{MUT}{mean user throughput}
\newacronym{multi-RTT}{multi-RTT}{multiple round trip time}
\newacronym{NEF}{NEF}{network exposure function}
\newacronym{NG-RAN}{NG-RAN}{next generation-radio access network}
\newacronym{NI-LR}{NI-LR}{network induced location request}
\newacronym{NLoS}{NLoS}{non-line-of-sight}
\newacronym{NR}{NR}{new radio}
\newacronym{NRPPa}{NRPPa}{new radio positioning protocol a}
\newacronym{MCS}{MCS}{modulation and coding scheme}
\newacronym{NWDAF}{NWDAF}{Network Data Analytic Function}
\newacronym{OSI}{OSI}{Open System Interconnection}
\newacronym{OFDM}{OFDM}{orthogonal frequency division multiplexing}
\newacronym{OSS}{OSS}{Operations Support System}
\newacronym{OTDoA}{OTDoA}{observed time difference of arrival}
\newacronym{PAPR}{PAPR}{peak-to-average power ratio}
\newacronym{PEB}{PEB}{position error bound}
\newacronym{PFL}{PFL}{positioning frequency layers}
\newacronym{PL}{PL}{protection level}
\newacronym{POI}{POI}{point of interest}
\newacronym{PCRB}{PCRB}{posterior CRB}
\newacronym{PNT}{PNT}{positioning, navigation and timing}
\newacronym{PoTi}{PoTi}{position and time}
\newacronym{PPP}{PPP}{precise point positioning}
\newacronym{PPP-RTK}{PPP-RTK}{precise point positioning real-time kinematic}
\newacronym{PRB}{PRB}{physical resource block}
\newacronym{PRS-RSRP}{PRS-RSRP}{PRS reference signal received power}
\newacronym{PRS}{PRS}{positioning reference signal}
\newacronym{PHY}{PHY}{physical layer}
\newacronym{PTS}{PTS}{Power Traffic Sharing}
\newacronym{QAM}{QAM}{quadrature amplitude modulation}
\newacronym{QoS}{QoS}{quality of service}
\newacronym{QPSK}{QPSK}{quadrature phase shift keying}
\newacronym{RADAR}{RADAR}{radio detection and ranging}
\newacronym{RAN}{RAN}{radio access network}
\newacronym{RAT}{RAT}{radio access technology}
\newacronym{RE}{RE}{resource element}
\newacronym{RedCap}{RedCap}{reduced capability}
\newacronym{REM}{REM}{radio environment map}
\newacronym{RCS}{RCS}{radar cross section}
\newacronym{RF}{RF}{radio frequency}
\newacronym{RIS}{RIS}{reconfigurable intelligent surface}
\newacronym{RMSE}{RMSE}{root-mean-square error}
\newacronym{RNN}{RNN}{recurrent neural network}
\newacronym{RRC}{RRC}{radio resource control}
\newacronym{RRM}{RRM}{radio resource management}
\newacronym{RS}{RS}{reference signal}
\newacronym{RSRP}{RSRP}{reference signal received power}
\newacronym{RSRQ}{RSRQ}{reference signal received quality}
\newacronym{RSS}{RSS}{received signal strength}
\newacronym{RSSI}{RSSI}{received signal strength indicator}
\newacronym{RSTD}{RSTD}{received signal time difference}
\newacronym{RTK}{RTK}{real time kinematics}
\newacronym{RTT}{RTT}{round-trip-time}
\newacronym{SAE}{SAE}{society of automotive engineers}
\newacronym{SBA}{SBA}{service based architecture}
\newacronym{SCI}{SCI}{soft context information}
\newacronym{SCS}{SCS}{subcarrier spacing}
\newacronym{SDK}{SDK}{software development kit}
\newacronym{SFI}{SFI}{soft feature information}
\newacronym{SI}{SI}{soft information}
\newacronym{SINR}{SINR}{signal-to-interference-plus-noise ratio}
\newacronym{SIR}{SIR}{signal-to-interference ratio}
\newacronym{SL}{SL}{sidelink}
\newacronym{SLA}{SLA}{service-level agreement}
\newacronym{SLAM}{SLAM}{simultaneous localization and mapping}
\newacronym{SNR}{SNR}{signal-to-noise ratio}
\newacronym{SAPTS}{SAPTS}{social-aware PTS controller} 
\newacronym{SoTA}{SoTA}{state-of-the-art}
\newacronym{SPEB}{SPEB}{squared position error bound}
\newacronym{SR}{SR}{sensor radar}
\newacronym{SRN}{SRN}{sensor radar network}
\newacronym{SRS}{SRS}{sounding reference signal}
\newacronym{SRS-RSRP}{SRS-RSRP}{SRS reference signal received power}
\newacronym{SSB}{SSB}{single-sideband}
\newacronym{SSR}{SSR}{state space representation}
\newacronym{SS-RSRP}{SS-RSRP}{synchronization-signal-based RSRP}
\newacronym{SS-RSRQ}{SS-RSRQ}{synchronization-signal-based RSRQ}
\newacronym{STAN}{STAN}{simultaneous tracking and navigation}
\newacronym{SVE}{SVE}{single-value estimates}
\newacronym{TA}{TA}{timing advance}
\newacronym{TBS}{TBS}{terrestrial beacon system}
\newacronym{TDoA}{TDoA}{time-difference-of-arrival}
\newacronym{TLE}{TLE}{two-line element}
\newacronym{TIR}{TIR}{target integrity risk}
\newacronym{ToA}{ToA}{time-of-arrival}
\newacronym{ToF}{ToF}{time of flight}
\newacronym{TRP}{TRP}{transmit/receive point}
\newacronym{TTA}{TTA}{time-to-alert}
\newacronym{TTI}{TTI}{transmission time interval}
\newacronym{TTFF}{TTFF}{time to first fix}
\newacronym{UAV}{UAV}{unmanned aerial vehicle}
\newacronym{UDM}{UDM}{unified data manager}
\newacronym{UE}{UE}{user equipment}
\newacronym{UL}{UL}{uplink time}
\newacronym{UL-AoA}{UL-AoA}{uplink angle-of-arrival}
\newacronym{UL-RToA}{UL-RToA}{uplink relative time of arrival}
\newacronym{UL-SRS}{UL-SRS}{uplink sounding reference signal}
\newacronym{UL-TDoA}{UL-TDoA}{uplink time difference of arrival}
\newacronym{UMi}{UMi}{urban micro-cellular}
\newacronym{URLLC}{URLLC}{ultra reliable low latency communication}
\newacronym{UTDoA}{UTDoA}{uplink time difference of arrival}
\newacronym{UWB}{UWB}{ultra-wideband}
\newacronym{WIFI}{WIFI}{wireless fidelity}
\newacronym{V2I}{V2I}{vehicle-to-infrastructure}
\newacronym{V2P}{V2P}{vehicle-to-pedestrian}
\newacronym{V2V}{V2V}{vehicle-to-vehicle}
\newacronym{V2X}{V2X}{vehicle-to-everything}
\newacronym{VRU}{VRU}{vulnerable road user}
\newacronym{RSU}{RSU}{road side unit}
\newacronym{STL}{STL}{satellite timing and location}
\newacronym{XR}{XR}{extended reality}
\newacronym{Z-AoA}{Z-AoA}{zenith angle of arrival}
\newacronym{E-CID}{E-CID}{enhanced cell identity}
\newacronym{mmWave}{mmWave}{millimeter-wave}
\newacronym{THz}{THz}{terahertz}

\newacronym{EOTD}{EOTD}{enhanced observed time difference}
\newacronym{A-GPS}{A-GPS}{assisted-GPS}
\newacronym{A-GNSS}{A-GNSS}{assisted-GNSS}
\newacronym{SMLC}{SMLC}{serving mobile location center}
\newacronym{LMU}{LMU}{location management unit}
\newacronym{RNC}{RNC}{radio network controller}
\newacronym{PE}{PE}{positioning element}
\newacronym{LPHAP}{LPHAP}{low-power high-accuracy positioning}
\newacronym{SLPP}{SLPP}{sidelink positioning protocol}
\newacronym{PDOA}{PDOA}{phase difference of arrival}
\newacronym{NTN}{NTN}{non-terrestrial networks}
\newacronym{TN}{TN}{terrestrial networks}

\newacronym{ESPIRIT}{ESPIRIT}{etimation of signal parameters via rotational invariance techniques}
\newacronym{GOSPA}{GOSPA}{generalized optimal subpattern as-
signment}

\newacronym{KF}{KF}{Kalman filter}
\newacronym{EKF}{EKF}{extended Kalman filter}
\newacronym{UKF}{UKF}{unscented Kalman filter}
\newacronym{PF}{PF}{particle filter}
\newacronym{BNN}{BNN}{Bayesian neural network}

\newacronym{AGV}{AGV}{automated guided vehicle}

\newacronym{LAN}{LAN}{local area network}
\newacronym{MAN}{MAN}{metropolitan area network}
\newacronym{PAN}{PAN}{personal area network}
\newacronym{BAN}{BAN}{body area network}
\newacronym{WAN}{WAN}{wide area network}

\newacronym{SSID}{SSID}{service set identifier}
\newacronym{DSRC}{DSRC}{directed short-range communications}

\newacronym{E911}{E911}{enhanced 911}

\newacronym{kNN}{kNN}{k-nearest neighbors}
\newacronym{LC}{LC}{loosely-coupled}
\newacronym{TC}{TC}{tightly-coupled}
\newacronym{UTC}{UTC}{ultra-tightly-coupled}
\newacronym{SVM}{SVM}{support vector machine}
\newacronym{RL}{RL}{reinforcement learning}

\newacronym{NRTK}{NRTK}{network real-time Kinematic}
\newacronym{PDR}{PDR}{pedestrian dead reckoning}
\newacronym{SF-RTK}{SF-RTK}{single-frequency real-time kinematic}
\newacronym{VIO}{VIO}{visual-inertial odometry}
\newacronym{WLAN}{WLAN}{wireless local area network}

\newacronym{ML}{ML}{machine learning}
\newacronym{RFID}{RFID}{radio frequency identification}
\newacronym{CPP}{CPP}{carrier phase positioning}
\newacronym{DFT}{DFT}{discrete Fourier transform}
\newacronym{RFS}{RFS}{random finite sets}
\usetikzlibrary{positioning,decorations.markings,calc}

\usetikzlibrary{shapes}

\usetikzlibrary{automata,positioning}
\usetikzlibrary{backgrounds,automata}
\usetikzlibrary{arrows, arrows.meta}
\usetikzlibrary{positioning,fit,calc}
\definecolor{LightCyan}{rgb}{0.9,1,1}
\definecolor{Gray}{gray}{0.85}

 \definecolor{mygray}{gray}{0.6}
 \definecolor{darkolivegreen}{rgb}{0.33, 0.42, 0.18}

\newcommand{\rev}[1]{\textcolor{black}{#1}}

\newcommand{\rmtx}{{\rm{T}}}
\newcommand{\rmrx}{{\rm{R}}}

\newcommand{\Ntx}{N_\rmtx}
\newcommand{\Nrx}{N_\rmrx}

\newcommand{\thetab}{ \boldsymbol{\theta} }
\newcommand{\phib}{ \boldsymbol{\phi} }

\newcommand{\phiazl}{ \phi_{\rm{az},\ell} }
\newcommand{\phiell}{ \phi_{\rm{el},\ell} }

\newcommand{\thetaazl}{ \theta_{\rm{az},\ell} }
\newcommand{\thetaell}{ \theta_{\rm{el},\ell} }

\newcommand{\arx}{\mathbf{a}_\rmrx}
\newcommand{\atx}{\mathbf{a}_\rmtx}

\newcolumntype{L}{>{\raggedright\arraybackslash}X}
\renewcommand{\arraystretch}{2}

\usepackage{colortbl}
\definecolor{pastel1}{HTML}{F7914D}  
\definecolor{pastel2}{HTML}{FFE699}  
\definecolor{pastel3}{HTML}{A9D18E}  
\definecolor{pastel4}{HTML}{BB99DD}  
\definecolor{pastel5}{HTML}{BDD7EE}  

\makeglossaries
\begin{document}
\title{Vehicular Wireless Positioning -- A Survey} 
\author{Sharief Saleh,~\IEEEmembership{Member,~IEEE},
Satyam Dwivedi,~\IEEEmembership{Member,~IEEE},
Russ Whiton,~\IEEEmembership{Member,~IEEE},\\
Peter Hammarberg,
Musa Furkan Keskin,~\IEEEmembership{Member,~IEEE},
Julia Equi,
Hui Chen,~\IEEEmembership{Member,~IEEE},\\
Florent Munier,
Olof Eriksson,
Fredrik Gunnarsson,~\IEEEmembership{Senior Member,~IEEE},\\
Fredrik Tufvesson,~\IEEEmembership{Fellow,~IEEE},
and~Henk Wymeersch,~\IEEEmembership{Fellow,~IEEE}

\thanks{This work is supported by the Vinnova B5GPOS Project under Grant 2022-01640 and by the Swedish Research Council (VR) under Grants 2022-03007 and 2024-04390. Sharief Saleh's work was supported by the European Union’s Horizon 2020 research and innovation programme under Marie Skłodowska Curie Grant 101152275.}

\thanks{Sharief Saleh, Musa Furkan Keskin, Hui Chen, and Henk Wymeersch are with the Department of Electrical Engineering, Chalmers University of Technology, 41296 Gothenburg, Sweden (e-mail: sharief@chalmers.se; furkan@chalmers.se; hui.chen@chalmers.se; henkw@chalmers.se).}

\thanks{Satyam Dwivedi, Peter Hammarberg, Julia Equi, Florent Munier, and Fredrik Gunnarsson are with Ericsson Research, Sweden (e-mail: satyam.dwivedi@ericsson.com; peter.hammarberg@ericsson.com; julia.equi@ericsson.com; florent.munier@ericsson.com; fredrik.gunnarsson@ericsson.com).}

\thanks{Russ Whiton is with the European Space Agency, Noordwijk, The Netherlands (e-mail: russ.whiton@esa.int).}

\thanks{Olof Eriksson was with Magna Electronics Sweden AB, 412 50 Gothenburg, Sweden.}

\thanks{Fredrik Tufvesson is with Lund University, 223 63 Lund, Sweden (e-mail:fredrik.tufvesson@eit.lth.se).}
}
\maketitle 

\begin{abstract}
The rapid advancement of connected and autonomous vehicles has driven a growing demand for precise and reliable positioning systems capable of operating in complex environments. Meeting these demands requires an integrated approach that combines multiple positioning technologies, including wireless-based systems, perception-based technologies, and motion-based sensors. This paper presents a comprehensive survey of wireless-based positioning for vehicular applications, with a focus on satellite-based positioning (such as global navigation satellite systems (GNSS) and low-Earth-orbit (LEO) satellites), cellular-based positioning (5G and beyond), and IEEE-based technologies (including Wi-Fi, ultrawideband (UWB), Bluetooth, and vehicle-to-vehicle (V2V) communications). \rev{First, the survey reviews a wide range of vehicular positioning use cases, outlining their specific performance requirements.} Next, it explores the historical development, standardization, and evolution of each wireless positioning technology, providing an in-depth categorization of existing positioning solutions and algorithms, and identifying open challenges and contemporary trends. Finally, the paper examines sensor fusion techniques that integrate these wireless systems with onboard perception and motion sensors to enhance positioning accuracy and resilience in real-world conditions. This survey thus offers a holistic perspective on the historical foundations, current advancements, and future directions of wireless-based positioning for vehicular applications, addressing a critical gap in the literature.
\end{abstract}

\bstctlcite{IEEEexample:BSTcontrol}

\section{Introduction}  
In the early days of automobiles, accessible and precise forms of maps and navigation tools were pioneered, including the hand-cranked rolled paper maps of the Iter Avto in the 1930s~\cite{liu2020high}. The 1980s saw the advent of manually initialized and dead-reckoning systems for navigation, together with cassette tapes for storing map data, inspired by Cold War inertial navigation systems for aviation~\cite{wang2008strategic}. Civilian \gls{GPS} receivers for automotive navigation emerged in the 1990s with systems like Mazda's Eunos Cosmos and GM's Guidestar \cite{ndrivenavhistory}. These early systems established satellite navigation as the workhorse of automotive navigation for the decades that followed.
Since then, the navigation task for land vehicles has been the focus of intense work, with the vision that both the navigation and the vehicle control tasks could be automated. The main goal behind automation is improving transportation safety and efficiency, which will bring a plethora of societal, economic, and environmental benefits~\cite{bagloee2016autonomous}. To motivate researchers to achieve those goals, the DARPA Grand Challenges of 2004-2013 offered large cash prizes for vehicles that could autonomously complete various navigation challenges~\cite{seetharaman2006unmanned}. However, technical, societal, and legislative barriers have prevented the more ambitious visions of DARPA and its related challenges from being realized~\cite{lin2014emerging}.
From the technical perspective, concerns relate mainly to the perception system in general and the absolute positioning system in particular. 
Perception involves positioning the vehicle and sensing the other traffic participants and static objects (e.g., vehicles, pedestrians, cyclists, buildings, roads, and lane markings). Such information is then used for the planning and control of vehicles. Clearly, errors in positioning and sensing will lead to errors in perception, which will propagate to planning and control, causing potentially unsafe situations. 
Given the essential role of perception in automated and assisted driving, a wide range of positioning and sensing technologies are employed to provide vehicles with an accurate and timely estimate of their position and view of their surroundings. The information gathered from those technologies can be exchanged between vehicles via \gls{V2V} communication to enable cooperative perception, effectively expanding each vehicle’s field of view beyond the reach of its onboard sensors \cite{kim2014multivehicle}. Additionally, the information provided by multiple technologies can be fused to provide more robust and accurate position estimates. In the following, we further categorize these vehicular positioning and sensing technologies.

\begin{figure*}
\includegraphics[trim=4pt 8pt 6pt 8pt,clip,width=1\textwidth]{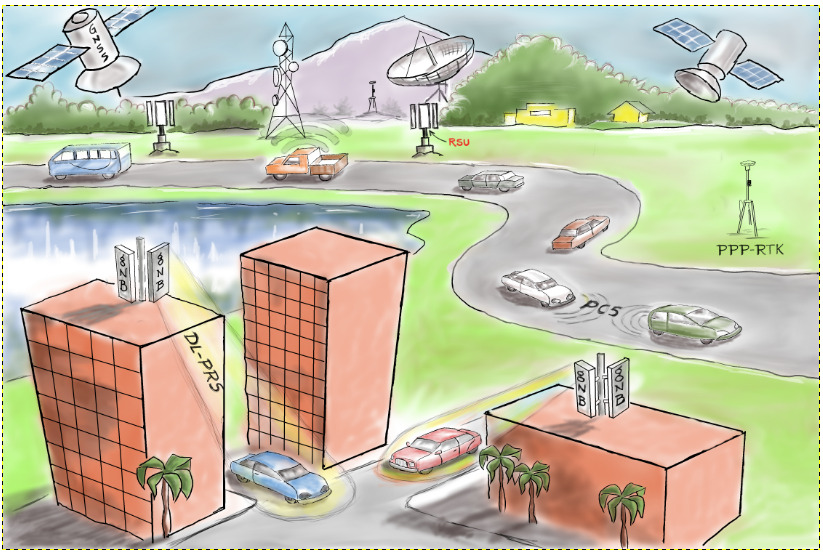}
\caption{\rev{A few scenarios of vehicular positioning: This figure illustrates key vehicular positioning scenarios, including urban, highway, and rural environments, highlighting the use of various radio technologies such as \gls{GNSS}, \gls{PPP-RTK}, 5G cellular base stations (gNB) using \gls{DL} \glspl{PRS}, \glspl{RSU}, and \gls{V2X}/PC5 and \gls{V2V} communications. The illustration sketch is courtesy of Prof. Sujit Kumar Chakrabarti, IIIT Bangalore.}}
\label{scenario_pic}
\end{figure*}

\subsection{Categories of Vehicular Positioning Technologies}
The individual positioning technologies can be broadly categorized into those that provide \textit{relative} positioning information and those that offer \textit{absolute} positioning information. In the following, we detail the two categories of positioning technologies.

\subsubsection{Relative Positioning Technologies}
In the first category, we count perception sensors such as cameras, radars, and lidars, which allow the vehicle to localize itself relative to its passive surroundings. Additionally, we count wheel odometers and \glspl{IMU}, which house accelerometers and gyroscopes, as they deliver relative information over time with respect to the initial position of the vehicle. Finally, any communication technology that supports peer-to-peer links (such as \gls{UWB}, Wi-Fi, and cellular sidelink) also yields relative positioning information, in the form of distances and angles between the connected vehicles. Generally, the relative information of all these technologies is accurate and can be provided at a high rate, but inevitably suffers from drifts, biases, and ambiguities in terms of global rotations and translations~\cite{heo2018ekf}. These can be corrected by sensors that provide absolute information. 

\subsubsection{Absolute Positioning Technologies}
The lion's share of absolute position information comes from radio signals, \rev{illustrated in Fig.~\ref{scenario_pic}}, among which we include satellite systems, like \gls{GNSS} and \gls{LEO} satellites, cellular radio systems, predominantly composed of 5G networks, and IEEE-based technologies, such as Wi-Fi, \gls{UWB}, and Bluetooth. \gls{GNSS} and \gls{LEO} satellites provide absolute position information, with worldwide coverage but varying degrees of accuracy, depending on the signal processing, side information, multipath, and satellite visibility~\cite{prol2022position}. Cellular networks and Wi-Fi can provide absolute position information in the form of distance and angle measurements with respect to cellular base stations or Wi-Fi anchors. These measurements come nearly for free, as the infrastructure is already deployed for communication services. On the other hand, \gls{UWB} provides extremely accurate ranging information at short distances, with excellent delay resolution, supporting vehicular positioning in cluttered and even indoor parking environments~\cite{alarifi2016ultra}. However, \gls{UWB} does require significant infrastructure investment, due to the short range. Finally, we also count high-definition (HD) maps that can be used in conjunction with cameras, radars, or lidars, to match the perceived surrounding landmarks to objects in the map, thereby providing an absolute position estimate \cite{liu2020high}.

\begin{table*}[htbp]
\centering
\caption{\rev{Comparison of General-Application, Indoor, and IoT-Focused Positioning Surveys.}}
\label{tab:survey_comparison_general}
\newcommand{\cmark}{\checkmark}
\newcommand{\xmark}{\times}
\renewcommand{\arraystretch}{1.2} 
\small
\begin{tabular}{|l|c|l|c|c|c|c|c|}
\hline
\textbf{Reference} & \textbf{Year} & \textbf{Application/Specialty} & \textbf{Satellite} & \textbf{Cellular} & \textbf{IEEE-based} & \textbf{V2X} & \textbf{Sensor Fusion} \\
\hline
\hline
\cite{davidson_survey_2017} & 2017 & Indoor, smartphone & $\xmark$ & $\xmark$ & $\cmark$ & $\xmark$ & $\cmark$ \\ \hline
\cite{liu_survey_2020} & 2020 & Indoor, general & $\xmark$ & $\xmark$ & $\cmark$ & $\xmark$ & $\xmark$ \\ \hline
\cite{mazhar_precise_2017} & 2017 & Indoor, general & $\xmark$ & $\xmark$ & $\cmark$ & $\xmark$ & $\xmark$ \\ \hline
\cite{elsanhoury_precision_2022} & 2022 & Indoor, logistics & $\xmark$ & $\xmark$ & $\cmark$ & $\xmark$ & $\cmark$ \\ \hline
\cite{motroni_survey_2021} & 2021 & Indoor, vehicular & $\xmark$ & $\xmark$ & $\cmark$ & $\xmark$ & $\cmark$ \\ \hline
\cite{saeed2019state} & 2019 & IoT, wireless sensor networks & $\xmark$ & $\xmark$ & $\cmark$ & $\xmark$ & $\xmark$ \\ \hline
\cite{han2016survey} & 2016 & IoT, wireless sensor networks & $\xmark$ & $\xmark$ & $\xmark$ & $\cmark$ & $\xmark$ \\ \hline
\cite{del_peral-rosado_survey_2018} & 2018 & General, 1G-5G & $\xmark$ & $\cmark$ & $\xmark$ & $\xmark$ & $\cmark$ \\ \hline
\cite{chen2022tutorial} & 2022 & General, THz & $\xmark$ & $\cmark$ & $\xmark$ & $\xmark$ & $\xmark$ \\ \hline
\cite{zhuang_survey_2018} & 2018 & General, light-based & $\xmark$ & $\xmark$ & $\cmark$ & $\cmark$ & $\cmark$ \\ \hline
\cite{Quoc2016survey} & 2016 & Outdoor, fingerprinting & $\xmark$ & $\xmark$ & $\cmark$ & $\xmark$ & $\cmark$ \\ \hline
\cite{laoudias_survey_2018} & 2018 & Indoor, multiple technologies & $\xmark$ & $\cmark$ & $\cmark$ & $\xmark$ & $\cmark$ \\ \hline
\cite{shastri2022review} & 2022 & Indoor, mmWave & $\xmark$ & $\cmark$ & $\xmark$ & $\xmark$ & $\xmark$ \\ \hline
\end{tabular}
\end{table*}

\begin{table*}[htbp]
\centering
\caption{Comparison of Vehicular-Focused Positioning Surveys.}
\label{tab:survey_comparison_vehicular}
\newcommand{\cmark}{\checkmark}
\newcommand{\xmark}{\times}
\renewcommand{\arraystretch}{1.2} 
\small
\begin{tabular}{|l|c|l|c|c|c|c|c|}
\hline
\textbf{Reference} & \textbf{Year} & \textbf{Application/Specialty} & \textbf{Satellite} & \textbf{Cellular} & \textbf{IEEE-based} & \textbf{V2X} & \textbf{Sensor Fusion} \\
\hline
\hline
\cite{zhu_gnss_2018} & 2018 & Vehicular, integrity & $\cmark$ & $\xmark$ & $\xmark$ & $\xmark$ & $\xmark$ \\ \hline
\cite{bresler_gnss_2016} & 2016 & Vehicular, NLoS & $\cmark$ & $\xmark$ & $\xmark$ & $\xmark$ & $\cmark$ \\ \hline
\cite{prol_position_2022} & 2022 & Vehicular, LEO & $\cmark$ & $\xmark$ & $\xmark$ & $\xmark$ & $\cmark$ \\ \hline
\cite{jin_survey_2024} & 2024 & Vehicular, cooperative & $\cmark$ & $\xmark$ & $\xmark$ & $\cmark$ & $\cmark$ \\ \hline
\cite{zhou20245g} & 2024 & Vehicular, cooperative & $\xmark$ & $\cmark$ & $\xmark$ & $\cmark$ & $\cmark$ \\ \hline
\cite{adegoke_infrastructure_2019} & 2019 & Vehicular, Wi-Fi & $\cmark$ & $\xmark$ & $\cmark$ & $\cmark$ & $\cmark$ \\ \hline
\cite{de_ponte_muller_survey_2017} & 2017 & Vehicular, relative positioning& $\cmark$ & $\xmark$ & $\cmark$ & $\cmark$ & $\cmark$ \\ \hline
\cite{kuutti_survey_2018} & 2018 & Vehicular, onboard sensors & $\cmark$ & $\xmark$ & $\cmark$ & $\cmark$ & $\cmark$ \\ \hline
\cite{kumar_survey_2023} & 2023 & Vehicular, onboard sensors & $\cmark$ & $\xmark$ & $\xmark$ & $\cmark$ & $\cmark$ \\ \hline
\cite{chalvatzaras_survey_2023} & 2023 & Vehicular, onboard sensors & $\cmark$ & $\xmark$ & $\xmark$ & $\xmark$ & $\cmark$ \\ \hline
\cite{shan_survey_2023} & 2023 & Vehicular, onboard sensors & $\cmark$ & $\xmark$ & $\cmark$ & $\cmark$ & $\cmark$ \\ \hline
\cite{lu_real-time_2022} & 2022 & Vehicular, real-time performance & $\cmark$ & $\xmark$ & $\cmark$ & $\cmark$ & $\cmark$ \\ \hline
\cite{laconte_survey_2022} & 2022 & Vehicular, high-way scenarios & $\cmark$ & $\xmark$ & $\xmark$ & $\xmark$ & $\cmark$ \\ \hline
\cite{yeong_sensor_2021} & 2021 & Vehicular, perception sensor fusion & $\xmark$ & $\xmark$ & $\xmark$ & $\xmark$ & $\cmark$ \\ \hline
\cite{wang_multi-sensor_2020} & 2020 & Vehicular, sensor fusion & $\cmark$ & $\xmark$ & $\cmark$ & $\cmark$ & $\cmark$ \\ \hline
\cite{fayyad_deep_2020} & 2020 & Vehicular, deep learning fusion & $\cmark$ & $\xmark$ & $\xmark$ & $\xmark$ & $\cmark$ \\ \hline
\cite{ounoughi_data_2023} & 2023 & Vehicular, sensor fusion & $\cmark$ & $\sim$ & $\sim$ & $\cmark$ & $\cmark$ \\ \hline
\cite{butt_integration_2022} & 2022 & Vehicular, sensor fusion & $\sim$ & $\sim$ & $\cmark$ & $\cmark$ & $\cmark$ \\ \hline
\rowcolor{gray!25}\textbf{This Survey} & 2025 & Vehicular & $\cmark$ & $\cmark$ & $\cmark$ & $\cmark$ & $\cmark$ \\ \hline
\end{tabular}
\end{table*}

\subsection{Previous Surveys and Contributions}
Due to the critical role of positioning in the perception, guidance, and planning of autonomous vehicles, a comprehensive survey on this topic is both timely and relevant. Numerous surveys have documented the vast landscape of positioning technologies for various applications. Here, we divide these surveys into general positioning surveys, \rev{summarized in Table~\ref{tab:survey_comparison_general}}, and vehicular-focused surveys, summarized in Table~\ref{tab:survey_comparison_vehicular}. 

\subsubsection{General Positioning Surveys}
A significant portion of general positioning surveys addresses the challenges of \textit{indoor} positioning, where \gls{GNSS} signals are unavailable. These surveys provide extensive reviews of various technologies such as Wi-Fi \cite{davidson_survey_2017, liu_survey_2020}, \gls{UWB} \cite{mazhar_precise_2017, elsanhoury_precision_2022}, and \gls{RFID} \cite{motroni_survey_2021}. Similarly, surveys on wireless sensor networks, like \cite{saeed2019state,han2016survey}, often focus on \gls{IoT} applications, which have different constraints and objectives than vehicle positioning. While foundational for understanding specific technologies, these indoor- and \gls{IoT}-centric surveys do not address the unique challenges of the vehicular domain, such as high speeds and complex, rapidly changing environments.

\begin{figure*}[t!]
    \centering
\includegraphics[width=0.9\textwidth]{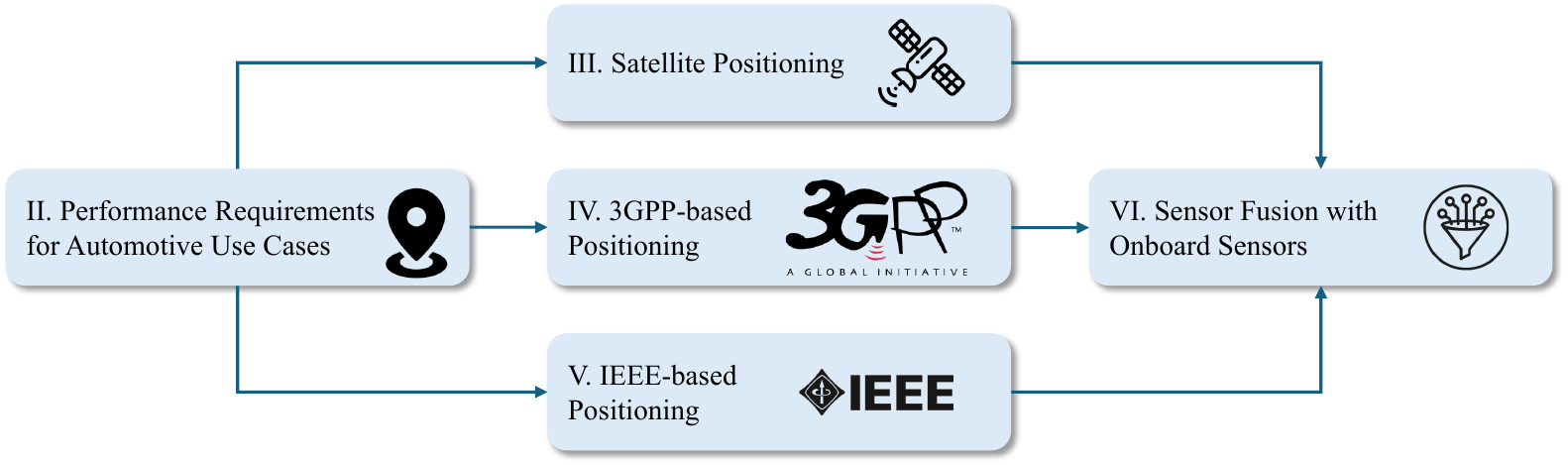}
    \caption{\rev{Overview of the main sections of the paper.}}
    \label{fig:structure}
\end{figure*}

As the focus shifts to outdoor applications, the literature covers wireless technologies that have wider coverage. For instance, \cite{del_peral-rosado_survey_2018} offers a thorough overview of cellular positioning technologies up to 5G but does not specialize in vehicular-specific needs. More recent works on 6G focus on terahertz bands suited for short-range applications \cite{chen2022tutorial}. Other technologies have also been reviewed individually; for instance, visible light positioning is explored in \cite{zhuang_survey_2018}, but its reliance on consistent lighting conditions limits its utility for general vehicular use. Likewise, outdoor Wi-Fi-based fingerprinting localization methods were surveyed in \cite{Quoc2016survey}, among other perception-based fingerprinting methods, without focusing on the diverse requirements of vehicular applications. More comprehensive surveys can be found in \cite{laoudias_survey_2018} and \cite{shastri2022review}, which cover a broad spectrum of radio-based positioning technologies (cellular and IEEE-based) and mmWave-specific positioning technologies (5G and radars), respectively. However, these works also lack a specific focus on vehicular applications and their unique challenges.

\subsubsection{Vehicular-Focused Surveys}
A growing number of surveys directly address the topic of vehicular localization. They can be generally divided into wireless-focused surveys \cite{zhu_gnss_2018, bresler_gnss_2016,prol_position_2022, jin_survey_2024, zhou20245g,adegoke_infrastructure_2019}, onboard/perception sensors-focused surveys \cite{de_ponte_muller_survey_2017, kuutti_survey_2018, kumar_survey_2023, chalvatzaras_survey_2023, shan_survey_2023, lu_real-time_2022, laconte_survey_2022}, and sensor fusion-focused surveys \cite{yeong_sensor_2021, wang_multi-sensor_2020,fayyad_deep_2020, ounoughi_data_2023, butt_integration_2022}. 

Wireless-focused vehicular positioning surveys mostly comprise satellite-based positioning surveys, and they cover a broad range of topics. For instance, critical issues of \gls{GNSS} integrity and \gls{NLoS} conditions in urban environments were covered in \cite{zhu_gnss_2018, bresler_gnss_2016}. Moreover, works on \gls{LEO} satellite \gls{PNT} were covered in \cite{prol_position_2022}. Cooperative \gls{GNSS} techniques for vehicular networks were covered in \cite{jin_survey_2024}. Likewise, cooperative localization within 5G networks was also covered in \cite{zhou20245g}. Furthermore, Wi-Fi-based positioning for vehicular applications was reviewed in \cite{adegoke_infrastructure_2019}. However, none of these surveys focuses on works that combine multiple wireless technologies.

Onboard sensor-focused surveys tackle various sensors and positioning aspects and scenarios. For instance, \cite{de_ponte_muller_survey_2017}, \cite{kuutti_survey_2018}, and \cite{kumar_survey_2023} cover relative positioning using perception sensors and \gls{V2V} communications. Authors in \cite{chalvatzaras_survey_2023} reviewed works on map-based localization techniques using perception sensors that sometimes utilize \gls{GNSS} signals. In \cite{shan_survey_2023}, the authors reviewed the performance of various wireless, perception-based, and inertial positioning systems. Likewise, \cite{lu_real-time_2022} performed a similar performance analysis on real-time positioning techniques and technologies. Finally, positioning techniques and technologies for highway scenarios were covered in \cite{laconte_survey_2022}. It is worth noting that none of the aforementioned surveys covered all wireless-based positioning technologies, especially cellular positioning, which was missing in most works.

Finally, as sensor fusion was identified as an indispensable component for achieving the accuracy and robustness required for autonomous operation, numerous research works and surveys covered the topic. In \cite{yeong_sensor_2021}, authors primarily reviewed sensor fusion techniques for perception-based positioning systems, with a very minor presence of wireless technologies. The review found in \cite{wang_multi-sensor_2020} is more comprehensive, as it covers sensor fusion of other sensors, including wireless technologies, except for cellular positioning. \cite{fayyad_deep_2020}, on the other hand, covered deep learning approaches for perception-based positioning, which included fusion with satellite-based positioning. Other surveys, like \cite{ounoughi_data_2023} and \cite{butt_integration_2022}, provide a broader perspective on data fusion for \gls{ITS}-based services in general, which includes vehicular positioning. However, none of these works covered all of the major wireless positioning technologies, especially cellular positioning.

This overview highlights a clear gap: surveys in this field either (i) focus on general application of positioning without explicit focus on vehicular applications and their unique demands; (ii) focus on a single wireless technology while neglecting other technologies; or (iii) focus on standalone solutions without discussing sensor fusion with other technologies. To address these gaps, \rev{this paper first presents an overview of vehicular use cases and their specific positioning requirements.} It then provides a comprehensive survey of the major existing wireless positioning solutions for vehicles, covering both proprietary technologies (e.g., satellite-based) and standardized solutions, including \gls{3GPP}-based and IEEE-based technologies. For each technology, this paper presents its historical evolution and standards, outlines its positioning fundamentals, reviews its contemporary advancements, and identifies its key open challenges and emerging trends. Finally, a dedicated section explores sensor fusion approaches that integrate wireless positioning with onboard vehicular sensors to enhance positioning accuracy and robustness in real-world scenarios.

\subsection{Survey Outline}
The remainder of the paper is structured as follows \rev{(see also Fig.~\ref{fig:structure}). Section \ref{Sec: requirements} presents key usage scenarios and performance requirements for vehicular positioning systems, with a focus on accuracy, reliability, and integrity in safety-critical applications.} Section \ref{Sec: satellite} delves into satellite-based positioning technologies, including the role of \gls{GNSS}, \gls{RTK}, \gls{PPP}, and the emerging use of  \gls{LEO} satellites for vehicular positioning. Section \ref{Sec: cellular} explores cellular network-based positioning, with a focus on \gls{mmWave} 5G systems and their potential for high-accuracy positioning in vehicular environments. Section \ref{Sec: IEEE} examines IEEE-based positioning technologies, including Wi-Fi, \gls{UWB}, and Bluetooth, highlighting their application in niche vehicular scenarios. Section \ref{Sec: sensor fusion} discusses sensor fusion techniques, integrating individual radio technologies with onboard perception and motion sensors like cameras, lidars, radars, and \glspl{IMU}, to improve positioning accuracy and robustness. Finally, Section \ref{Sec: outlook} summarizes the general research directions in the field of wireless vehicular positioning.

\section{\rev{Performance Requirements for Automotive Use Cases}} \label{Sec: requirements}
\rev{Modern vehicles are equipped with a wide range of positioning technologies that enable diverse use cases aimed at enhancing both \textit{safety} and \textit{efficiency} in modern transportation systems. Safety-oriented use cases primarily focus on reducing traffic accidents and lowering fatality rates, while efficiency-oriented use cases aim to minimize carbon emissions and alleviate traffic congestion, thereby improving economic and environmental outcomes. These applications can be implemented either by directly controlling autonomous vehicles or by broadcasting and sharing positional and sensing information with other vehicles and users. In the first case, ego positioning is essential for services like navigation \cite{CLAUSSEN1991}, active safety applications \cite{piao2010vehicle}, and vehicle automation \cite{kuutti2018survey,yurtsever2020survey}. In the second case, cooperative traffic operations are enabled through \gls{C-ITS} using \gls{V2X} communication \cite{Sjoberg2017}. These cooperative systems support services such as emergency braking alerts, vehicle platooning, and location-based traffic information \cite{V2X_use_cases_5gaa}. The performance requirements of positioning systems—such as accuracy, uncertainty, and latency—vary significantly across these applications. Safety-critical use cases, for instance, demand higher precision and lower latency than efficiency-oriented ones. Similarly, applications that control autonomous vehicles have stricter positioning requirements than those that rely on broadcasting event information. This section provides an overview of the various vehicular use cases and their positioning requirements. Specifically, general key performance indicators are discussed in Section~\ref{sec:KPIs}, followed by a deep dive into positioning integrity metrics in Section~\ref{sec:integrity}, and an examination of relevant requirements for selected use case examples in Section~\ref{sec:scenarios}.}

\subsection{\rev{Positioning Performance Indicators}} \label{sec:KPIs}
\rev{When assessing the performance of a positioning system, several \glspl{KPI} are crucial for ensuring accuracy, reliability, and efficiency. Good summaries can be found in the literature, e.g., for \gls{GNSS} \cite{GSA2021} and cellular positioning \cite{22872, hexax2022d31, V2X_position_metrics_5gaa}, and a summary is provided in Table~\ref{tab:kpi}. We divide these \glspl{KPI} into four broad sets, namely performance-related, reliability-related, timeliness-related, and scalability-related \glspl{KPI}.}

\begin{table}[t]
    \centering
    \caption{\rev{Positioning Performance Indicators.}}
    \label{tab:kpi}
    \footnotesize
    \begin{tabular}{
            |m{0.14\textwidth}|
            m{0.24\textwidth}|
        }
        \hline
        \textbf{KPI Set} & \textbf{Performance Indicators} \\
        \hline
        \rowcolor{pastel2}
        Performance KPIs & \vspace{1mm}
        \begin{itemize}[leftmargin=*]
            \item Accuracy (RMSE, CEP)
            \item Precision (GDOP)
            \item Resolution \vspace{-3mm}
        \end{itemize} \\
        \hline
        \rowcolor{pastel3}
        Reliability KPIs & \vspace{1mm}
        \begin{itemize}[leftmargin=*]
            \item Availability
            \item Continuity
            \item Robustness
            \item Integrity\vspace{-3mm}
        \end{itemize} \\
        \hline
        \rowcolor{pastel1}
        Timeliness KPIs & \vspace{1mm}
        \begin{itemize}[leftmargin=*]
            \item Latency
            \item Time to First Fix
            \item Measurement and Update Rate\vspace{-3mm}
        \end{itemize} \\
        \hline
        \rowcolor{pastel5}
        Scalability KPIs & \vspace{1mm}
        \begin{itemize}[leftmargin=*]
            \item Radio Resource Utilization
            \item Power Consumption
            \item Computational Complexity
            \item Cost Effectiveness\vspace{-3mm}
        \end{itemize} \\
        \hline
    \end{tabular}
\end{table}

\rev{The first set of \glspl{KPI}, arguably the most focused on \gls{KPI} set in the literature, encompasses accuracy, precision, and resolution. Accuracy measures how closely the system's position estimates match the true geographical coordinates. This can be quantified using metrics like the \gls{RMSE} and the circular error probability. Ultimately, the achievable accuracy decides the applicability of the system for a specific service or use case. In this review, we divide positioning accuracy into the following levels: centimeter-level ($\varepsilon<$10 cm), decimeter-level (10 cm $\leq \varepsilon <$30 cm), sub-meter-level (30 cm $\leq \varepsilon <$1 m), meter-level (1 m $\leq \varepsilon <$2 m), few/several meters (2 m $\leq \varepsilon <$10 m), and tens of meters (10 m $\leq \varepsilon <$100 m), where $\varepsilon$ is the accuracy reported. Precision refers to the consistency of position estimates when measurements are taken under the same conditions. How precision of specific measurements impacts the positioning accuracy can be expressed in the \gls{GDOP} \cite{langley1999dilution}, which, for example, is impacted by the geometry of the reference nodes in radio-based positioning. Resolution refers to the smallest distinguishable unit or the smallest change the system can detect. This metric is more relevant when discussing specific measurements, e.g., the delay or the angle of an incoming radio signal.}

\rev{The second set of \glspl{KPI}, which captures the reliability of a positioning system, encompasses the system's availability, continuity, robustness, and integrity. Availability represents the proportion of time the positioning system is operational and can provide position estimates. High availability is essential for applications requiring continuous position tracking, and can be measured by calculating the up-time percentage and tracking the duration of system outages.
Continuity is a related concept, as it measures how often the system is interrupted. Robustness captures the system's resilience towards external interferences like jammers and spoofers. 
Integrity ensures the trustworthiness of position information. 
This includes the system's ability to provide timely warnings when it cannot meet accuracy requirements. Integrity is particularly important for liability and safety-critical applications, including automated driving. Metrics for integrity include protection levels and integrity risk, which measure the probability that the system provides incorrect position estimates without warning. Due to the importance of integrity to vehicle navigation and safety, this topic is discussed in further detail in \ref{sec:integrity}.}

\rev{The third set of \glspl{KPI}, relating to timeliness, incorporates latency, \gls{TTFF}, and measurement/update rate. Latency can have various definitions based on the context. For instance, latency can measure the delay between the positioning request and the time it is reported by the positioning system. Also, it can measure the processing time between acquiring the geometric measurements and delivering the positioning estimate to the system. Low latency is crucial for real-time applications like navigation and tracking, where outdated position information can lead to errors or inefficiencies. 
\gls{TTFF} indicates how fast the system can acquire a position after being turned on. This is crucial for applications where quick position acquisition is needed. 
The measurement rate is the frequency at which geometric measurements are aquired, like ranges and angles. On the other hand, update rate refers to the frequency at which position estimates are delivered. Both rates are important when vehicles are expected to experience a high rate of dynamics (e.g., taking sharp turns or driving on highways).}

\rev{The fourth set of \glspl{KPI} focuses on the scalability of the positioning system, evaluating its efficiency and practicality across various dimensions. Radio resource utilization assesses how effectively the system uses available spectrum and manages interference. Power consumption metrics examine the energy required for positioning operations, crucial for battery-powered and electric vehicles. Computational complexity analyzes the processing overhead and algorithmic efficiency of positioning techniques. Cost considerations include infrastructure deployment expenses, technology costs, and radio resource costs per device and in total, as well as the long-term economic viability of the positioning solution.}

\subsection{\rev{Position Integrity}}
\label{sec:integrity}
\rev{Positioning accuracy, timely delivery of position information, and scalability of positioning systems are all fundamental for enabling positioning services in general. However, for safety and liability-critical applications, in particular, great attention needs to be placed on the tail of the underlying position error distribution \cite{Cosmen2008integrity}. For safety-critical services, a failure in detecting a too-large erroneous position estimate could put lives at risk, and a failure in case of a liability-critical service can have severe monetary or legal implications. Hence, emphasis needs to be placed on detecting and mitigating potentially rare error events.}

\begin{figure}
    \centering
    \includegraphics[width=0.95\linewidth]{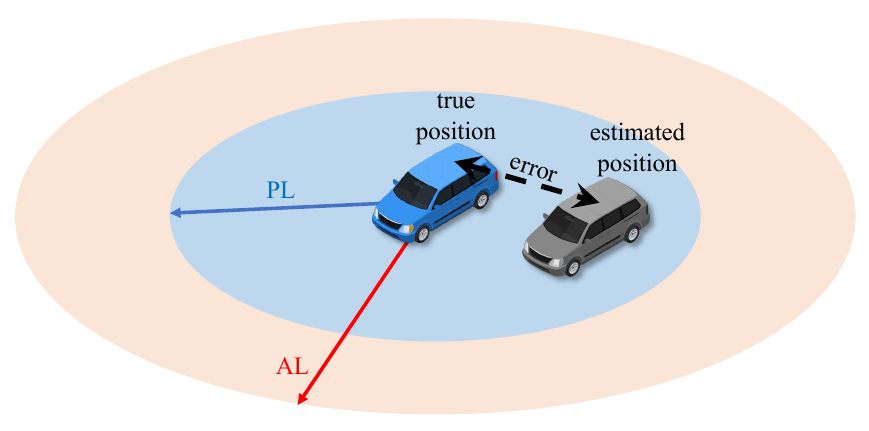}
    \caption{\rev{The relationship between the integrity parameters PL and AL.}}
    \label{fig:integrity_PL_AL}
\end{figure}

\begin{figure}
    \centering
    \includegraphics[width=0.95\linewidth]{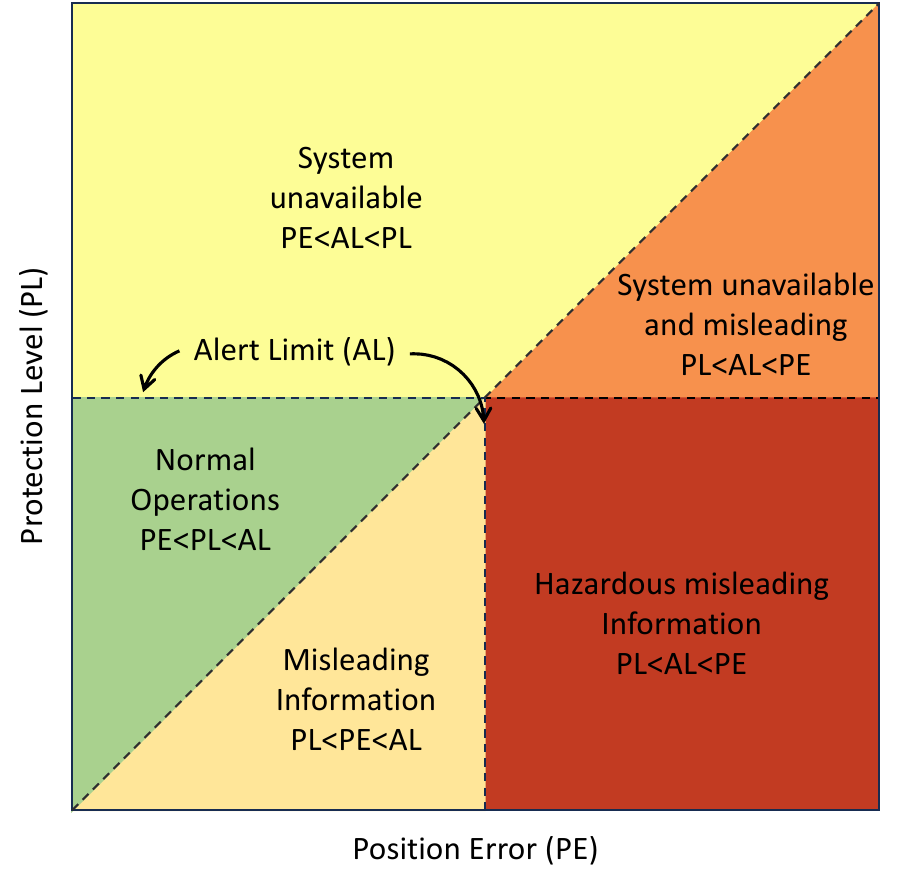}
    \caption{\rev{The Stanford diagram showing the relationship between the system availability, positioning error, and integrity parameters PL and AL.}}
    \label{fig:Stanford_diagram}
\end{figure}

\rev{To handle very strict functional safety aspects of positioning, the concept of position integrity was developed and formalized for \gls{GNSS} in the area of aviation \cite{larson2018gaussian,Zabalegui2020}. For aviation landing systems, the concept of error overbounding is used for \textit{approach and landing} systems \cite{larson2018gaussian}. Monitoring and error overbounding are needed to secure an acceptably low fail-rate of the system \cite{Rife2012,Shao2024}. To do that, an integrity framework was introduced, which enables the assessment of the reliability and trustworthiness of the information provided by a positioning system and allows a vehicle to make informed decisions based on the output. The position integrity framework relies on two key concepts, namely the \gls{AL} and \gls{PL}, illustrated in Fig.~\ref{fig:integrity_PL_AL}. The \gls{AL} indicates the maximum allowable positioning error (PE) produced by the positioning system, and the \gls{PL} represents the margin of safety that must be maintained to ensure that the target integrity risk remains below acceptable limits given the current \gls{AL}.
The target integrity risk would be the output of a hazard analysis and risk assessment, taking all potentially dangerous outcomes and their individual risks and probabilities into account \cite{Khastgir2017,whiton2022cellular}.
The relationship between the position system availability and the above integrity parameters is visualized through the Stanford diagram shown in Fig.~\ref{fig:Stanford_diagram}.
Further reading on integrity can be found in \cite{38857, Reid2019requirements,zhu_gnss_2018,whiton2022cellular,whiton2024dude}. 
While aviation integrity monitoring has a decades-long history \cite{hegarty2015rtca}, more general sensor integrity monitoring has become a subject of more intense research in recent years \cite{de2022recent}.
This also includes cellular positioning, where an integrity framework was adopted as part of the \gls{3GPP}'s Release 18 specification \cite{38.859}.}

\subsection{\rev{Use Case Positioning Requirements}}
\label{sec:scenarios}
\rev{Setting requirement values or even quantifying the \gls{KPI} attributes listed in Table~\ref{tab:kpi} is a difficult task, which ultimately depends on how the position estimate provided by the system is integrated into the vehicle overall positioning solution.
Still, efforts have been made in both industrial collaboration and standardization organizations, as well as in various academic projects and publications.
As discussed at the beginning of this section, there are a number of vehicular applications related to both traffic safety and efficiency requiring estimates of the vehicle position.
Naturally, requirements are dependent on the specific service that is being targeted, and the different individual components of the system will face different requirements depending on the overall systematization and on the performance of the other involved components being part of the positioning system.}

\rev{For traffic safety, \cite{Williams2021requirements} discusses requirements for a number of applications, including collision warnings, vehicle approaching indication, and restricted lane warnings. In this work, the positioning accuracy is categorized into three levels: coarse, lane-level, and where-in-lane, with accuracy requirements ranging from 0.1-10 m (95th percentile). The update-rate requirements are between 0.1-10 Hz. Most stringent requirements are observed for collision warnings, where the relative distance between vehicles is needed. 
Safety-related applications are also discussed in \cite{HEIGHTS2018requirements}, covering, e.g., vulnerable road user positioning. Requirements for vulnerable road user positioning accuracy is 0.2 m (95th percentile), while integrity requirements for \gls{AL} is 0.4 m, and the \gls{TTA} requirement is 1 s.}

\rev{For applications related to traffic efficiency, information on requirements can be found in \cite{V2X_use_cases_5gaa,Williams2021requirements,Stephenson2016requirements}. In \cite{V2X_use_cases_5gaa}, a large set of applications is reviewed and service level requirements are provided, including those for group start, cooperative lane merge, and automated intersection crossing. Requirements are expressed in position accuracy with reliability, where cooperative procedures pose strict requirements of 0.15 m ($3\sigma$).  Similarly, \cite{Williams2021requirements} elaborates on the requirements of a number of applications, including vehicle platooning, which requires similar accuracy at a positioning reporting rate of 100 Hz.}

\rev{Finally, the area of vehicle automation is covered in detail in \cite{Reid2019requirements}, with emphasis on integrity. Longitudinal, lateral, and vertical localization error bounds (alert limits) and 95\% accuracy requirements are derived for different road and vehicle types. Requirements include, for a mid-size vehicle operating on a freeway, lateral accuracy of 0.24 m, with an alert limit of 0.72 m, delivered at a fail-rate of $10^{-8}$/hour.}

\section{Satellite Positioning}\label{Sec: satellite}
\gls{GNSS} receivers have been the backbone of automotive navigation systems since the early 1990s and have transitioned from a luxury feature to mandatory on all vehicles in major markets to comply with emergency responder systems including eCall in the European Union \cite{ecall_eu} and \gls{E911} in the United States \cite{e911_usa}. The ability to correct long-term biases of other sensors in a global reference frame \cite{GrovesPrinciples} was fundamental in enabling the transition from attempts at bespoke dead reckoning-based navigation systems to making it a standard feature for mid-tier and premium segment vehicles today \cite{liu2020high}.
The major \gls{GNSS} constellations, operating in \gls{MEO}, are the dominant source of all \gls{PNT} today. However, the most exciting development in satellite positioning is closer to Earth, in \gls{LEO}. In this section, the history, opportunities, and challenges for each are addressed. Finally, references are given to give some intuition for the performance of modern commercial satellite positioning systems for vehicular use cases.

\subsection{History of Satellite Positioning} The first man-made satellite, Sputnik 1, was launched into \gls{LEO} in 1957 by the Soviet Union. Observers in the United States monitoring the changing Doppler frequency during the orbital passes quickly realized that users on Earth could use artificial satellites with known orbital parameters as references for determining their own location on the Earth. The TRANSIT system (with the first satellite launched already in 1959) provided this service for American Naval ships and submarines through cumbersome observations of Doppler frequency \cite{psiaki2021navigation}. Components of other American Department of Defense programs, including the Navy's Timation project and the Air Force's ``Project 621B", were integrated into a new system, the \gls{GPS}, which launched its first satellite in 1978 and quickly rendered TRANSIT irrelevant \cite{parkinson1997origins}. Users could determine position significantly more quickly and more accurately by using ranging to multiple satellites rather than by taking long readings of Doppler measurements. This was especially relevant for \gls{MEO}, which had geometries that changed much more slowly than the original \gls{LEO} orbits of the first satellite systems.

\subsubsection{MEO Satellites} Although \gls{GPS} is at its core a military system, civilian signals have been broadcast since the beginning\footnote{Not after the tragic 1983 downing of Korean Air Flight 007, as is commonly erroneously reported \cite{korean_air}.} and the other constellations have followed suit in providing civilian signals.  The second major system to come online was GLONASS,  initiated by the Soviet Union in the late 20th century and completed by Russia to achieve global coverage in 2011. BeiDou-3 achieved complete constellation status by the Chinese military in 2020.  The European Space Agency's civilian system, Galileo, is widely used, though it has not formally reached the status of full operational capability (FOC) as of this writing. 

\subsubsection{Regional Systems} In addition to these four systems with global coverage, there are regional complementary satellite-based augmentation systems (SBAS) visible only regionally for ground users. SBAS can provide additional ranging references, higher integrity operation, and other data transfer services. These include the Japanese quasi-zenith satellite System (QZSS), American wide area augmentation system (WAAS), European geostationary navigation overlay service (EGNOS), among others in operation and in development.

\subsubsection{LEO Satellites} The introduction of \gls{LEO} megaconstellations for communication, including Starlink and Iridium, has resulted in an explosion of interest in utilizing these satellites for navigation \cite{prol_position_2022}.  There are obvious benefits compared to the \gls{MEO} used by the dedicated \gls{GNSS} constellations, such as the reduced expense of launching \gls{LEO} satellites, significantly smaller path loss, and the opportunity to deploy signals more flexibly and rapidly.  These advantages have even inspired commercial interest in developing dedicated \gls{LEO} constellations for navigation \cite{reid_leo_commercial}, and Iridium has even deployed the \gls{STL} services in order to offer some back-up positioning and timing services to users, and to offer some satellite functionality to users in attenuated environments \cite{reid2020navigation}. In this manner, research and development in satellite navigation have come back down to the orbits of the systems of the 1950s and 1960s after half a century of focus on \gls{MEO}.

\subsection{Satellite Positioning Fundamentals}
\begin{figure}
\begin{center}
\includegraphics[width=0.4\textwidth]{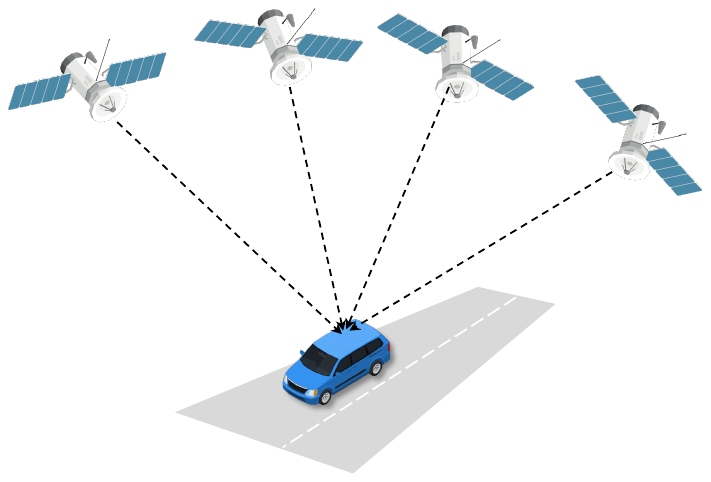}
\caption{\rev{Multilateration with satellite references, the method behind \gls{GNSS} positioning, the receiver estimates ranges to three or more satellites. In practice, receiver clock drift necessitates four or more references to solve a four-plus variable problem with ``pseudoranges", including clock offset.}}
\label{fig:multilateration_gnss}
\end{center}
\end{figure}

Single-point \gls{MEO} \gls{GNSS} receiver-derived positions are based on the method of multilateration. Observations of distances to three or more reference objects at known locations in a global frame are sufficient to solve for user position \cite{borre2012algorithms}, \rev{as illustrated in Figure~\ref{fig:multilateration_gnss}}. Each satellite broadcasts a low data-rate description of where it is (the navigation message), multiplexed with ranging codes, pseudorandom noise sequences (PRNs)\footnote{GLONASS is the exception in that it uses frequency division multiple access (FDMA), though the constellation has moved towards using code division multiple access (CDMA) as the other constellations do \cite{urlichich2011glonass}.} known a priori at the receiver to allow for identification and ranging \cite{doberstein2011fundamentals}.
In practice, determination of distance is done based on clock synchronization and \gls{ToA}. However, users typically employ inexpensive crystal oscillators that drift at the equivalent light speed distance of hundreds of meters per second, meaning that time synchronization is not nearly good enough to estimate position with any reasonable accuracy. To negate this hardware limitation, a fourth satellite must be used, and the unknown state of the receiver is formulated with four variables: three-dimensional position and an associated clock drift. The clock drift is represented in all the observable range measurements from the satellites, and it is for this reason that the range measurements are called ``pseudoranges".

For \gls{LEO} positioning, this pseudorange model can also be employed, but observations of Doppler shift are also commonly used, which are larger in absolute terms than for \gls{MEO} \cite[Section III.A]{stock2024survey}. This represents more closely the original TRANSIT system, see \cite[Section 2]{psiaki2021navigation} for an explicit mathematical description of Doppler positioning techniques. \rev{Readers interested in a generalized channel model for wireless technologies and how delays and Dopplers affect the received signals are referred to Appendix \ref{appendix}.}

\subsection{MEO-based Positioning}
\subsubsection{MEO Satellite Challenges}
\begin{figure}
\begin{center}
\includegraphics[width=0.45\textwidth]{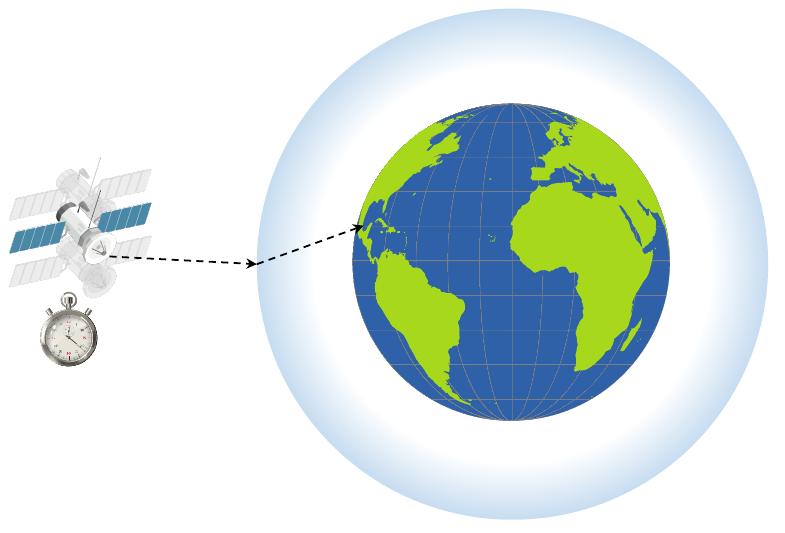}
\caption{\rev{Major sources of \gls{GNSS} position error - Satellite orbital and clock errors, atmospheric errors, and multipath propagation (not pictured).}}
\label{fig:physical_errors_gnss}
\end{center}
\end{figure}

There are some well-studied and modeled physics challenges that limit real-time performance of standalone \gls{GNSS} receivers even for high-end hardware, \rev{which are illustrated in Figure~\ref{fig:physical_errors_gnss}}. Solving these problems is fundamental for precision operation, whether it be for scientific applications or commercial users with high accuracy requirements in fields like surveying or precision agriculture. The navigation message broadcast by satellites includes estimations of these parameters, but the frequency of updates and the accuracy of the models have bottlenecks in uploading the information to the satellite as well as communicating to users owing to the low data rate of the navigation message.  

\textbf{Orbital Inaccuracies}: Although satellite orbits are predicted and modeled, they are not perfectly known and require continual estimation for precise operation \cite{hugentobler2017satellite}.  Inaccuracies in position references translate to inaccuracies in user position. The information in the navigation message is not always recent, and the description of the orbits is limited, which means that a standalone user will have limited ability to compensate for any orbital errors.

\textbf{Clock Drift}: Despite satellites having atomic clocks onboard, they are subject to drift. \rev{In addition to the individual satellite biases, there are also biases between different signals from the satellites' phase variations across their different transmit antenna directions. To counter that,} \gls{MEO} satellites use multiple atomic clocks to provide highly stable time references; but they still do not completely eliminate satellite clock drifts.\footnote{Offsets are also included in the navigation message together with orbital parameters.} However, just as with satellite orbital errors, the standalone receiver will have limited ability to independently estimate individual satellite clock drifts from the system time base.

\textbf{Atmospheric Effects}: Electron content in the ionosphere, as well as water vapor and dry particles in the troposphere, introduce additional delays, which have temporal and regional correlation. These effects, too, can only be mitigated in part by a standalone user, through the use of multi-frequency receivers or parametric models for estimating atmospheric delays, which permit receivers to compensate to achieve moderate levels of accuracy \cite{teunissen2017springer}.

\subsubsection{MEO Satellite Solutions}\label{sec:MEO solutions}
Precise positioning in \gls{GNSS} entails utilizing different strategies for mitigating the physical error sources described in the previous subsection, which scientific users have worked on for decades.  A symbiotic relationship between scientific and real-time users has helped bring precise positioning to mass market applications.

\textbf{RTK}:
Before \gls{GPS} was launched, scientists had begun developing ``reconstructed carrier phase" methods that would allow for highly precise estimates of differential position from \gls{GPS} observations \cite{bossler1980using}. Using the wavelength itself (19 centimeters at the L1 frequency) as the measuring stick rather than the 300-meter code chip equivalent distance, a much more precise measurement could be made. A well-surveyed proximate location (a ``base") experiences most of the major error sources impacting the observations in the same manner as a user (the ``rover"), as long as the distance between the two was not beyond the decorrelation distance for error sources like atmospheric effects. By double differencing observations from the base on the rover side, highly precise, centimeter-level position estimates could be attained.
This solved the problem of the coarse precision of the civilian code observable. Not only that, but at that time, artificial stochastic noise was also added to the code phase, known as selective availability (SA).\footnote{This was removed in the year 2000, in part because differential positioning methods were so common anyway but also as a rival constellation from Russia was coming online.} This significantly limited the accuracy that any standalone civilian user could achieve using code measurements.
Surveyors saw the utility of applying these to their discipline, but some practical limitations required new innovations to make them suitable for their needs.  It was desirable to enable operation for a moving or ``kinematic" surveyor, as well as to perform calculations in real-time rather than through a long period of data collection followed by post-processing and analysis in the lab \cite{remondi1984using}. \gls{RTK} surveying was developed to achieve centimeter-level performance in real-time through differenced measurements from proximate receivers, and subsequent expansion to ``network \gls{RTK} (NRTK)". NRTK utilizes a network of reference stations to interpolate virtual reference stations closer to the user receiver, allowing for longer baselines \cite{janssen2009comparison}. This is the standard method employed for surveying or other use cases where a proximate base station can reliably be employed.

\textbf{PPP}:
Another family of highly accurate positioning methods known as \gls{PPP} was developed based on undifferenced observations from receivers distributed over a much larger geographical area, capable of jointly estimating many of the primary physical error sources, \cite{zumberge1997precise}.  Such undifferenced observations were shown to be suitable for estimating tectonic plate velocities on a millimeter level \cite{larson1997global}.  \gls{PPP} too was developed into a real-time technology for commercial users in applications including agriculture, with distribution of error source information over the internet for decimeter-level real-time performance \cite{kechine2004experimental}. Commercial operations today support other applications demanding high precision with limited mobility and unobstructed sky views, such as precision agriculture.

\begin{figure}
\begin{center}
\includegraphics[width=0.45\textwidth]{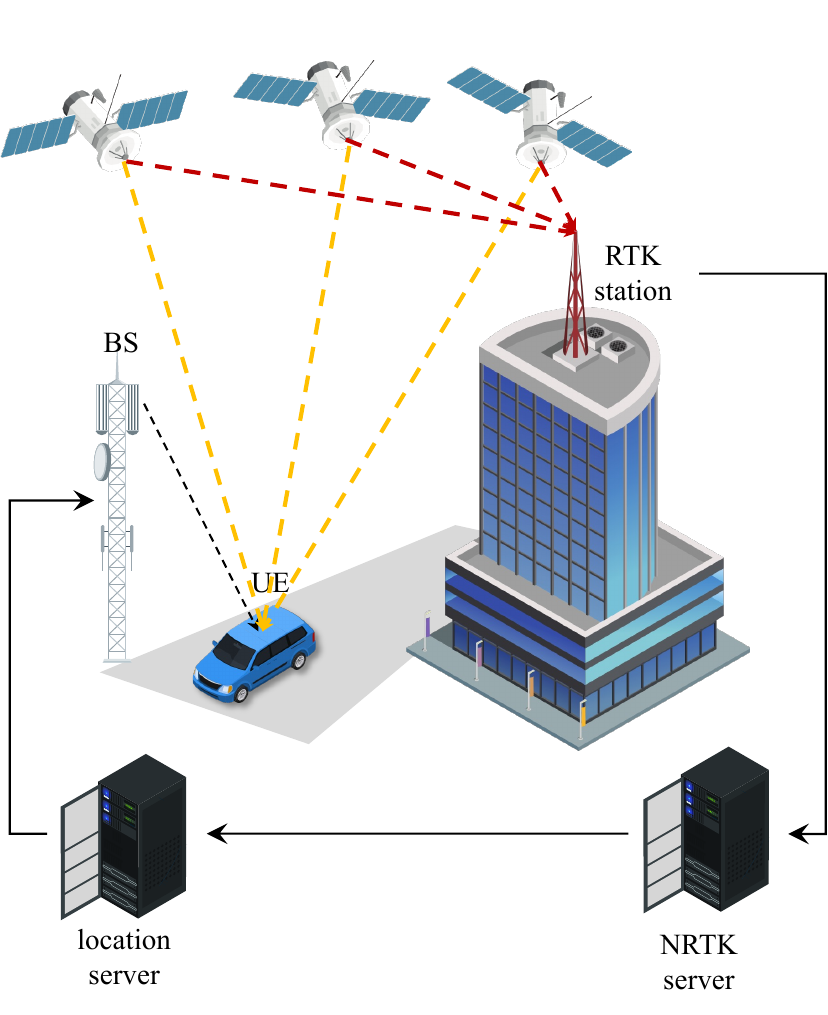}
\caption{\rev{Scalable \gls{3GPP} high accuracy \gls{GNSS} correction data distribution.}}
\label{fig:3gpp_correction_data_dist}
\end{center}
\end{figure}

\textbf{PPP-RTK}:
The concept was quickly expanded to attain better accuracy and convergence times by using denser local networks to estimate atmospheric errors locally. These networks are known as \gls{PPP-RTK} networks, as they combine the global properties of \gls{PPP} with more localized consideration of atmospheric errors \cite{wabbena2005ppp}. The atmospheric errors are addressed through smaller reference station spacing, similar to the approach used in \gls{RTK}.
\gls{PPP-RTK} is promising for vehicular use cases, combining the fast convergence times of \gls{RTK} with the lower density reference network requirements of \gls{PPP} \cite{li2022review}. Massive investment in \gls{PPP-RTK} networks, combined with multi-constellation commercial chipsets, has brought decimeter-level performance to automotive applications \cite{Swift_GNSS_developments}, and standardization of correction format is ongoing to optimize for bandwidth, update rate, and geographical coverage \cite{vana2019analysis}. To facilitate the provisioning of \gls{PPP-RTK} and NRTK corrections, \gls{3GPP} introduced functionalities to authorize and distribute such corrections to a device mass market \cite{36305}\rev{, as shown in Fig. \ref{fig:3gpp_correction_data_dist}}.

\subsection{LEO-based Positioning}
\subsubsection{LEO Satellite Challenges}
The prospect of \gls{LEO} navigation is not without challenges. Many of the error sources for \gls{MEO} constellations are worse for \gls{LEO} systems \cite[Section IV]{stock2024survey}. 

\textbf{Orbital Inaccuracies}: \gls{LEO} orbits are typically described by \gls{TLE} files, which are offered by only a few sources with limited update rates. These files also fail to model many of the physical effects that impact orbits in reality \cite{vallado2012two}, which means that older orbital information can be off by kilometers over the course of a day. \gls{MEO} \gls{GNSS} systems are deployed from the beginning with a sophisticated control segment which performs important tasks like precise orbit and time estimation. These segments also upload data to satellites so they can describe their own status to user receivers operating in standalone mode \cite{teunissen2017springer}. \gls{LEO} systems do not yet have any such infrastructure for orbit determination, and they are even more impacted by the Earth's gravitational field owing to the closer proximity, adding to the difficulty of orbital estimation \cite{reid2016leveraging}. 

\textbf{Atmospheric Effects and Clock Stability}: One of the most effective techniques for atmospheric delay compensation, the use of multi-frequency receivers, may not be applicable for \gls{LEO} systems if they transmit observable signals at only a single frequency. Clock stability likely will not be achieved to the same degree either, owing to the need to have more satellites with less expensive components to create a fully operational constellation \cite{reid2020navigation}.

\textbf{Size of Footprint on Earth}: \gls{LEO} satellites have smaller footprints than \gls{MEO} systems, limiting their coverage and necessitating the deployment of a greater number of \gls{LEO} satellites and ground stations for precise operation \cite{eissfeller2024comparative}. The coverage issue is further compounded by the use of the Ku band for downlink data communication, which operates at frequencies an order of magnitude higher than those of \gls{MEO} \gls{GNSS} systems. Communication over these higher frequencies requires beamforming, which further narrows the coverage~\cite{del2019technical}. As a result, a conflict arises between providing high data rates via phase-steered beams and ensuring broad coverage to support trilateration for users across large geographical areas.

\textbf{Commercial Viability}: Finally, dedicated \gls{LEO}-\gls{PNT}, if not financed by state actors in the same manner as the major \gls{MEO} constellations, may encounter difficulty in establishing plausible business models in the face of ``free" competitors in the \gls{MEO} constellations unless compelling performance is achieved or other functionality is offered.

\subsubsection{LEO Satellite Solutions}
There is no consensus regarding how the \gls{LEO} architecture for end users should be structured. Early simulations and measurements have used numerous different approaches to address the challenges identified in the previous section. They can be split into two categories, dedicated \gls{LEO}-\gls{PNT}, in which system resources are dedicated specifically for navigation or timing, and opportunistic \gls{LEO}-\gls{PNT}, in which communication signals are used ``\textit{opportunistically}" as a basis for navigation.

\textbf{Dedicated LEO-PNT}:
Numerous concepts and projects of dedicated \gls{LEO}-\gls{PNT} are under development in the USA, China, and Europe. These systems are either communication-based systems with dedicated \gls{PNT} signals or constellations built entirely for the purpose of \gls{PNT}. See \cite[Table 1]{eissfeller2024comparative}.
An example of communication system re-purposing is the Iridium constellation's \gls{STL} feature. This is designed to provide an independent \gls{PNT} source that can penetrate through attenuating objects by virtue of having higher signal strength. However, the constellation is not designed to have multiple satellites visible at all times, so the \gls{STL} feature is intended to be used with longer observation times, primarily as a redundant backup for fixed infrastructure for timing applications \cite{gutt2018recent}.
An example of a dedicated constellation is Xona's PULSAR system. By piggybacking on \gls{MEO} constellations for orbit and time determination \cite{reid_leo_commercial}, this system presents a \gls{LEO} solution that is not fully independent, especially when trying to establish guarantees for long-term time synchronization.

\textbf{Opportunistic LEO-PNT}:
Opportunistic \gls{LEO} is a field of intense study. See \cite{stock2024survey} for a thorough survey of opportunistic \gls{LEO}. Theoretical frameworks have been developed, both with analyses of estimable states based on various observables as well as looking at constellation sizes \cite{psiaki2021navigation}.
Physical measurements have also been conducted. Starlink communication signals have been used for navigation purposes, and a framework has been developed for this \cite{starlink_positioning_kassas}. Unlike \gls{SLAM}, in which static references are mapped, simultaneous tracking and navigation frameworks attempt to generate refined estimates of satellite trajectories simultaneously with estimating ego-user navigation states \cite{kassas2019new}. More advanced concepts for precise operation have even been demonstrated, including a surveyed baseline to perform differencing \cite{starlink_positioning_kassas}.

\subsection{Contemporary Commercial Solutions and Open Problems}
Although \gls{GNSS}-based satellite geodesy enables millimeter-level performance for fixed terrestrial stations on Earth \cite{larson1997global}, numerous practical limitations prohibit anything resembling this kind of performance for vehicular applications. Commercial-grade receivers use far simpler receiver architectures and narrower front-end bandwidths \cite{braasch1999gps}, resulting in significantly higher noise for both code and carrier phase measurements \cite{de2018precise}.
However, commercial systems deployed recently have achieved impressive performance in vehicular environments using consumer-grade receivers and antennas. Trimble's RTX service showed a horizontal error level of less than 0.5 m for 95\% and sub-meter protection levels with no integrity errors \cite{rodriguez2021protection}. Similar sub-meter level accuracy was demonstrated for a cross-US drive using Swift Navigation's Skylark service and Starling positioning engine \cite{goparaju2020project}. Similar accuracy and integrity results were achieved using Hexagon's Terrastar service \cite{norman2019integrity}. Note that verifying the integrity performance and confirming the rate of integrity failures requires a sample size beyond what is practically testable, at least for one person \cite{walter2003integrity}. Hence, protection levels have a degree of modeling assumptions built into them.   
\gls{LEO} performance in vehicular environments is not thoroughly tested. The Iridium satellite timing and location signals demonstrated 20-meter accuracy in one urban environment \cite{reid2020navigation}, but there is limited data on performance for \gls{LEO} systems in general.

\section{3GPP-based Positioning}\label{Sec: cellular}
Cellular networks have traditionally been developed with the primary goal of facilitating communication, with each successive generation bringing enhancements in bandwidth, connectivity, and overall functionality. While the initial focus of these networks was solely on communication, positioning has gradually emerged as an important feature, evolving from an opportunistic byproduct to a critical component of modern and future cellular networks. Cellular positioning technology was initially driven by regulatory mandates, with the U.S. \gls{FCC} requiring \gls{E911} emergency call capabilities in 1996, and the European Council following suit in 2000 \cite{del_peral-rosado_survey_2018}. Over time, cellular networks evolved and advanced to meet these requirements, eventually reaching a level of maturity that enabled their use in vehicular applications \cite{wang_recent_2023}. In this section, we first explore the evolution of cellular network standards to support vehicular positioning. \rev{Next, we delve into the fundamentals of 5G-\gls{NR}'s millimeter-wave (mmWave), also known as frequency range 2 (FR2), and sub-6 GHz (FR1)} positioning and examine their contemporary research directions, architectures, and solutions. \rev{The discussion then broadens to consider potential 6G positioning technologies such as \glspl{RIS}, the usage of 7-24 GHz (cmWave) and the sub-terahertz (sub-THz) frequency bands, and privacy/security aspects, offering insights into the future trajectory of cellular positioning systems.} Finally, we present contemporary research challenges and open problems in cellular positioning in general.

\subsection{History of Positioning in Cellular Standards}
\begin{figure}[t]
    \centering
    \includegraphics[width=0.99\linewidth]{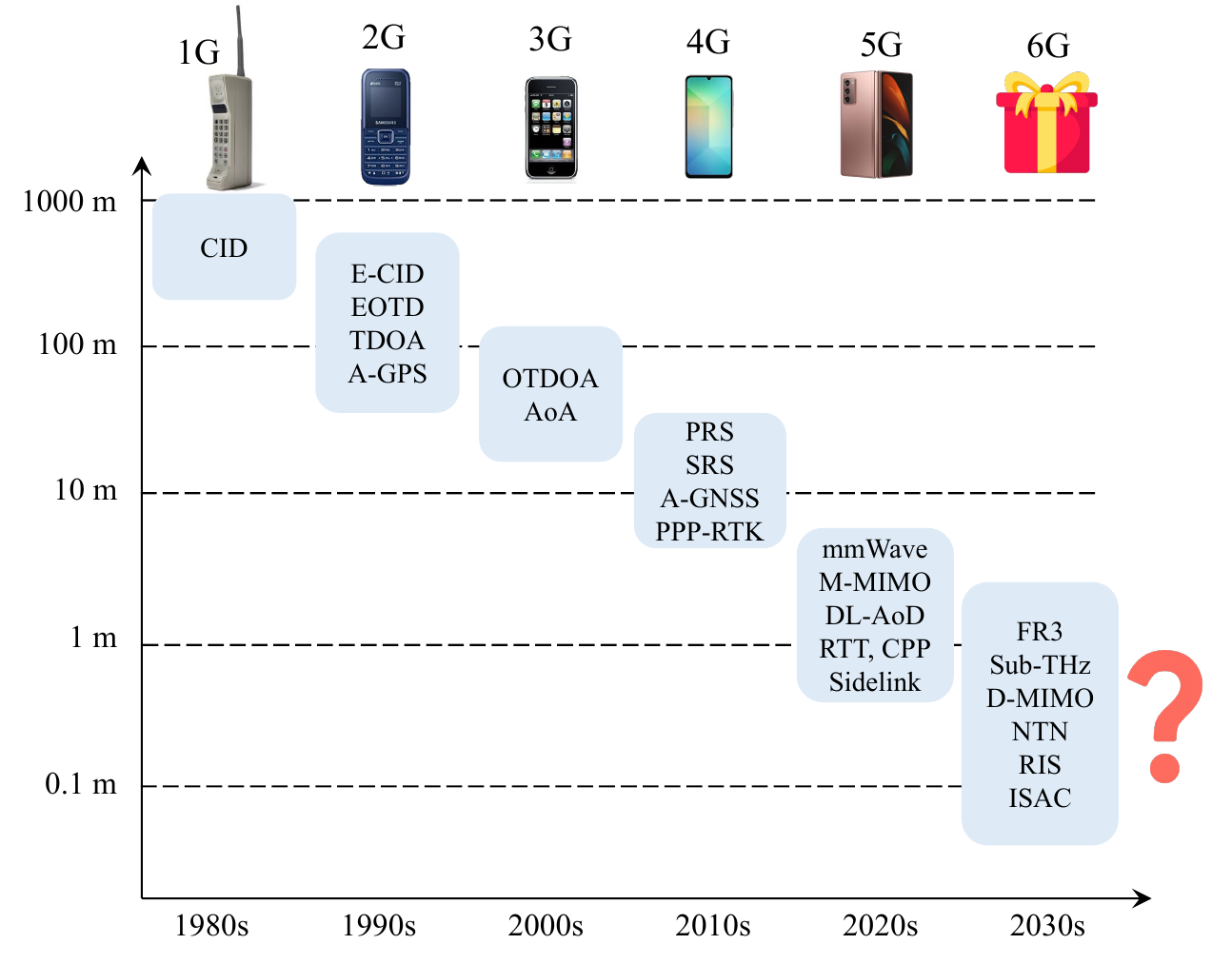}
    \caption{Illustration of the evolution of cellular positioning capabilities over decades and generations. The height of the boxes illustrates the positioning accuracy of the given generation, and the text inside the box presents the new measurements and enablers for the given generation.}
    \label{fig:evolution_new}
\end{figure}

The evolution of cellular networks to support positioning has been extensively documented in various studies, each covering different periods and Releases of the standards \cite{del_peral-rosado_survey_2018, campos_evolution_2017, yang_overview_2022, razavi_positioning_2018, dwivedi_positioning_2021, ren_progress_2021, wang_recent_2023}. For example, \cite{del_peral-rosado_survey_2018} explores positioning technologies from 1G to 5G's Release 15, while \cite{campos_evolution_2017} examines developments from 2G to 4G. Coverage of the evolution from 2G to 5G's Release 16 is presented in \cite{yang_overview_2022}, whereas \cite{razavi_positioning_2018} focuses on the transition from 4G to 5G's Release 15. More recent advancements in 5G are highlighted in \cite{dwivedi_positioning_2021} and \cite{ren_progress_2021}, which discuss Releases 16 and 17, respectively. The latest developments in 5G standards, from Release 15 to 18, are documented in \cite{wang_recent_2023}. In the following, we present a summary of positioning in cellular standards from 2G to the latest frozen \gls{3GPP} Release (Release 18), and the currently ongoing releases (Releases 19 and 20). We cover the evolution of the targeted positioning use cases, their positioning requirements, and the methods, protocols, and functions introduced in the standards to serve those use cases, with more emphasis on 5G. A summary of the evolution of \gls{3GPP}'s cellular positioning performance and capabilities over generations is depicted in Fig.~\ref{fig:evolution_new}.

\textbf{2G}:
Before \gls{3GPP} existed, rudimentary positioning services were supported by 2G-GSM networks via timing advance/\gls{E-CID}, \gls{EOTD}, \gls{UL}-\gls{TDoA}, and assisted-\gls{GPS} \cite{campos_evolution_2017,del_peral-rosado_survey_2018,yang_overview_2022}. These positioning services were aimed towards localizing emergency calls, which did not have high positioning requirements \cite{del_peral-rosado_survey_2018}. To support such services, new network elements were introduced, including the \gls{SMLC},  the \gls{LMU}, and the \gls{GMLC} \cite{campos_evolution_2017,del_peral-rosado_survey_2018,yang_overview_2022}. The introduction of these network elements and localization methods serves as the foundation for future standardization, making positioning an integral part of upcoming cellular networks.

\textbf{3G}:
The \gls{3GPP} standards continued the support for location services in 3G, a tradition that will ``\textit{evolve}" over time. In 3G, the \gls{GMLC}, \gls{SMLC}, and LMUs were incorporated into the \gls{RNC} \cite{campos_evolution_2017}. Additionally, a new \gls{PE} was added to the table, which greatly enhanced the \gls{OTDoA} measurements \cite{campos_evolution_2017}. Moreover, \gls{3GPP} added uplink \gls{AoA} estimation to the cellular positioning arsenal, thanks to the use of adaptive (smart) antennas. Yet, as the number of antenna array elements was low at the time, \gls{AoA} measurements were not of high quality. In addition to that, 3G communications predominantly operated under \gls{NLoS} conditions, which caused high \gls{AoA} errors. Nonetheless, 3G's \gls{AoA} measurements enhanced the capability of the existing \gls{E-CID} methods \cite{campos_evolution_2017}. Finally, 3G witnessed an increase in bandwidth, and hence enhanced time resolution, but that was not enough to meet the needs of vehicular positioning.

\textbf{4G}:
Following in the footsteps of its predecessors, 4G standards refined the existing positioning methods and network elements. Such improvements included (i) the introduction of assisted-\gls{GNSS}, an enhanced form of assisted-\gls{GPS} that makes use of other \gls{GNSS} constellations; (ii) the introduction of dedicated \gls{TDoA} positioning signals like the \gls{DL}-\gls{PRS} and the \gls{UL}-\gls{SRS}; (iii) hybrid assisted-\gls{GNSS} and \gls{TDoA} positioning; and (iv) the \gls{LPP} \cite{campos_evolution_2017, razavi_positioning_2018}. Although these improvements were huge in terms of their impact on the cellular positioning future, they were not enough to provide the positioning solution needed by vehicular applications. This was mainly due to the low bandwidth ($20$-$100$ MHz) and number of antennas, and the high latency.

\textbf{5G}:
Until this point, cellular positioning was driven by the regulatory requirements for emergency calls. In 5G, however, this trend changes. The following is a brief survey on how the \gls{3GPP}'s 5G standardization evolved to tackle the vehicular positioning problem in each 5G \gls{3GPP} Release.

\textit{Release 15} was the first of \gls{3GPP}'s 5G standardization releases. In Release 15, the aim was set from the beginning to provide localization services to commercial applications such as factory automation, railways, and \glspl{UAV}, as well as mission-critical applications like first responders \cite{22261}. These localization services can be provided by (i) \gls{RAT}-dependent solutions based on LTE positioning architecture; or (ii) non-\gls{RAT}-dependent solutions such as assisted-\gls{GNSS}, \rev{Wi-Fi, Bluetooth}, barometric pressure, and dead-reckoning sensors (i.e., accelerometers and gyroscopes) \cite{38305}.\footnote{This split is not new and was already implemented in previous generations.} To facilitate the 5G positioning process, Release 15 introduced the \gls{NRPPa}, which works in tandem with the existing \gls{LPP}, as well as the \gls{LMF} \cite{38305, yang_overview_2022, wang_recent_2023}. Here, \gls{NRPPa} is used to enable communication between the \gls{LMF} and \gls{BS}, while \gls{LPP} allows the \gls{LMF} and the \gls{UE} to communicate; more details can be found in \cite{38455}. Additionally, the \gls{BS} and \gls{UE} can directly communicate using the \gls{RRC} protocol, which is not specific to positioning but is also used by all 5G functions. Finally, Release 15 introduced support for unicast distribution of \gls{RTK}-\gls{GNSS} and real-time \gls{PPP}-\gls{GNSS} corrections in assisted-\gls{GNSS} operations as mentioned in Sec.\ref{sec:MEO solutions}.

\textit{Release 16} pushed the limits of cellular localization by expanding the targeted use cases, \cite{22872}, and aiming to achieve horizontal localization accuracy of $<10$ m ($80\%$ of the time) for outdoor users, $<3$ m ($80\%$ of the time) for indoor users, and vertical accuracy of $<3$ m ($80\%$ of the time) for both indoor and outdoor users for commercial applications, in addition to a maximum latency of $1$ s \cite{38855}.\footnote{Regulatory requirements are less stringent in Release 16, compared to commercial requirements, which demanded horizontal accuracy of $<50$ m ($80\%$ of the time) and vertical accuracy of $<5$ m ($80\%$ of the time), and a maximum latency of $30$ s \cite{38855}.} To deliver on these promises, Release 16 provided many solutions. For instance, Release 16 expanded the bandwidth of the \gls{DL}-\gls{PRS} and \gls{UL}-\gls{SRS} signals to $400$ MHz, which enabled precise time-based ranging measurements. These measurements included \gls{UL}-\gls{TDoA}, \gls{DL}-\gls{TDoA}, and multi-\gls{RTT}. Release 16 also added \gls{DL}-\gls{AoD} measurements to the cellular positioning toolkit. 5G's \gls{DL}-\gls{AoD} and \gls{UL}-\gls{AoA} measurements have exceptionally high accuracy and resolution, thanks to the high number of antennas used in 5G's massive \gls{MIMO} systems. Finally, Release 16 added the Chinese BeiDou constellations and \gls{GNSS} \gls{PPP-RTK} to its assisted-\gls{GNSS} framework and introduced 5G cellular broadcast of corrections. The correction services can now be provided via either unicast or broadcast, and devices can request on-demand non-broadcast correction data~\cite{38305, dwivedi_positioning_2021, wang_recent_2023, yang_overview_2022}.

\textit{Release 17} added industrial \gls{IoT} use cases to the table, which requires \gls{LPHAP} services \cite{38857}. To provide such services, Release 17 focused on enhancing the 5G performance in terms of accuracy, latency, and power efficiency. Hence, the new targeted horizontal and vertical accuracy for commercial use cases was set to $<1$ m and $<3$ m, respectively, ($90\%$ of the time) \cite{38857}. For industrial \gls{IoT} applications, however, the targeted horizontal and vertical accuracy are $<0.2$ m and $<1$ m, respectively, ($90\%$ of the time) \cite{38857}. The targeted latency for either application is now $100$ ms. Raising the bar for 5G positioning requirements brought 5G one step closer to realizing vehicular positioning requirements. Towards achieving those goals, Release 17 proposed various enhancements to (i) mitigate multi-path effects; (ii) mitigate \gls{BS}-\gls{UE} synchronization issues; (iii) improve accuracy for \gls{UL}-\gls{AoA} and \gls{DL}-\gls{AoD} measurements; (iv) enable various ``idle'' and ``inactive'' states to enhance power efficiency; and (v) reduce measurement gaps to reduce latency \cite{38857, wang_recent_2023, ren_progress_2021}.

\textit{Release 18} is the first release of \gls{3GPP}'s 5G-Advanced standardization effort. Release 18 marked \gls{3GPP}'s first commitment to provide positioning services that are specifically tailored to vehicular applications and \gls{RedCap} \glspl{UE} \cite{38859, wang_recent_2023}. To further address vehicular positioning, relative positioning in terms of \gls{SL} positioning between vehicles was introduced. Release 18 had two sets of requirement targets, lane level accuracy, i.e. $<1.5$ m and $<3$ m of horizontal and vertical accuracy, respectively, for $90$\% of the time, and sub-meter accuracy, i.e., $<0.5$ m and $<2$ m of horizontal and vertical accuracy, respectively, for $90$\% of the time \cite{38859}. To achieve such requirements, Release 18 proposed a myriad of solutions including (i) \gls{SL}-\gls{SRS}; (ii) \gls{SL}-\gls{TDoA}, \gls{SL}-\gls{RTT}, \gls{SL}-\gls{AoD}, and \gls{SL}-\gls{AoA} measurements; (iii) \gls{SLPP}; (iv) \gls{CPP} measurements including uplink and downlink \gls{PDOA}; (v) bandwidth aggregation for reference signals; (vi) \gls{NTN} for country verification; (vii) \gls{AI} for \gls{NLoS} positioning; (viii) sidelink operation in both licensed and unlicensed spectrum; and (ix) focus on positioning integrity computation \cite{38859, wang_recent_2023}. \rev{\gls{RAT}-independent enhancements include \gls{GNSS} phase center offset and variations representation, \gls{GNSS} \gls{LoS}/\gls{NLoS} indications per satellite, and support for Bluetooth \gls{AoA} measurements.} Further reading on Release 18's enhancements in architectural aspects \cite{23700A, 23586A}, procedural aspects \cite{24514Pd}, protocols \cite{38355P}, and security aspects \cite{38893S} can be found in the references herein. 

\rev{\textit{Releases 19} is the latest frozen installment of 5G-Advance, frozen in December 2025.} 
Release 19 focuses on specific components and measurements related to \gls{AI}/\gls{ML}-based positioning, and indicates further investigation of the usage of \gls{NTN} and sidelink positioning in 5G-Advanced. Release 19 also introduces support for the Indian Constellation (NavIC), India's regional positioning satellite constellation, adding it to 5G's and 4G's A-\gls{GNSS} arsenal. Additionally, Release 19 marks the official emergence of \gls{ISAC} in cellular networks, which triggered the study of its geometric channel models \cite{22837,22137}. Release 19 also introduced new use-cases that require tight sensing and positioning performance, like metaverse services \cite{22856}. 

\rev{\textit{Release 20}, as of the time of writing, is yet to be frozen/finalized by June 2027.} Release 20 is divided into two parallel streams focusing on (i) finalizing 5G-Advanced standards and (ii) conducting initial studies for the development of the 6G standards. Release 20's 5G-Advanced studies will focus on cellular sensing results for vertical applications and \gls{ISAC} in general, metaverse services, enhancing 5G-based \gls{NTN} localization services. On the other hand, Release \rev{20}'s initial 6G studies focused on 6G-specific use-cases, scenarios, and their requirements, which heavily featured \gls{ISAC}-based services and integrated \gls{TN} and \gls{NTN}-based positioning scenarios \cite{22870}. In particular, section 7 of the report highlights 20 unique \gls{ISAC}-based use-cases, including vehicular positioning, and section 8 highlights various \gls{NTN} positioning scenarios. It is worth noting here that \gls{TN}-\gls{NTN} integration holds a great promise for future cellular positioning, especially for vehicular applications~\cite{saleh_integrated_2025}. Such studies will help guide Release 21's normative work towards standardizing 6G in the near future.

\begin{figure}
    \centering
    \includegraphics[width=0.99\linewidth]{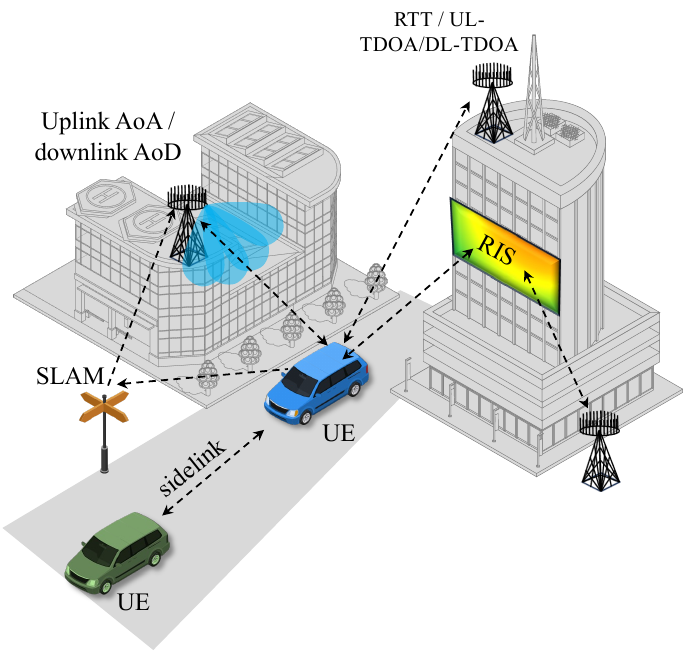}
    \caption{\rev{\gls{3GPP}-based positioning: overview of 5G and beyond 5G positioning enablers.}}
    \label{fig:3GPP_new}
\end{figure}

\subsection{5G Positioning Fundamentals}
A typical cellular localization system consists of three types of entities: anchors (e.g., \gls{BS}, roadside unit), environment (e.g., map information or incidence points such as reflectors, scatterers, etc.), and the vehicle (or \gls{UE})\rev{, shown in Fig.~\ref{fig:3GPP_new}}. Each of these entities has its own state, either as prior information or unknowns to be estimated. For example, a \gls{BS} equipped with a planar array has the state of position, orientation, and velocity (if the \gls{BS} is mounted on a mobile platform like a \gls{UAV} or a \gls{LEO} satellite). An incidence point that reflects or scatters the signals forming an \gls{NLoS} path has the state of positions and possibly that of a reflection coefficient. The state of the \gls{UE} may consist of position, orientation (if equipped with a planar array), clock offset (asynchronous with the anchors), and velocity (if under mobility). The unknown states can be formulated as deterministic unknowns in single-snapshot localization and are modeled with specific distributions that update with time in tracking scenarios. The goal of cellular localization is to estimate the state of the \gls{UE} based on the anchor state and the map state (if any) using uplink, downlink, and/or sidelink signals~\cite{chen2022tutorial}. To achieve these goals, cellular localization is performed in three main stages: (i) system optimization; (ii) channel parameter estimation; and (iii) positioning and tracking. All of these stages are supported by knowledge of the underlying channel models \rev{(shown in Appendix \ref{appendix})} and theoretical error bounds \cite{sallouha2024ground}. Although the survey will primarily focus on the last phase of positioning and tracking, we glance, in this section, through the fundamentals of all stages and bounds for completeness.

\subsubsection{System Optimization}\label{pipeline_step_1}
System optimization in cellular networks can be categorized into long-term and short-term processes. Long-term optimization involves decisions that are largely permanent and challenging to modify once implemented. Examples include determining the placement of \glspl{BS} \cite{saleh2021evaluation}, as well as selecting the number and types of antennas \rev{(e.g., $\Ntx$ and $\Nrx$ in \eqref{eq_rec}, Appendix \ref{appendix})} and radio frequency chains. These choices define the baseline capabilities of the system and are mainly dictated by the minimum communication performance and economic constraints. In contrast, short-term optimization is dynamic and adaptable over shorter timescales. This includes \rev{designing the transmit signal $s(t)$ in \eqref{eq_rec}, Appendix \ref{appendix}}, i.e., configuring radio resources (e.g., transmission power, time, and frequency allocations) and designing waveforms and precoder/combiner codebooks. For instance, in 5G NR, time-frequency resource allocation and signal design are managed through the \gls{PRS} and the \gls{SRS}, which provide staggered pilot signals to improve delay estimation performance~\cite{38855}. The signal design also extends to the spatial domain, where precoder optimization at the transmitter and combiner optimization at the receiver aim to minimize positioning errors \cite{garcia2018optimal,keskin2022optimal}. Short-term signal optimization can be achieved by minimizing the \gls{CRB} when prior information is available~\cite{keskin2022optimal}. In cases where no prior information about the states is available, random signals or fixed codebooks are typically employed.

\subsubsection{Channel Parameters Estimation}
To estimate the state of the \gls{UE}, we need to first estimate the channel from the observed signals. Since the designed pilot signals are known to the receivers, the channel matrix between the user and the \gls{BS} during a coherence time can be estimated (e.g., using \gls{LS}).  To further exploit the sparsity of channels for localization, channel parameter estimation algorithms, compressive sensing, and generalized approximate message passing can be used to estimate the geometric parameters of each path contained in the channel matrix~\cite{wymeersch2022radio1}. As mmWave propagation is characterized by a small number of paths or clusters of paths, compressive sensing methods are favored for channel estimation \cite{overview_mmwave_JSTSP_2016}. By further decomposing the channel into geometrical parameters (of each resolvable path) such as complex channel gain, angle of arrival, angle of departure, delay, and Doppler, the localization problem can be solved based on the geometrical relationship between the state and channel parameters~\cite{sallouha2024ground, wymeersch2022radio1}. 

\subsubsection{Positioning and Tracking}
Based on the channel geometric parameters, direct~\cite{garcia2017direct} or multi-stage~\cite{shahmansoori2017position} localization algorithms can be designed to obtain the \gls{UE} state. In addition, the positions of incidence points of reflection can also be estimated as a by-product, which is usually referred to as mapping~\cite{chen2022tutorial} or \gls{SLAM}~\cite{rastorgueva2024millimeter, kim2023ris}. In cases where historical localization results are available, tracking algorithms such as the \gls{KF}, and its various flavors~\cite{ge2022computationally}, and particle filters can be implemented with the aid of transition models~\cite{guerra2021near}. When in scenarios with an unknown number of clutters, more advanced filters that consider random finite sets and hypothesis densities can be adopted~\cite{kaltiokallio2023multi}. Eventually, the localization results can be obtained, and the distribution of the \gls{UE}/map state will be used as prior information to serve signal design in the next instances. \rev{With the above in mind, 3GPP-based vehicular positioning methods can be broadly grouped into four methodological families, namely, model-based, learning-based, Bayesian-based, and graph-based—each with distinct strengths and limitations. \textit{Model-based approaches} use geometric relationships (\gls{ToA}/\gls{TDoA}, \gls{AoA}/\gls{AoD}, Doppler) to perform multilateration, maximum likelihood, or least square estimation, offering interpretability, low complexity, and strong performance under \gls{LoS} or well-modeled channels, but they degrade under \gls{NLoS}, hardware impairments, or model mismatch. \textit{Learning-based methods} exploit \gls{CSI}, beam patterns, or I/Q samples to learn nonlinear mappings or augment geometric estimators, capturing complex propagation effects that are analytically intractable; however, they require extensive labeled data, have limited interpretability, and struggle with generalization across environments. \textit{Bayesian-based techniques} (\gls{EKF}/\gls{UKF}, particle filters, \gls{SLAM}, random finite sets) incorporate temporal correlation, mobility models, and sensor fusion to enhance robustness in dynamic vehicular scenarios, providing principled uncertainty handling but at higher computational cost and with sensitivity to modeling assumptions. \text{Graph-based approaches} formulate positioning as inference on factor graphs using constraints from \gls{ToA}/\gls{AoA}, sidelink \gls{V2V} ranging, or virtual anchors, enabling cooperative localization and spatial consistency, yet they rely on accurate constraint construction and may incur communication and coordination overhead. Together, these methodologies form complementary tools: model-based methods serve as clean benchmarks, learning-based approaches improve performance in complex propagation, Bayesian methods provide temporal continuity and robustness, and graph-based formulations excel in cooperative and SLAM-like scenarios.}

\subsubsection{Analysis Tools}
Based on Fisher information analysis, the \gls{CRB}s of unknown state parameters can be derived~\cite{kay1993fundamentals}. This bound provides the best performance, in terms of \gls{RMSE}, that a system can achieve. Hence, \gls{CRB} could serve as (i) a benchmark of efficiency of the proposed localization algorithms; (ii) an objective function for system optimization purposes; or (iii) a measurement weighting metric in weighted positioning algorithms. Other types of \gls{CRB} can also be adopted, such as the constrained \gls{CRB} (C\gls{CRB})~\cite{nazari2023mmwave} (when 3D orientation estimation is involved), misspecified \gls{CRB} (M\gls{CRB})~\cite{chen2022mcrb} (when model mismatches exist), and Bayesian \gls{CRB} (B\gls{CRB})~\cite{italiano2024tutorial} (when state tracking is performed). Regarding \gls{SLAM} algorithms, the generalized optimal subpattern assignment distance is usually used to quantify the mapping and data association performance~\cite{kim2024set, rastorgueva2024millimeter}.

\subsection{5G Positioning Solutions}
Unlike \gls{GNSS} and other conventional positioning technologies, 5G mmWave (FR2) positioning can be performed using a single \gls{BS} through \gls{SLAM} algorithms, thereby reducing the deployment costs for accurate positioning solutions. To provide redundancy and increased degrees of freedom, measurements from multiple \glspl{BS} can also be utilized to improve accuracy. To further enhance positioning accuracy and reliability, cooperation with other vehicles/users (through \gls{V2V} sidelinks) can be exploited. \rev{In addition to mmWave positioning, 5G sub-6 GHz (FR1) positioning can also aid with vehicular positioning. The following focuses mainly on providing a detailed review of mmWave vehicular positioning deployment scenarios and their associated algorithms, with a brief treatment of sub-6 GHz positioning works in the end.} It is worth noting that some indoor works were also included, as their methodology can be easily applied in an outdoor vehicular setting. \rev{A summary of some of the selected works is provided in Table~\ref{tab_mmwave_solutions}.} 

\begin{table*}
    \centering
    \caption{\rev{Summary of selected cellular mmWave vehicular localization works, categorized by number of BSs (green and yellow for single- and multi-BS, respectively) and the usage of cooperative positioning (orange).}} 
    \vspace{2mm}
    \label{tab_mmwave_solutions}
    \scriptsize
    \renewcommand{\arraystretch}{1.2} 
    \begin{tabular}{|p{0.1\textwidth}|p{0.2\textwidth}|p{0.2\textwidth}|p{0.18\textwidth}|p{0.19\textwidth}|}
        \hline
        \multicolumn{1}{|c|}{\textbf{Technology}} & \multicolumn{1}{c|}{\textbf{Environment and Coverage}} & \multicolumn{1}{c|}{\textbf{Measurements and Techniques}} & \multicolumn{1}{c|}{\textbf{Accuracy}} & \multicolumn{1}{c|}{\textbf{Validation}}       
\\              \hline
\rowcolor{pastel2} Single-BS \cite{kakkavas2019performance} & LoS+NLoS, 30 m & AoA, AoD, ToA & Decimeter to meter-level  & Theoretical
\\              \hline
\rowcolor{pastel2} Single-BS \cite{henkDownlink_2018} & LoS+NLoS, 100 m & AoA, AoD, ToA, belief propagation on factor graphs & Decimeter to meter-level  & Theoretical
\\              \hline
\rowcolor{pastel2} Single-BS \cite{wen20205gb} & LoS+NLoS, 30 m & AoA, AoD, ToA, tensor-ESPRIT & Decimeter to sub-meter-level  & Theoretical
\\              \hline
\rowcolor{pastel2} Single-BS \cite{Chen2022SAM} & Urban canyon, 100 m & AoA, AoD, TDoA, OMP+DNN & Decimeter to several meter  & Simulation, Wireless Insite
\\              \hline
\rowcolor{pastel2} Single-BS \cite{gante2020dethroning} & Urban canyon, 400 m & PDP, DNN & Several meters  & Simulation, Wireless Insite
\\              \hline
\rowcolor{pastel2} Single-BS \cite{bistatic_doubleBounce_2024} & Indoor, 10 m & AoA, AoD, ToA, snapshot SLAM using multi-bounce reflections  & Decimeter-level  & Experimental setup
\\              \hline
\rowcolor{pastel2} Single-BS \cite{bader2023step}  & Dense urban canyon, 125 m & AoA, AoD, ToA, RSS  & Decimeter-level  & Simulation, Siradel’s S5GChannel
\\              \hline
			  \hline
\rowcolor{pastel3} Multi-BS \cite{ko2021v2x} & Dense urban canyon, [100 300] m & E-CID, multi-RTT, ToA, TDoA, AoA, AoD, RSRP & Decimeter to several meters & Theoretical + simulation + experimental
\\              \hline
\rowcolor{pastel3} Multi-BS \cite{saleh2021vehicular} & Dense urban canyon, 100 m & TDoA, EKF by excluding BSs with high linearization errors & Decimeter-level & Simulation, Siradel’s S5GChannel
\\              \hline
\rowcolor{pastel3} Multi-BS \cite{saloranta2018novel} & LoS environment, 50 m & AoA, AoD, ToA & Decimeter to sub-meter-level & Theoretical 
\\              \hline
\rowcolor{pastel3} Multi-BS \cite{vehPos5G_2016} & LoS environment & ToA & Centimeter to sub-meter-level m & Theoretical
\\              \hline
\rowcolor{pastel3} Multi-BS \cite{heimann2020cross} & Indoor, 10 m & AoA, AoD & Centimeter-level & Experimental setup
\\              \hline
\rowcolor{pastel3} Multi-BS \cite{rastorgueva2020networking} & Indoor factory, [100 300] m & Beam RSRP converted to AoD, EKF & Decimeter to meter-level  & Theoretical
\\              \hline
\rowcolor{pastel3} Multi-BS \cite{6G_BNN_Pos_2024} & Urban, inter-site distance of 200 m with 19 sites & ToA, AoA, RSS, Bayesian NN & Sub-meter-level & Simulation, Wireless Insite
\\              \hline
\rowcolor{pastel3} Multi-BS \cite{karfakis2023nr5g} & Outdoor agricultural environment, robot tracking, 500 m & ToA, EKF & Decimeter to sub-meter-level & Simulation, 5G Matlab Toolbox
\\              \hline
                \hline
\rowcolor{pastel1} Cooperative \cite{kim20205g} & LoS+NLoS, 100 m & AoA, AoD, ToA, Probability hypothesis density (PHD) filter and map fusion & Decimeter to sub-meter-level  & Theoretical
\\              \hline
\rowcolor{pastel1} Cooperative \cite{ge2024v2x} & LoS+NLoS, 140 m & AoA, ToA, 2D ESPRIT & Meter-level & Theoretical
\\              \hline
\rowcolor{pastel1} Cooperative \cite{tedeschini2023cooperative} & LoS+NLoS 1x1 km$^2$, 19 RSUs & AoA, ToA, DL & Meter-level, LoS & Theoretical
\\              \hline

			  \hline
    \end{tabular}
\end{table*}

\subsubsection{\rev{FR2 (mmWave)} Single-BS Solutions and SLAM} 
While conventional positioning requires multiple \glspl{BS}, there has been a focused effort on providing positioning capabilities using a single \gls{BS}, by harnessing the natural multipath present in the environment. The idea of positioning by \emph{multipath exploitation} can be traced back to \cite{miao2007positioning,shen2009use}. These works derive performance bounds and algorithms for multi-\gls{BS} positioning with multipath exploitation, either without \gls{LoS} \cite{miao2007positioning} or with \gls{LoS} \cite{shen2009use}. This idea was later adopted for single-\gls{BS} positioning (see \cite{kakkavas2019performance} and references therein) in a 5G mmWave context, utilizing the combined effect of three properties: (i) the ability to estimate \gls{AoA}, \gls{AoD}, and \gls{ToA}; (ii) the high degree of multipath resolvability in delay and angle domains; and (iii) the fact that each multipath component is characterized by more observations than unknowns \cite{henkDownlink_2018,nuria_Proc_IEEE_2024}. This effect in turn provides the means to localize a vehicle's \gls{UE}, synchronize it to the \gls{BS}, and determine the vehicle's heading, all with a single \gls{BS} \cite{kim20205g}. Furthermore, since the incidence points of the multipath components can be determined, the combined problem of \gls{UE} localization and scatter point detection and localization, is a classic \gls{SLAM} problem, often termed channel-\gls{SLAM}  \cite{ulmschneider2020multipath} or radio-\gls{SLAM}~ \cite{leitinger2019belief}. 

Variations of the single-\gls{BS} positioning problem have focused on (i) generalizing to more realistic channel models or (ii) improving the \gls{SLAM} methodology. In the first category, we count \cite{nazari2023mmwave,wen20205gb}, which considers both specular and diffuse multipath, showing that both types of multipath convey geometric information, useful for \gls{SLAM}, based on models from \cite{wen20205g}. In the second category, model-based algorithmic refinements were introduced in \cite{Chen2022SAM}, showing how efficient snapshot positioning can be performed by a combination of \gls{TDoA} and difference-of-direction measurements. As a step further, \cite{kaltiokallio2024robust} proposes a robust snapshot radio \gls{SLAM} algorithm that can cope with outlier measurements originating from multi-bounce reflections in mixed \gls{LoS}-\gls{NLoS} conditions, which are typical in deep urban navigation scenarios. In \cite{general_SLAM_2024}, both simulation and indoor experimental results have been provided to evaluate the performance of various single-\gls{BS} \gls{SLAM} algorithms under realistic conditions. Furthermore, moving beyond the standard way of using first-order interactions, \cite{bistatic_doubleBounce_2024} exploits multi-bounce reflections to map the environment with mmWave signals. Finally, machine learning approaches for single-\gls{BS} positioning were explored in \cite{gante2020dethroning, bader2023step}, \rev{\cite{lv2022deep, whiton2024wiometrics,ML_loc_lidar_5g_2021,gao2023single,talebian2025robust,yuhao_unilocpro_2025}}, as they can harness non-geometric information in rich channels and thus remove the need for many antennas and large bandwidth. Here, \cite{lv2022deep} demonstrated sub-meter accuracy (90\% percentile) using a deep learning end-to-end algorithm, while \cite{gante2020dethroning} introduced the concept of beamformed fingerprints, used to position even in harsh \gls{NLoS} environments with high energy efficiency. Moreover, \cite{bader2023step} explored ensemble learning techniques to classify reflection orders of multipath signals, optimizing positioning accuracy in dense urban environments, and validated the approach via a ray-tracing-based 5G simulator.  \rev{In \cite{whiton2024wiometrics}, an artificial \gls{NN}-based fingerprinting method was proposed to localize a vehicle equipped with a massive \gls{MIMO} receiver mounted on its roof using downlink LTE signals. The work in \cite{ML_loc_lidar_5g_2021} introduces a convolutional \gls{NN}-based architecture for localization using lidar and 5G mmWave measurements, including \gls{RSS}, \gls{AoA}, \gls{AoD} and \gls{ToA}, while \cite{gao2023single} and \cite{talebian2025robust} develop a \gls{ML}-based \gls{LoS} identification method. In \cite{yuhao_unilocpro_2025}, a hybrid single-BS method is proposed that combines model-based estimation and channel charting-based unsupervised learning (the reader is referred to \cite[Sec.~IV]{zhou20245g} for a broader review of \gls{ML}-based 5G localization methods).} Finally, \cite{angleSLAM_5G_2023} presents an angle-based \gls{SLAM} approach designed to work within the constraints of the 5G NR framework, utilizing a single \gls{BS}. The study utilizes a 28 GHz mmWave platform and employs an angle-only \gls{SLAM} algorithm, achieving sub-meter localization accuracy without strict synchronization requirements, even in complex indoor environments.

\subsubsection{\rev{FR2 (mmWave)} Multi-BS Solutions} 
Vehicular positioning with multiple \glspl{BS} in 5G mmWave systems has recently attracted significant attention due to the benefits it offers in terms of localization accuracy and robustness. Approaches in this domain are generally categorized into snapshot-based and tracking-based methods. 

In \textit{snapshot-based positioning}, \gls{ToA} and \gls{TDoA} based \textit{trilateration} are commonly employed. For instance, \cite{saleh2021evaluation} investigates the effect of 5G cell densification on accuracy using \gls{ToA} measurements from multiple \glspl{BS} in a realistic vehicular scenario, which indicates sub-meter accuracy with an inter-cell spacing of $160 \, \rm{m}$ for a vehicle moving at $35 \, \rm{km/h}$. Similarly, \cite{vehPos5G_2016} performs \gls{ToA}-based trilateration positioning with multiple \glspl{BS}. Moreover, \cite{perera2023gdop} proposes a \gls{GDOP} based \gls{BS} selection algorithm in \gls{TDoA} based multi-\gls{BS} positioning in mixed \gls{LoS} and \gls{NLoS} environments, showing more than an order-of-magnitude improvement over the case without \gls{BS} selection in the 90\% percentile accuracy metric in urban scenarios. Finally, \cite{rt_sub6_mmWave_comp} carries out a comparison of \gls{TDoA}-based sub-6 GHz and mmWave positioning in industrial environments characterized by dense clutter, using ray-tracing data, showing 2D 90\% percentile positioning error of 1.2 m in both bands in a static scenario with four \glspl{BS} deployed at the corners of a 29 m x 25 m room.

\textit{Triangulation}-based techniques, that solely rely on angle-based measurements, are also prevalent in the vehicular snapshot positioning literature. For instance, \cite{heimann2020cross} performs angle-only mmWave positioning using \gls{UL}-\gls{AoA} and \gls{DL}-\gls{AoD} measurements with multiple \glspl{BS} either using wide or narrow beams. In \cite{spatial_signal_5g_learning_2023}, an end-to-end learning approach for joint \gls{BS} beamforming and \gls{UE}-side receiver optimization is proposed for 5G \gls{AoD}-based downlink localization with multiple \glspl{BS}, showing robustness against hardware impairments including mutual coupling and element spacing perturbations. 

\textit{Hybrid} methods, which combine multiple measurement types, are also explored in snapshot settings. In \cite{saloranta2018novel}, mmWave \gls{MIMO} downlink scenario with multiple \glspl{BS} is considered, where \gls{AoA}, \gls{AoD} and delay measurements of the \gls{LoS} links are used for positioning. An optimal power allocation strategy among multiple \glspl{BS} and beams is developed to minimize the \gls{CRB} on position estimation, indicating that deploying more \glspl{BS} yields more energy-efficient and accurate solutions than increasing the power of each beam. \rev{In \cite{DL_MultiTask_2024}, a \gls{DL}-based multi-\gls{BS} positioning algorithm is introduced, where the \gls{CSI} fingerprints of multiple \glspl{BS} are combined in either early fusion (fusion of the \glspl{CIS}) or late fusion (fusion of per-BS position estimates).}  \cite{5G_urban_sim_VTC_2023} assesses the positioning performance of 5G mmWave in urban scenarios using ray-tracing data, comparing single- and multiple-\gls{BS} configurations. The study confirms the overall effectiveness of multi-\gls{BS} approaches while pointing out the associated challenges for time-based algorithms such as stringent network synchronization requirements (i.e., sub-ns asynchronicity for a sub-meter accuracy). In scenarios with suboptimal synchronization, hybrid positioning strategies that integrate \gls{RAT}-internal time and/or angle measurements with \gls{RAT}-external \glspl{GNSS} data in single-\gls{BS} setups have been found to surpass multi-\gls{BS} solutions. 

In \textit{tracking}-based approaches, multi-\gls{BS} configurations are designed to maintain real-time localization of moving vehicles, often employing adaptive algorithms to handle dynamic channel conditions. In \cite{saleh2021vehicular}, the authors propose an \gls{EKF} approach for 5G mmWave \gls{TDoA} based positioning, where the \gls{EKF}'s measurement covariance matrix is dynamically tuned to exclude \glspl{BS} inducing high linearization errors.  An alternative approach in multi-\gls{BS} positioning and tracking is to employ a distance ratio-based positioning method \cite{DRSS_5G_2024}, which extracts the difference of received signal strength measurements to achieve an accuracy of 43.8 m (90\% percentile) in an outdoor experimental setup with six \glspl{BS}. In addition, multi-\gls{BS} positioning found applications in industrial 5G mmWave deployments \cite{rastorgueva2020networking}, where \gls{DL} \gls{RSRP} measurements at the \gls{UE}, represented by an automated guided vehicle, were utilized to simultaneously localize the \gls{UE} and estimate orientation uncertainties of \glspl{BS}. \rev{Furthermore, a Bayesian \gls{NN}-based multi-\gls{BS} tracking system is proposed in \cite{6G_BNN_pos_ICC} for localization in complex urban scenarios, exploiting the full CSI in a MIMO-OFDM setting.} 

Tracking approaches using hybrid measurements are also widely adopted in the multi-\gls{BS} positioning literature. In \cite{5G_multiEpoch_Sync}, authors introduce a multi-epoch hybrid positioning algorithm using \gls{ToA} and \gls{AoD} measurements by exploiting the temporal correlation of clock offsets of multiple \gls{BS}, achieving sub-meter positioning \gls{RMSE} in a scenario with four \glspl{BS} placed 100 m away from one another. To cope with large path loss and signal blockages at mmWave, \cite{6G_BNN_Pos_2024} introduces a real-time Bayesian \gls{NN} approach for positioning and tracking with multiple \glspl{BS} in urban environments to estimate both positions and uncertainties using \gls{ToA}, \gls{AoA}, and \gls{RSS} measurements. By employing a teacher-student Bayesian \gls{NN} framework, the method enables robust, real-time location tracking and achieves sub-meter accuracy, outperforming traditional deep learning and geometric-based techniques under challenging signal conditions such as \gls{LoS} blockage to all \glspl{BS}. In \cite{talvitie2019radio}, an \gls{EKF}-based tracking method is proposed that uses \gls{TDoA} and \gls{AoA} 5G measurements obtained from the \gls{UL}-\gls{SRS} signals transmitted by a train to track its position, velocity, and clock offsets. The proposed approach achieves 2.8 m positioning accuracy with 95\% of availability. In \cite{karfakis2023nr5g}, a novel 5G NR \gls{SLAM} framework is introduced for mobile robot localization with several \glspl{BS}. Relying on the factor graph \gls{SLAM} algorithm, the framework combines Bayesian filters with downlink \gls{ToA} and \gls{RSSI} measurements, the latter being used for correction purposes, to estimate robot states, including position and heading.

\subsubsection{Cooperative and Sidelink Solutions}
As more and more vehicles become connected, \gls{CP} becomes a natural solution for enabling and enhancing vehicular positioning. Such a view is further reinforced and amplified with the introduction of sidelink communication in 5G-\gls{NR}, which enables sidelink relative measurements as mentioned earlier~\cite{kim20205g, ge2024v2x}. The cooperation between vehicles in such networks can be categorized into two forms, namely \textit{explicit} and \textit{implicit} cooperation. Through a vehicular network, explicit \gls{CP} methods share the explicit position of the vehicles and inter-vehicle geometry measurements (e.g., distance)~\cite{wang2020three}. On the other hand, implicit \gls{CP} methods localize non-cooperative physical features (such as people, traffic lights, or inactive cars) in the surrounding areas and use them as common noisy reference points~\cite{barbieri2023implicit}. The information acquisition and sharing phase can also be realized by other standard communication technologies such as Bluetooth, IEEE 802.15.4, Zig-Bee, Wi-Fi-Direct, and 4G LTE, which have been evaluated in~\cite{severi2018beyond}. The \gls{CP} solutions can be further classified into (i) learning-based~\cite{tedeschini2023cooperative} and (ii) model-based approaches, depending on the problem formulation~\cite{wymeersch2009cooperative}. Model-based approaches can categorized further into non-Bayesian multidimensional scaling~\cite{chen2020distributed}, \gls{MLE}-based methods~\cite{chen2024multi}, and Bayesian methods like expectation-maximization~\cite{wang2022cooperative}. A more detailed summary can be found in~\cite{zhou20245g}.

Cooperative positioning does not come without its challenges, though. Chief among these challenges are resource allocation, security, and privacy aspects, which cannot be ignored. By default, sidelink communication increases the complexity of the communication network, making resource allocation (e.g., power allocation and scheduling) a critical factor in \gls{CP}, especially for applications with limited resources. To increase the information gain from other vehicles through cooperation, decentralized resource allocation is needed, which can be solved via convex optimization~\cite{dai2014distributed} and deep reinforcement learning~\cite{peng2019decentralized}. During the information exchange phase, a large volume of data occupies the communication resources. In addition, users may tend to refuse to upload private location data. As a consequence, distributed learning methods, such as federated learning~\cite{kong2021fedvcp}, are required to provide \gls{CP} with much-needed robustness, security, and reduced computational costs.

\subsubsection{\rev{Positioning in FR1 (Sub-6 GHz)}}
Despite the recent booming interest in mmWave \cite{liu2022integrated}, sub-6 GHz localization still has great potential in vehicular scenarios \cite{bartoletti2021positioning,ko2021v2x}. When the \gls{LoS} link can be resolved (e.g., through the use of large antenna arrays), high-accuracy positioning becomes possible (e.g., over sidelink \cite{ge2024v2x,v2x_2024}). Sub-6 GHz localization has certain crucial advantages over mmWave, including improved coverage and less blockage due to obstacles \cite{blanco2022augmenting}. At sub-6 GHz bands, atmospheric attenuation and absorption by gases will generally be lower than mmWave bands, leading to significantly enhanced link budgets \cite{bjornson2019massive}. In addition, lower frequencies induce slower channel variations \cite{bjornson2019massive}, which means that sub-6 GHz can support much longer coherence integration times than mmWave bands, enabling localization of vehicles with higher mobility. Hence, sub-6 GHz provides favorable propagation conditions that can facilitate the localization of high-mobility vehicles, possibly far away from \glspl{BS} or \glspl{RSU}.
Moreover, integrated mmWave and sub-6 GHz vehicular networks offer a promising approach that can combine the benefits of sub-6 GHz (better coverage, mobility support, less vulnerability to blockage) and mmWave (high directivity, large bandwidths, sparse channel) bands \cite{semiari2019integrated,yan2019machine}. For example, localization information obtained through sub-6 GHz transmission can be exploited to shorten the beam training interval at mmWave \cite{semiari2019integrated}, leading, in turn, to high-quality and low-latency location estimates for vehicular users. Moreover, \gls{CSI} at sub-6 GHz bands can be used to localize vehicles by employing machine learning algorithms, which assist sub-6 GHz-to-mmWave handover mechanisms \cite{yan2019machine}. \rev{Furthermore, self-attention and channel attention mechanisms can be employed to improve \gls{CSI}-based sub-6 GHz deep learning-aided positioning \cite{dl_attention_2024}.}

\subsection{\rev{A Glimpse into 6G Positioning Technologies}} 
\rev{While 5G's mmWave has received a great deal of attention for accurate positioning, 6G positioning is expected to leverage various technologies like \glspl{RIS}, new frequency ranges, particularly cmWave (7–24 GHz) and sub-THz bands (100–300 GHz), to improve resolution, robustness, and environmental awareness significantly. In addition to those technologies, security and privacy aspects of positioning are also taking a much needed attention. In the following, we briefly review works in these domains that pertain to vehicular positioning.}

\subsubsection{RIS-based Solutions}
A \gls{RIS} is a two-dimensional surface made of meta-materials with reconfigurable impedance, allowing it to control electromagnetic wave interactions such as scattering, absorption, reflection, and diffraction. This enables the \gls{RIS} to precisely direct reflected signals, effectively extending wireless communication coverage beyond \gls{LoS} limitations~\cite{bjornson2022reconfigurable}. In addition to that, \glspl{RIS} are cost-effective and are less power-hungry compared to classical \glspl{BS}. Hence, it is envisioned that \gls{RIS} will play a crucial role in beyond 5G and 6G communication systems. However, \gls{RIS} benefits do not stop at communication services, as recent research works on \gls{RIS} have shown great benefits for localization and mapping in terms of performance, energy consumption, and cost~\cite{wymeersch2020radio}. For instance, localizing a vehicle with a simple communication scenario that involves a single-antenna \gls{BS} and a single-antenna \gls{UE} is traditionally not possible. However, with the aid of \gls{RIS}, additional \gls{RIS}-based \gls{AoD} and \gls{ToA} measurements can now be acquired (given sufficient bandwidth for delay estimation), which enables the localization and synchronization of the \gls{UE}~\cite{keykhosravi2021siso}. Likewise, localizing and synchronizing a \gls{UE} with a single snapshot using a single cellular-based \gls{LEO} satellite is not possible. However, adding a single \gls{RIS} solves the problem, as it provides extra \gls{AoD} and \gls{ToA} measurements as shown in~\cite{saleh_6g_2024}. It was also shown that with an antenna array equipped at a terrestrial \gls{BS}, the bandwidth requirement can be relieved due to the \gls{AoD} estimation~\cite{fascista2021ris}. It is worth noting that the \gls{UE} is expected to have high mobility in a practical vehicular communication system. Such mobility will cause Doppler-induced fast-time phase rotation and slow-time phase progressions across consecutive OFDM symbols. Those effects should be considered in the system model to ensure accurate results~\cite{keykhosravi2022ris}. A more thorough study on both vehicular localization and tracking by leveraging \gls{RIS} to mitigate multipath fading, Doppler effects, and tracking delays can be found in~\cite{RIS_B5G_ITS_2024}, and its open challenges are highlighted in~\cite{6g_irs_vtm_2024}.

\gls{RIS} comes in many shapes and forms, one of which is transparent \gls{RIS}, reported in~\cite{kitayama2021transparent}. With the aid of transparent \gls{RIS}, \glspl{RIS} can now be mounted on the windows and ceilings of buildings and vehicles without blocking the view through them. As a consequence, the wavefront curvature of the signal can be exploited for localization, even in challenging scenarios where hardware impairments~\cite{ozturk2024ris} and \gls{LoS} blockage~\cite{dardari2021nlos} are present. This form of localization is known as near-field localization and builds on the same principles as for physically large or distributed antenna structures~\cite{chen20246g}. It is worth noting here that, unlike distributed \gls{MIMO} systems, \gls{RIS} can be configured by a low-cost control unit instead of the dedicated hardware design of a \gls{BS}.

In the above-mentioned works, we mainly discussed how \gls{RIS} could support localization in addition to the existing \glspl{BS} (with or without \gls{LoS}). Another use case is zero access points, where no \gls{BS} is involved and the \gls{UE} is equipped with a full-duplexing array (e.g., a vehicular radar)~\cite{kim2023ris}. In that case, the \gls{UE} transmits signals and then receives the reflected signal from the RIS, based on which the \gls{UE} can localize itself. When multiple \glspl{RIS} or multiple \glspl{UE} are available, cooperative RIS-aided positioning is also possible~\cite{chen2023riss, chen2024multi}.

Despite the fact that \gls{RIS} can greatly benefit localization, the design of the \gls{RIS} coefficients is not an easy task. For a large \gls{RIS} with thousands of elements, the \gls{RIS} coefficients could be optimized based on communication or localization performance, where a tradeoff is needed~\cite{luo2022uav}. Additionally, interference becomes a real challenge when multiple \glspl{RIS} are considered in the system for coverage extension. However, this challenge can be addressed by utilizing orthogonal beams~\cite{keykhosravi2021semi}. Finally, the calibration of RIS anchors (position and orientation) must be taken into account, which leads to joint localization and calibration approaches as shown in~\cite{zheng2023jrcup, ghazalian2024joint}.

\subsubsection{\rev{Positioning in cmWave Range (7-24 GHz)}}
\rev{The emerging cmWaves have recently been identified as a key resource for 6G networks because it provides a favorable balance between coverage, penetration, and resolution, unlike the higher-loss mmWave FR2 band~\cite{FR3-1}. As shown in recent analyses of upper-mid-band cellular operation~\cite{FR3-2}, cmWaves offers wider coverage and better obstacle penetration than FR2 while still supporting meaningful delay and angular resolution when used with massive \gls{MIMO} or distributed-\gls{MIMO} architectures. Measurement campaigns conducted at 16.95 GHz in an indoor factory environment further reveal that cmWaves exhibit reduced path loss under \gls{LoS} but significantly higher attenuation in \gls{NLoS} scenarios due to material interaction and dense scattering~\cite{FR3-3}. Such frequency-dependent propagation behavior directly impacts sensing quality and positioning robustness.}

\subsubsection{Positioning in Sub-THz (100-300 GHz)}
From a communication perspective, the high data rate requirement (e.g., processing data in the cloud) pushes the central frequency to the THz band~\cite{sarieddeen2020next}. From the localization point of view, ultra-massive \gls{MIMO} and even larger bandwidth can provide unparalleled localization performance~\cite{chen2022tutorial}. In addition, the near-field scenarios are able to capture the curvature-of-arrival of signals and perform beam focusing, which increases localization accuracy with mitigated interference~\cite{chen20246g, abu2021near}. 
Initial positioning experiments at 142 GHz have been performed to achieve a mean accuracy of 24.8 cm~\cite{kanhere2021outdoor}. Design and experimental validation of radio \gls{SLAM} at the THz band has been performed in~\cite{lotti2023radio}. \rev{The work in \cite{chiu20246g} proposes a \gls{DL}-based sub-THz indoor localization method using convolutional-\gls{NN} with and without attention mechanisms and demonstrates cm-level accuracy using ray-tracing data.}
However, we should be clear that these benefits do not come for free. For instance, the high path loss at sub-THz bands limits the localization service area. Although a large array can combat such a loss, the resulting narrow beamwidth and wideband effect pose challenges in beam training~\cite{tan2021wideband} and mobile scenarios~\cite{azari2022thz}. Also, the processing of a large volume of data requires extra effort to perform the localization tasks. Furthermore, when we seek high-accuracy localization, the effect of model mismatch cannot be ignored. Additionally, the phenomenon of performance saturation has been observed in several works, such as channel model mismatch~\cite{chen2022channel}, and hardware impairments~\cite{chen2023modeling}. Hence, there is a need to develop more accurate models for sub-THz bands to reduce the effect of model mismatch and use data-driven approaches~\cite{fan2021siabr} to learn these unknowns.

\subsubsection{\rev{Security and Privacy Aspects}}
\rev{Both security and privacy threats must be considered in cellular-based localization, as they affect the system in fundamentally different ways. Security threats aim to manipulate or disrupt the localization process, for example, through jamming, spoofing, or injecting false reflections, allowing adversaries to corrupt positioning results~\cite{Sec-1}. Attackers may also manipulate RIS reflections to degrade localization accuracy or force incorrect beam alignment~\cite{Sec-2}. In contrast, privacy threats do not modify the localization output. Instead, they exploit the fact that cellular pilots and waveforms inherently encode geometric features, such as range, angle, and Doppler, to infer sensitive information, including user location, trajectory, and behavioral patterns~\cite{Sec-3}. Such passive sensing enables identity inference, profiling, and long-term user tracking, creating privacy risks beyond traditional communication confidentiality. To enable secure and privacy-aware localization, countermeasures include secure beamforming, artificial-noise injection, and sensing-aware waveform control, all of which help reduce both attack impact and information leakage~\cite{Sec-4}.}

\subsection{Challenges and Open Problems}
Cellular network positioning, especially with 5G and beyond wireless systems, faces several key challenges. Multi-bounce reflections and environmental clutter may impede accurate signal detection and target association~\cite{kaltiokallio2023multi, ge2022computationally, bistatic_doubleBounce_2024}, while hardware impairments such as non-linearities and phase noise further degrade signal quality and channel estimation performance, resulting in severe degradations in localization performance \cite{chen2023modeling,HWI_TVT_2023,5G_sim_HWI_2023}. Effective calibration of antenna arrays at \glspl{BS} and \glspl{RIS} can be critical~\cite{ULA_failures_2020,calibration_5g_2023,ozturk2024ris}, particularly in multi-device setups. Additionally, achieving precise synchronization among \glspl{BS} as well as between \glspl{UE} and \glspl{BS} proves to be essential for localization relying on time- and phase-based measurements~\cite{RIS_Alessio_JSTSP_2022, chen2024multi, keykhosravi2022ris,fascista2024joint,5G_sim_HWI_2023}. An effective remedy against poor synchronization is to employ hybrid positioning techniques that combine \gls{RAT}-based single-\gls{BS} time and angle measurements as well as \gls{RAT}-external measurements from technologies such as \gls{GNSS} \cite{5G_urban_sim_VTC_2023}. Additionally, balancing accurate location data with user privacy poses a significant challenge~\cite{feng2024framework}, while poor trustworthiness of data shared among vehicles over 5G \gls{V2V} links can jeopardize the localization process, potentially leading to accidents \cite{trust_5G_loc_2020}. Another challenge with cellular positioning pertains to coordination in heterogeneous networks involving devices with a wide range of capabilities, which adds complexity due to varying standards and technologies \cite{hetMulti_2024}. Furthermore, improving positioning and tracking performance in dynamic environments necessitates the use of accurate mobility models \cite{vehicleTrack_2024_TVT}. Integrating relative position information from cooperative and sidelink measurements with global information originating from \glspl{BS} can complicate vehicular positioning \cite{soatti2018implicit}. Lastly, to better capture the characteristics of different propagation environments, developing accurate channel models for near-field~\cite{chen20246g,giovannetti2024performance} and extended targets~\cite{garcia2022cramer} is essential for designing powerful localization and velocity estimation algorithms.

\section{IEEE-based Positioning} \label{Sec: IEEE}
\begin{figure}[t]
\begin{center}
\includegraphics[width=0.5\textwidth]{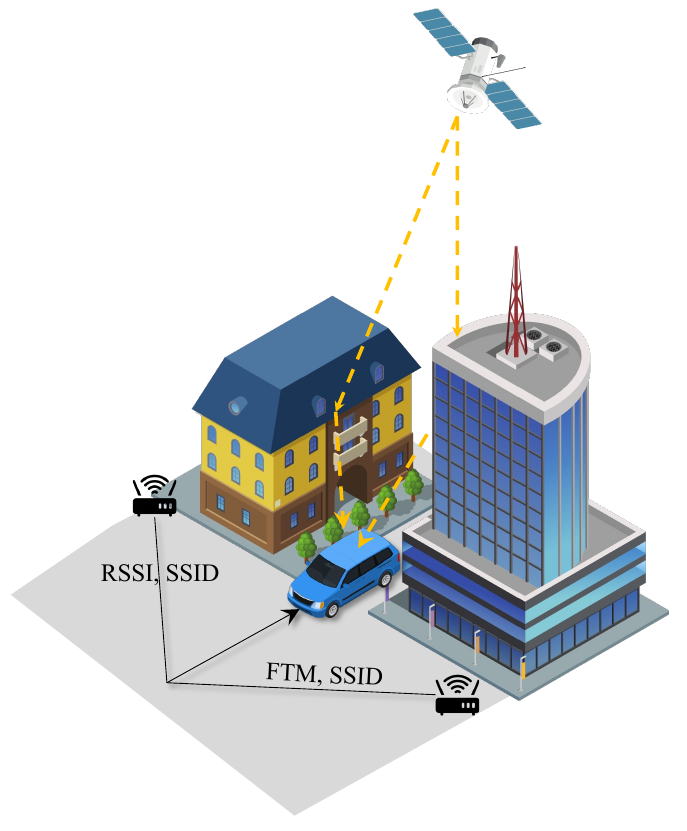}
\caption{\rev{Urban canyon and Wi-Fi-based positioning solutions.}}
\label{fig:WiFicanyon}
\end{center}
\end{figure}

IEEE has standardized several wireless technologies that can support positioning, including Wi-Fi, \gls{UWB}, and Bluetooth. These technologies, designed for short-range communication, are primarily suited for indoor positioning. However, they also show promise in vehicular applications, especially in dense urban environments and parking lots\rev{, as shown in Fig.~\ref{fig:WiFicanyon},} or when communication with surrounding vehicles is possible. Direct IEEE-based positioning is facilitated through various geometric measurements, which can be fused with other sensors to enhance vehicular positioning accuracy. Additionally, IEEE-based \gls{V2V} communications also enable the exchange of positioning and perception data with other vehicles, which facilitates cooperative sensor fusion. In this section, we provide an overview of how IEEE-based wireless technologies evolved to support vehicular positioning through a plethora of IEEE standards and amendments. We also outline the fundamentals of positioning using IEEE-based technology. Finally, we present the contemporary research that utilizes these technologies for vehicular positioning.

\subsection{History of Positioning in IEEE Standards}\label{IEEEevol}
IEEE 802 is an IEEE standard family that comprises 24 sub-families of network standards (IEEE 802.1 through 802.24). These standards cover a wide range of network types, up from small-scale \glspl{PAN}, standardized in IEEE 802.15 (e.g., Bluetooth, \gls{UWB}, \gls{RFID}, Zigbee, and visible light communications), to larger \glspl{LAN}, standardized in IEEE 802.11 (e.g., Wi-Fi), and \glspl{MAN}, standardized in IEEE 802.16 (e.g., WiMAX, the main competitor to \gls{3GPP}'s LTE).\footnote{The other 21 sub-families are out of the scope of this survey.} Each sub-family can have groups within it, focusing on different aspects/technologies. For instance, IEEE 802.15 (the \gls{PAN} sub-family) includes IEEE 802.15.1, which focuses on Bluetooth standards; IEEE 802.15.4, which focuses on low-rate wireless \gls{PAN} technologies like Zigbee in the original IEEE 802.15.4 standard, \gls{UWB} in IEEE 802.15.4a and IEEE 802.15.4z amendments, and \gls{RFID} in IEEE 802.15.4f amendment; and IEEE 802.15.7, which focuses on visible light communications. Here, the letters at the end (e.g., a, f, and z) refer to amendments made over time to the original standards. In the following, we present the standardization history of IEEE 802.11 and IEEE 802.15.

\subsubsection{Wi-Fi Positioning and Sensing Standards (IEEE 802.11)} 
The \gls{LAN}'s Wi-Fi IEEE 802.11 standard was established in 1997 (in parallel with \gls{3GPP}'s development of 3G), and has evolved over the years through a series of amendments (e.g., 802.11a, 802.11b, etc.). A commercial branding name is assigned to some of the amendments. For instance, Wi-Fi 4, 5, 6, 7, and 8 are designated to IEEE 802.11 amendments n, ac, ax, be, and bn, respectively. The noteworthy IEEE 802.11 amendments that relate to our topic are IEEE 802.11p and bd, which relate to vehicular communications, and IEEE 802.11mc, az, bk, and bf, which relate to Wi-Fi positioning and sensing in general. However, prior to any Wi-Fi-based positioning standards, researchers have already been utilizing Wi-Fi \glspl{AP} opportunistically for positioning purposes using \gls{RSS}-based fingerprinting techniques since the late 90s, which did not require any specific standardization. This is usually called the first generation of Wi-Fi positioning; among four generations. 

The introduction of \glspl{FTM} in IEEE 802.11mc (also known as IEEE 802.11REVmc) in 2016 marked the start of the second generation of Wi-Fi positioning~\cite{802.11_2016}. \gls{FTM} is a protocol that enables the measurement of the \gls{RTT} between the user and the Wi-Fi \gls{AP}. The initial maximum bandwidth allocated to \gls{FTM} was 80 MHz in 2016. The first generation of Wi-Fi devices that incorporated these capabilities were Wi-Fi 5 devices manufactured after 2016. 

In 2023, IEEE 802.11az, also known as next-generation positioning (NGP), marked the start of the third generation of Wi-Fi positioning~\cite{802.11az}. The new standard increased the bandwidth of \gls{FTM} signals to 160 MHz (based on the IEEE 802.11ax Wi-Fi 6 standard developed earlier in 2021) and enabled \gls{FTM}-based \gls{TDoA} measurements. It also added support to \gls{MIMO} setups, which enabled angle-based measurements like \gls{AoA} and \gls{AoD}, albeit with low accuracy due to the low antenna count on Wi-Fi access points and user equipment, due to the operation in 2.4, 5, and 6 GHz bands. Finally, IEEE 802.11az brought much-needed security and privacy measures to counter spoofing and other types of attacks. 

The fourth generation of Wi-Fi positioning has just started in May 2025, with the introduction of IEEE 802.11bk, named 320 MHz positioning.\footnote{Although the standard was approved and is currently active, the standard document is yet to be published. Hence, information in this paragraph is taken from available online drafts of the standard document. The same goes for IEEE 802.11bf.} As the name suggests, the standard increases the bandwidth of \gls{FTM} signals to 320 MHz, making them close in bandwidth to 5G's \gls{PRS} signals. The increase to 320 MHz bandwidth in the 6 GHz band comes as a consequence of the increase of communication bandwidth to 320 MHz in Wi-Fi 7 (IEEE 802.11be, rolled out in 2024). In parallel, IEEE 802.11bf, named enhancements for wireless local area network (WLAN) sensing, introduced sensing capabilities to Wi-Fi systems below 7.125 GHz and above 45 GHz in 2025 \cite{sahoo_sensing_2024, WiFiSensing}. These standards mirror the \gls{ISAC} standardization efforts led by \gls{3GPP} in late 5G-Advanced and early 6G (i.e., Releases 19 and 20). 

\subsubsection{Wi-Fi Vehicular Communications Standards} \textit{IEEE 802.11p}, named wireless access in vehicular environments (WAVE), was developed in 2010 to enable \gls{ITS} through \gls{V2V} and \gls{V2I} communications~\cite{802.11p}. Since IEEE 802.11p was not explicitly designed for positioning, it lacks dedicated localization features. Nevertheless, IEEE-based wireless systems have occasionally been referenced for their potential to support outdoor vehicular positioning \cite{arena2020review, eichler2007performance}. Some of the notable technologies that were standardized based on IEEE 802.11p are \gls{DSRC} in the US and \gls{ITS}-G5 in Europe. The allocated frequency band for IEEE 802.11p communications is around 5.8-5.9 GHz (total bandwidth of 30-75 MHz), with a per channel bandwidth of 10 MHz.\footnote{This band is also close to, and sometimes contested by, cellular-\gls{V2X} (C-\gls{V2X} PC5) bands. The regulation and allocation of these bands differ slightly from one continent/country to the other.} \textit{IEEE 802.11bd}, named enhancements for next generation V2X, is the successor of IEEE 802.11p, and was developed in 2022~\cite{802.11bd}. This amendment increased the bandwidth of \gls{V2X} communications to 20 MHz, and introduced support for \gls{MIMO} and mmWave (60 GHz) setups.

\subsubsection{UWB and Bluetooth Standards (IEEE 802.15)}
The earliest attempt towards standardizing positioning-capable IEEE 802 technologies was through the introduction of \gls{UWB} in 802.15.4a in 2007~\cite{9810941,802.15.4}. With bandwidths of up to 500 MHz, \gls{UWB} significantly improved short-range positioning accuracy through \gls{RTT} and \gls{TDoA} measurements. Commercial \gls{UWB} devices emerged around 2010 and have gained prominence, particularly in consumer products like Apple’s AirTag~\cite{UWB_wiki}. The latest \gls{UWB} amendments were in IEEE 802.15.4z in 2020, which added \gls{AoA} measurements to the \gls{UWB} toolkit in addition to other functionalities to increase \gls{UWB}'s accuracy and integrity~\cite{802.15.4z}. Although the \gls{UWB} standard gained popularity for its high ranging and positioning accuracy, restrictions in the US and EU limit its use as a fixed outdoor device \cite{ETSI_UWB}, allowing it only for ranging between moving objects in outdoor environments. This makes \gls{UWB} suitable for cooperative vehicular positioning.

Bluetooth, named after Harald Bluetooth Gormsson, a 10th-century King of Denmark, was invented in 1989 by Ericsson. It was standardized in 1999 by the Bluetooth Special Interest Group (SIG) and first commercialized in 2000. Early versions of Bluetooth were also adopted by IEEE as part of the 802.15.1 standard, which covered Bluetooth up to version 1.2 \cite{802.15.1}. However, after 2005, IEEE discontinued updates to 802.15.1, and Bluetooth has since been fully maintained by SIG (covering versions 2.x through 6.x). Although the original IEEE-based Bluetooth standards did not include explicit positioning capabilities, they introduced support for Bluetooth-based \gls{RSSI} measurements, which have since been used extensively for coarse proximity estimation. Later, SIG added direction finding in Bluetooth 5.1 (2019), enabling coarse \gls{AoA} and \gls{AoD} measurements~\cite{BT5.1}. More recently, Bluetooth 6.0 (2024) introduced channel sounding, enabling two new types of high-accuracy ranging: (i) high accuracy distance measurement (HADM), an \gls{RTT} measurements leveraging ~80 MHz of bandwidth, and (ii) phase-based ranging (PBR), which uses carrier-phase measurements at 2.4 GHz~\cite{BT6.0}. 


\subsection{IEEE-based Positioning Fundamentals}
IEEE-based vehicular positioning methods can be generally divided into two categories: methods that utilize \gls{V2I} communications (anchor-based) and those that utilize \gls{V2V} communications (cooperative positioning). In the following, we briefly highlight the various positioning measurements and techniques used in each category. \rev{Readers interested in a generalized channel model for wireless technologies are referred to Appendix \ref{appendix}.}

\subsubsection{Anchor-based Positioning (V2I)}
Techniques in this category are either geometric-based (akin to satellite and cellular positioning methods) or fingerprinting-based (utilizing end-to-end \gls{ML} techniques). Geometric-based methods mainly utilize \gls{RTT}/\gls{TDoA} measurements, and rarely use \gls{AoA}/\gls{AoD} measurements. To use those measurements for positioning, we need to ensure (i) \gls{LoS} between the \gls{AP} and the user, (ii) resolvability of the \gls{LoS} path from the multipath components, and (iii) access to enough \glspl{AP} to solve the positioning problem. Unfortunately, most of these requirements are not met in vehicular positioning scenarios, where the \gls{AP} is placed indoors (e.g., Wi-Fi anchors). However, when those anchors are utilized as \glspl{RSU}, then this category of positioning becomes viable, especially when fused with other onboard sensors.

On the other hand, fingerprinting methods utilize raw \gls{SSID}, \gls{CSI}, or \gls{RSSI} measurements. Fingerprinting is performed in two main phases, an offline (training) phase and an online positioning phase. In the training phase, the positioning area is divided into a grid. The user is then placed at each grid tile, and the aforementioned measurements are collected and labeled with the position of the given tile. Each measurement–label tuple is considered a fingerprint and is fed to an \gls{ML} algorithm to train it. In the online phase, raw measurements are collected and processed by the trained \gls{ML} algorithm to find the closest fingerprint/position. The accuracy of fingerprinting methods depends on (i) the resolution of the training grid, (ii) the accuracy of the labeling process, (iii) the dynamic changes of the surrounding environment, and (iv) the \gls{ML} algorithm used. Additionally, unlike geometric-based methods, they do not require \gls{LoS} access to \glspl{AP}, and positioning is possible with access to a single \gls{AP}. Hence, fingerprinting methods are more suitable in stable indoor parking lots where only a single \gls{NLoS} \gls{AP} might be available. Likewise, fingerprinting methods are not suitable for highly dynamic environments like highways and urban environments.

\subsubsection{Cooperative Positioning (V2V)}
Techniques in this category can be categorized into relative and absolute positioning techniques. Both techniques conduct range, angle, and/or Doppler measurements between the user and the surrounding vehicles, mainly via \gls{DSRC} and \gls{UWB}. This enables the user to estimate their position, orientation, and/or velocity with respect to the other vehicles (relative positioning). Absolute positioning takes this a step further by incorporating absolute position estimates of one or more of the surrounding vehicles—usually performed by other absolute positioning onboard sensors—to estimate the user's own absolute position. Of course, positioning errors in the cooperative absolute position data of the other vehicles and relative measurement errors will propagate to the user's absolute positioning error. Those relative and/or absolute positioning estimates are usually fused with other onboard sensors to enhance the overall positioning accuracy.

\subsection{IEEE-based Positioning Solutions}
Before delving into the various research works on IEEE-based positioning techniques, it is worth noting three important characteristics of the IEEE positioning community. First, the IEEE positioning community is more focused on practical implementation and experimentation, as opposed to the 5G positioning community's focus on theoretical and simulation-based research. This is mainly because of the ease of access, openness, decentralization, wide commercialization, and low cost of IEEE-based anchor points and receivers, compared to their cellular counterparts. Hence, researchers worldwide can easily buy and modify these devices, leading to more field experimentation. Second, most IEEE-based positioning works are dedicated to indoor pedestrian scenarios, as opposed to the satellite and 5G positioning communities. Most of those works are not compatible with vehicular applications due to assumptions regarding the dynamics of the user and the surrounding environment. Third, most of the works that are indeed focused on vehicular positioning do integrate the IEEE-based technologies with other onboard sensors, which will be discussed in the next sections. Hence, only a few works remain to be discussed after excluding indoor-positioning and integrated positioning works. In the following, we present the various research works that utilized IEEE-based technologies for vehicular positioning, categorized by the technology used. \rev{A summary of the works is presented in Table~\ref{tab:ieee_localization_summary}.}

\begin{table*}
\centering
\caption{\rev{Summary of IEEE-based vehicular wireless localization works, categorized by technology used.}}
\label{tab:ieee_localization_summary}
\vspace{2mm}
\footnotesize
\renewcommand{\arraystretch}{1.2} 
\begin{tabular}{|m{0.16\textwidth}|m{0.14\textwidth}|m{0.35\textwidth}|m{0.2\textwidth}|}
\hline
\textbf{Technology} & \textbf{Environment} & \textbf{Measurements \& Techniques} & \textbf{Accuracy} \\ \hline
\rowcolor{pastel2} 802.11 Wi-Fi \cite{huang_multi-stage_2020} & Outdoor & RSS, PPRM + KF + LS-TSE + UKF & Meter-level \\ \hline
\rowcolor{pastel2} 802.11 Wi-Fi \cite{8355136} & Outdoor & RSS + SSID, kNN fingerprinting & Tens of meters \\ \hline
\rowcolor{pastel2} 802.11 Wi-Fi \cite{gomez_wifi-based_2023} & Parking & RSS, LSTM neural network fingerprinting& Several meters \\ \hline
\rowcolor{pastel2} 802.11 Wi-Fi \cite{ding_three-dimensional_2022} & Indoor & CSI, RNN + LSTM fingerprinting & Meter-level \\ \hline
\rowcolor{pastel2} 802.11 Wi-Fi \cite{zhang_vehicular_2021} & Toll stations & CSI, DFT + similarity degree fingerprinting & Decimeter-level \\ \hline \hline
\rowcolor{pastel3} 802.11p V2V \cite{liu_hybrid_2020} & Dense urban & Range-rate + pseudo-ranges, CKF & Several to tens of meters \\ \hline
\rowcolor{pastel3} 802.11p V2V \cite{liu_towards_2023} & Indoor parking & RSS, HMM + OTL fingerprinting & Several meters \\ \hline
\rowcolor{pastel3} 802.11p V2V \cite{soatti_implicit_2018} & Urban canyons & Sensed features (radar, lidar), Gaussian message passing & Sub-meter \\ \hline
\rowcolor{pastel3} 802.11p V2V \cite{watta_vehicle_2021} & Road & Range + angle, geometric modeling + NN & 99\% accuracy (detection rate) \\ \hline \hline
\rowcolor{pastel1} 802.15.4 UWB \cite{cao_improving_2022} & Underground & RTT, GMM + NNIMM + KF + MFO & Decimeter-level \\ \hline
\rowcolor{pastel1} 802.15.4 UWB \cite{wu_pedestrian_2024} & Outdoor & RTT, EKF & Sub-meter-level \\ \hline
\rowcolor{pastel1} 802.15.4 UWB \cite{huang_ultra-wideband_2021} & Indoor & TDoA, gradient descent-Taylor & Decimeter-level \\ \hline
\rowcolor{pastel1} 802.15.4 UWB \cite{jiang_practical_2022} & Indoor parking & TDoA, GDOP analysis & Centimeters to tens of meters \\ \hline
\rowcolor{pastel1} 802.15.4 UWB \cite{lee_uwb-based_2021} & Indoor & TDoA, LS & Decimeter-level \\ \hline
\rowcolor{pastel1} 802.15.4 UWB \cite{wen_automated_2020} & Highway/Tunnel & TDoA, LS & Decimeter to meter-level \\ \hline
\rowcolor{pastel1} 802.15.4 UWB \cite{zhang_uwb-based_2024} & Parking & TDoA, KF + EKF & Decimeter-level \\ \hline
\rowcolor{pastel1} 802.15.4 UWB \cite{hu_multipath-assisted_2025} & Underground parking & ToA + AoA, weighted iterative LS + raytracing & Decimeter-level \\ \hline
\rowcolor{pastel1} 802.15.4 UWB \cite{piavanini_experimental_2023} & Race track & TDoA + AoA, UKF + IMM & Decimeter-level \\ \hline
\rowcolor{pastel1} 802.15.4 UWB \cite{wang_uwb_2021} & Outdoor & RTT, non-linear LS + HOMO-LM & Decimeter-level \\ \hline
\rowcolor{pastel1} 802.15.4 UWB \cite{wang_development_2024} & Rail & RTT, direct ranging & Decimeter-level ranging \\ \hline \hline
\rowcolor{pastel5} 802.15.1 Bluetooth \cite{paulino_evaluating_2021} & Indoor warehouse & AoA, LS + KF & Sub-meter to meter-level \\ \hline
\rowcolor{pastel5} 802.15.1 Bluetooth \cite{paulino_self-localization_2023} & Outdoor & AoA, LMS & Sub-meter to meter-level \\ \hline
\end{tabular}
\end{table*}

\subsubsection{IEEE 802.11 (Wi-Fi)}
As mentioned above, Wi-Fi-based positioning works are divided into anchor-based and cooperative solutions. In the following, we present works in both categories.

\textbf{\textit{Anchor-based Wi-Fi Positioning (\gls{V2I}})}: Although most of the standalone IEEE 802.11 anchor-based vehicular positioning works utilize fingerprinting methods, due to the reasons highlighted above, one of the few works using geometric-based Wi-Fi positioning is found in \cite{huang_multi-stage_2020}. Although the authors' main goal is to track pedestrian users in outdoor urban scenarios, their method can be easily extended to vehicular positioning. The presented method utilizes multiple \gls{RSS}-based measurements to infer the range between the user and multiple Wi-Fi \glspl{AP}. The method operates in two main stages: an offline phase where a piecewise polynomial regression model (PPRM) is developed to establish a robust \gls{RSS}–distance relationship, which involves pre-processing \gls{RSS} values via a Gaussian filter to reduce signal fluctuations. In the subsequent online phase, a constant velocity model \gls{KF} first smooths the real-time \gls{RSS} measurements, which are then input into the calibrated PPRM to estimate the Euclidean distance between the target and Wi-Fi detectors. Finally, an \gls{LS} Taylor series expansion (LS-TSE) calculates a coarse position estimate, which is then further refined and smoothed by a \gls{UKF}. Field experiments in an urban road environment in Guangzhou, China, demonstrated a meter-level positioning accuracy. Authors in \cite{8355136} present a \gls{kNN}-based Wi-Fi fingerprinting methodology that can utilize \gls{RSSI}/\gls{SSID} measurements in an outdoor urban scenario. Field results show that the methods can sustain 30 m of accuracy for 50\% of the time. In \cite{gomez_wifi-based_2023}, authors proposed a Wi-Fi-based fingerprinting methodology that estimates a vehicle's absolute location in a parking space by utilizing \gls{RSS} measurements. The proposed method enhances accuracy by leveraging historical \gls{RSS} information over time, rather than relying solely on current \gls{RSS} fingerprints. The core of the system is a long short-term memory (LSTM)-based neural network architecture trained to estimate the device's position from these temporal sequences of Wi-Fi \gls{RSS}. A key contribution is that the site surveying for the training data is highly reduced, as it involves collecting \gls{RSS} samples and conventional low-precision \gls{GNSS}-based position estimates from mobile devices while driving. The proposed method achieved several meters of positioning accuracy. Authors in \cite{ding_three-dimensional_2022} developed a sensing-based \gls{CSI} fingerprinting algorithm to track indoor mobile users in a \gls{MIMO} system. This method can be easily extended to a parking lot positioning scenario. The \gls{CSI} here encodes both the channel's amplitude and phase shifts caused by delays and angles. Instead of training on \gls{CSI} measurements directly, the authors decompose the \gls{CSI} into multiple single-rank tensors to expedite the feature extraction process. Next, the extracted features undergo feature optimization and selection to remove features that are redundant and keep features that are highly related to the user's position. Finally, the authors trained a recurrent neural network (RNN) with LSTM to extract the position of the user from the optimized features, achieving meter-level accuracy. Authors in~\cite{zhang_vehicular_2021} proposed a \gls{CSI}-based fingerprinting method using a Wi-Fi-based \gls{RSU} with a modified 802.11p \gls{V2I} module to position vehicles in toll stations. The \gls{CSI} here is affected by both distance and angle of the user, as the \gls{AP} has access to an antenna array. The method employed discrete Fourier transform-based signal processing to extract the \gls{LoS} path \gls{CSI} components (i.e., resolving multipath). The filtered \gls{CSI} fingerprint is matched with already labeled fingerprints through a similarity degree algorithm to localize the user. Experimental results show decimeter levels of accuracy. Lastly, it is worth noting that Google vehicles are reported to improve their maps and positioning accuracy by collecting and sending raw Wi-Fi-based \gls{SSID} and \gls{RSSI} measurements to their cloud \cite{kiss_google_2010} \rev{(see Fig. \ref{fig:WiFicanyon})}. For further reading on Wi-Fi-based fingerprinting methods, the reader is directed to \cite{Quoc2016survey}, which surveys Wi-Fi-based fingerprinting works that utilize \gls{RSSI} measurements in various outdoor scenarios. It is worth noting that in all of the aforementioned works, knowledge about the exact position of the \glspl{AP} is needed. However, such knowledge is usually not readily available for all vehicles on the road. To remedy that, few works focused on estimating the position of these \glspl{AP} via knowledge of the position of the vehicle with the aid of other sensors, as shown in \cite{cheng2005accuracy,8007254}.

\textbf{\textit{Cooperative Wi-Fi Positioning (\gls{V2V})}}: Authors in \cite{liu_hybrid_2020} proposed a hybrid integrity monitoring technique to address limitations of GNSS-based vehicular positioning in challenging environments like dense urban areas. This method employs \gls{DSRC} range-rate measurements to assist integrity monitoring and fault detection by creating virtual satellite measurements that expand the observation vector, even when satellite visibility is limited. The system integrates real and simulated pseudo-ranges into a cubature Kalman filter (CKF) based estimator, enabling fault detection and exclusion. Simulations confirmed that the proposed solution outperforms conventional receiver autonomous integrity monitoring (RAIM) methods and achieves several meters to tens of meters of positioning accuracy. In \cite{liu_towards_2023}, the authors investigated robust Wi-Fi \gls{RSS}-based fingerprint-based vehicle tracking in dynamic indoor parking environments, proposing an online learning framework to continuously adapt the localization model to signal variations. The framework consists of a hidden Markov model (HMM)-based online evaluation (HOE) method to assess localization accuracy, and an online transfer learning (OTL) algorithm that combines batch and online classification models via weight allocation. OTL further enhances robustness by continuously updating the fingerprint database through instance-based transferring, resampling offline fingerprints based on their similarity to current real-time data. A comprehensive evaluation in real-world indoor parking environments demonstrated several meters of positioning accuracy. Authors in \cite{soatti_implicit_2018} proposed a relative cooperative positioning approach to enhance GNSS-based vehicle positioning in \gls{C-ITS}, particularly in urban canyons where GNSS signals are degraded. The method exploits \gls{V2V} connectivity by having vehicles jointly sensing non-cooperative physical features (e.g., people, traffic lights) using their on-board sensors like radar or lidar, which then serve as common noisy reference points. Information on these sensed features is fused through \gls{V2V} links via a consensus procedure nested within a distributed Gaussian message passing (GMP) algorithm. Simulation-based performance results showed that relative cooperative positioning methods can significantly improve vehicle location accuracy compared to stand-alone GNSS, achieving sub-meter accuracy in urban scenarios. Another relative positioning approach to prevent vehicular crashes was reported in  \cite{watta_vehicle_2021}. Authors presented Geo+NN, a geometric-based neural network \gls{V2V} localization framework that utilizes \gls{DSRC} to localize nearby vehicles. The system combines geometric modeling to extract key features (distance, perpendicular distance, relative angle) from \gls{V2V} data, and a neural network that uses these features to classify the remote vehicle's position into one of eight classes (e.g., ahead, behind, adjacent lane, etc.). Using real-world \gls{DSRC} driving data, the proposed neural networks achieved over 99\% accuracy for remote vehicle position detection. For prediction, the system achieved above 75\% accuracy for look-ahead times less than 0.7 s in 8-class prediction, and above 90\% accuracy for 6-class prediction within the same time frame.

\subsubsection{IEEE 802.15 (UWB and Bluetooth)}
In this section, we review works that utilize IEEE 802.15-based \gls{UWB} and Bluetooth technologies for positioning purposes. It is worth noting that \gls{UWB} is more popular for positioning purposes compared to Bluetooth due to its accurate ranging capabilities.

\textbf{\textit{\gls{UWB}}}: The restrictions on outdoor usage of \gls{UWB} in the EU and the US did not stop researchers worldwide from developing vehicular positioning solutions using \gls{UWB}. Those solutions can be generally categorized into range-based solutions, hybrid range and angle-based solutions, and cooperative solutions. Range-based solutions either utilize \gls{RTT} or \gls{TDoA} measurements. In the first category, authors in \cite{cao_improving_2022} proposed GMM-NNIMM-CLMFO, a novel localization scheme for complicated underground \gls{NLoS} environments using \gls{UWB} P440 sensors. The method has two main stages: (i) refining of range measurements by leveraging a Gaussian mixture model (GMM), a neural network-based interacting multiple model (NNIMM), and a variational Bayesian (VB)-based \gls{KF}, and (ii) localization using a Caffery localization (CL) method—a linear trilateration approach—refined by moth-flame optimization (MFO) technique. This approach achieved a decimeter level of accuracy and showed robustness by mitigating LOS and NLOS errors. In \cite{wu_pedestrian_2024}, authors introduced a \gls{UWB}-based solution that not only estimates the position of the center of the vehicle, but also its orientation, and the position of its surrounding pedestrians. The system is designed to alert users when pedestrians are in the vicinity of the vehicle. The system fuses multiple \gls{RTT} measurements from three \gls{UWB} anchors and four \gls{UWB} tags fixed on the vehicle and a single \gls{UWB} tag per pedestrian via an \gls{EKF}. Based on the position and orientation estimates, the system utilizes a dynamic threshold algorithm to trigger alerts. Experimental results demonstrated sub-meter level of accuracy. 

In the \gls{TDoA} camp, authors in \cite{huang_ultra-wideband_2021} designed and implemented an indoor autonomous disinfection vehicle that is localized using multiple \gls{UWB} transmitters. The system fuses multiple \gls{TDoA} measurements using a gradient descent (GD)-Taylor method for optimal position finding and a generalized traversal path planning procedure. Experimental demonstrations in a meeting room showed a decimeter level of accuracy. In \cite{jiang_practical_2022}, the authors proposed an economical \gls{UWB} anchor placement approach for indoor autonomous valet parking systems. The method assumes the usage of \gls{TDoA} measurements to estimate the vehicle's positioning accuracy via \gls{GDOP}. Experimental tests in a real underground parking lot showed a wide range of achievable accuracy, ranging from centimeters to tens of meters. In \cite{lee_uwb-based_2021}, authors proposed a \gls{UWB}-based positioning system for indoor automated guided vehicles using \gls{UWB}-mounted \glspl{UAV}. The proposed system fuses \gls{TDoA} measurements from the \gls{UWB} anchors using \gls{LS}. Experimental results show a decimeter level of accuracy while using five \glspl{AP}. In \cite{wen_automated_2020}, authors proposed a \gls{UWB}-based positioning solution for vehicles driving on highways and in tunnels by fusing \gls{TDoA} measurements via \gls{LS}. Experimental results show decimeter to meter-level accuracy, depending on the speed of the vehicle. Finally, authors in \cite{zhang_uwb-based_2024} proposed a \gls{UWB}-based positioning solution to track vehicles in a smart parking space. The proposed method first tracks the \gls{ToA}, clock offset, and the carrier frequency offset (CFO) of multiple \gls{UWB} \glspl{AP} via a \gls{KF} to synchronize the user. Next, the \gls{TDoA} measurements are fused using an \gls{EKF} with a constant acceleration model. Experimental results show consistent decimeter-level positioning accuracy.

In the hybrid positioning camp, authors in \cite{hu_multipath-assisted_2025} proposed a multipath-assisted \gls{UWB} vehicle localization framework for underground parking. The proposed method utilizes both \gls{LoS} and \gls{NLoS} \gls{ToA} and \gls{AoA} measurements to simultaneously localize the user and map the environment. The method utilizes both raytracing techniques to estimate the location of the multipath's virtual anchors, which are then filtered and fused using a weighted iterative \gls{LS} (W-IRLS) approach. The authors also investigated four localization modes, namely pure triangulation (\gls{AoA}), pure trilateration (\gls{ToA}), hybrid positioning using \gls{ToA} + \gls{AoA}, and hybrid positioning using \gls{TDoA} + \gls{AoA}. Experimental results show decimeter-level accuracy across the board, with higher performance in the hybrid \gls{ToA}-\gls{AoA} approach. In \cite{piavanini_experimental_2023}, the authors presented experimental validation of vehicle positioning in a race track using \gls{UWB}, specifically targeting unpredictable vehicle maneuvers. The core methodology involves a \gls{UKF} embedding an IMM (i.e., constant velocity, acceleration, and turn rate) that utilizes \gls{TDoA} and \gls{AoA} measurements. Experimental results demonstrated a decimeter-level positioning accuracy.

Among the cooperative \gls{UWB} works, authors in \cite{wang_uwb_2021} proposed a relative planar localization system (position and orientation) for outdoor vehicles, using three \glspl{UWB} modules on each vehicle. \gls{RTT} measurements between \gls{UWB} pairs on each vehicle are utilized. The system features a localization algorithm that solves a non-linear \gls{LS} problem via a homotopy Levenberg–Marquardt (HOMO-LM) algorithm with geometrical constraints. They also propose a self-calibration method to correct for the CFO in each \gls{UWB}. Extensive experimental tests demonstrated a consistent decimeter-level accuracy. In \cite{wang_development_2024}, the authors presented a \gls{UWB}-based cooperative \gls{V2V} ranging system for self-propelled rail vehicles, aiming for infrastructure-free operation. Each vehicle is equipped with four \glspl{UWB}; two masters and two slaves (one pair in the front corners and the other pair in the back corners of the vehicle). The master \gls{UWB} placed in the front of a given vehicle is responsible for obtaining \gls{RTT} ranging information by communicating with the slave \gls{UWB} installed in the back of the vehicle in front of it. Likewise, the front-installed slave \gls{UWB} is responsible for aiding the master \gls{UWB} installed in the back of the vehicle in front. Extensive experimental work demonstrated decimeter-level ranging capabilities.

\textbf{\textit{Bluetooth}}: Works that utilize Bluetooth for vehicular positioning are few, and can be categorized into two groups: (i) pure vehicular positioning works, and (ii) traffic monitoring works. An example of \gls{AoA}-based Bluetooth positioning works can be found \cite{paulino_evaluating_2021}, where the authors presented a low-cost indoor localization method for warehouse vehicles, using Bluetooth 5.1 \gls{AoA} measurements. In the proposed system, vehicles are equipped with a single directional antenna array that is connected to multiple omnidirectional \glspl{AP}. The \gls{AoA} measurements are fused using a \gls{LS} to provide a snapshot position estimate, which is then fed to a \gls{KF} to track the vehicle. Simulations based on realistic \gls{AoA} measurements show sub-meter to meter-level accuracy. This work was then expanded to experimental outdoor positioning in \cite{paulino_self-localization_2023}, where an 8-element uniform circular array antenna receivers were used (Bluetooth 5.1). The methodology computes \gls{AoA} measurements and positions the users using a least-mean-squares method. Experimental results on \gls{AoA} estimation performance in an anechoic chamber showed an \gls{RMSE} of 10.7°. The outdoor experimental results show a positioning accuracy of sub-meter to meter level.\footnote{Both of those works assume perfect knowledge of the orientation of the user, which is not practical in reality.} In traffic monitoring-focused works, estimating the position is not the main goal per se, but rather counting the number of vehicles in a given area/region. For instance, \cite{boudabous_vehicular_2021} utilized Bluetooth passive scanning data, including time-stamped records of detected devices’ identifiers and \gls{RSS} to analyze vehicular traffic in urban areas. The methodology involved a data-driven approach using statistical and machine learning models for traffic flow quantification, and an algorithm leveraging \gls{RSS} for average travel speed estimation.

\subsection{Challenges and Open Problems}
The main challenges that plagued IEEE-based positioning since its inception is multipath resolvability and short-range communications. We do not think that the short-range communication is going to be solved soon, as they are the defining feature of such networks. However, the long-lasting legacy of multipath resolvability issues is now shifting as the recent standards approved in May 2025 effectively increased the bandwidth of positioning pilot signals to 320 MHz (time/range-resolution), as highlighted in Sec.~\ref {IEEEevol}. Moreover, the introduction of WiGig's mmWave communications capabilities in IEEE 802.11ad (60 GHz, 2012) and IEEE 802.11ay (45 GHz, 2021) with bandwidths ranging from 540 MHz to 8.64 GHz has an immense opportunity for positioning resolvability in both range and angular domains.\footnote{IEEE 802.11aj was also introduced in 2018 to support operation in the Chinese 45 GHz and 60 GHz mmWave bands.} However, positioning standards for those bands are yet to be introduced. Having access to resolvable and accurate range and angle measurements will give rise to single-\gls{AP} geometric positioning and \gls{SLAM} techniques, as seen in the emerging 5G networks. Nevertheless, since the IEEE-based positioning research community is more experimental in nature, research on off-the-shelf anchors and receivers that reflect the latest standardization developments might take a while. Another practical challenge that faces IEEE-based positioning implementations is that of scalability and synchronization. Current frameworks are heavily reliant on \gls{RTT} measurements, which require a dedicated link between the transmitter and the receiver to circumvent synchronization issues. Such a requirement limits the scalability of IEEE-based positioning systems. Hence, a shift towards the usage of \gls{ToA} and/or \gls{TDoA} might be required to enhance the scalability of these systems.

\section{Sensor Fusion with Onboard Sensors}\label{Sec: sensor fusion}
\begin{figure}
    \centering
    \includegraphics[width=0.99\linewidth]{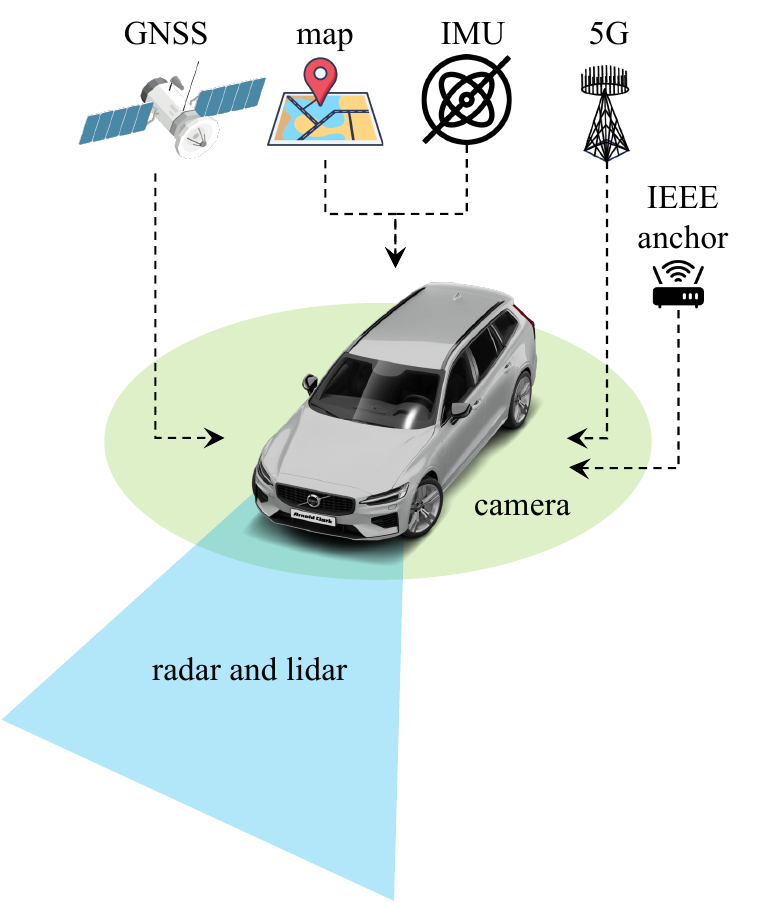}
    \caption{Illustration of vehicular localization technologies that can be fused to enhance performance.}
    \label{fig:sensor_fusion_new}
\end{figure}

As explored in the previous sections, no single wireless technology can provide an uninterruptible positioning solution in all environments due to inevitable signal blockage. Fortunately, modern vehicles are equipped with various sensors, illustrated in Fig.~\ref{fig:sensor_fusion_new}, that can bridge these gaps and enhance the positioning performance of wireless technologies \cite{saleh_5g_2023}. As mentioned before, these sensors can be categorized into (i) perception-based sensors like cameras, lidars, and radars; and (ii) motion sensors such as accelerometers, gyroscopes, and odometers. Like the case with wireless technologies, we cannot fully depend on these sensors as each has its own challenges. However, more often than not, onboard sensors and wireless technologies have complementary characteristics\rev{, as shown in Table \ref{tab:localization_comparison_survey_revised}.} Thus, by devising proper sensor fusion strategies, we can provide an uninterrupted and accurate positioning solution that fits the needs of vehicular applications. In this section, we start with a brief discussion on the characteristics of onboard perception and motion sensors. Then, we showcase and categorize the fundamentals of the various sensor fusion techniques and schemes available in the literature. Finally, we present state-of-the-art works on sensor fusion between radio technologies and onboard sensors.

\begin{table*}[!ht]
    \centering
    \caption{\rev{Comparative Analysis of Localization Technologies.}}
    \label{tab:localization_comparison_survey_revised}
    \footnotesize 
    \begin{tabularx}{\textwidth}{
        >{\bfseries}p{1.7cm}
        >{\raggedright\arraybackslash}X
        >{\raggedright\arraybackslash}X
        >{\raggedright\arraybackslash}X
        >{\raggedright\arraybackslash}X
        >{\raggedright\arraybackslash}X
        >{\raggedright\arraybackslash}X
        >{\raggedright\arraybackslash}X
        >{\raggedright\arraybackslash}X
        >{\raggedright\arraybackslash}X
    }
        \toprule
        & \textbf{GNSS} & \textbf{5G} & \textbf{Wi-Fi} & \textbf{UWB} & \textbf{Bluetooth} & \textbf{IMU/Odo.} & \textbf{Camera} & \textbf{Lidar} & \textbf{Radar} \\
        \midrule

        \rowcolor{gray!25}
        \multicolumn{10}{l}{\itshape\textbf{Performance \& Robustness}} \\
        Accuracy & RTK: cm-level, PPP: decimeter-level, Commercial: sub-meter  & mmWave: accurate range and angle measurements, sub-6 GHz: low-accuracy ranging &  Low-accuracy & Extremely accurate ranging & Low-accuracy  & High short-term accuracy & Rich measurements  & Rich point clouds  & Limited accuracy and resolution compared to camera/lidar  \\
        \rowcolor{gray!10}
        Vulnerabilities & Urban canyons, blockage, multipath, atmospheric effects & mmWave: blockage, Sub-6 GHz: multipath resolvability    & Range, blockage, and multipath resolvability & Blockage & Range, blockage, and multipath resolvability & Error accumulation over time (needs assisting absolute solution to reset errors) & Dark and imbalanced lighting, rain, fog, snow, featureless environments & Rain, fog, snow  & Resolution  \\
        \midrule

        \rowcolor{gray!25}
        \multicolumn{10}{l}{\itshape\textbf{Coverage \& Environment}} \\
        Coverage & Global  & Medium range (Sub-6 GHz is better) & Short range  & Short range  & Short range  & Local  & Local, sensor field-of-view  & Local, sensor field-of-view  & Local, sensor field-of-view  \\
        \rowcolor{gray!10}
        Environment & Outdoor (clear sky)  & Indoor and outdoor (Sub-6G Hz is better) & Indoor (restricted outdoors)  & Indoors (restricted outdoors)  & Indoors (restricted outdoors)  & All  & Well-lit environments  & All & All \\
        \midrule

        \rowcolor{gray!25}
        \multicolumn{10}{l}{\itshape\textbf{Timeliness \& Scalability}} \\
        Update Rate & Low  & Medium   & Medium & Medium  & Medium & High & Medium  & Medium  & Medium \\
        \rowcolor{gray!10}
        Sensor and infrastructure Cost & Low  & Low  & Receiver: Low, Infrastructure: High & Receiver: Low, Infrastructure: High  & Receiver: Low, Infrastructure: High  & Low  & Low-medium  & High  & Low-medium  \\
        \bottomrule
    \end{tabularx}
\end{table*}

\subsection{Characteristics of Onboard Sensors}
\subsubsection{Perception Sensors} Cameras, radars, and lidars can perform both absolute positioning, with the aid of HD-maps through map matching methods \cite{chao2020survey}, and relative positioning, through dead-reckoning techniques \cite{agostinho2022practical}. Map matching techniques are heavily dependent on the quality of the HD-map used and how up-to-date it is \cite{chao2020survey}. Hence, outages can be expected in areas HD-maps are either non-existent or outdated.  Moreover, perception outages can occur when scenes are ambiguous due to a lack of points of interest \cite{chao2020survey}. On the other hand, dead-reckoning-based solutions do not require existing maps to perform well. However, they are prone to accumulation of error due to their inherit design \cite{agostinho2022practical}. Hence, they need periodic corrections from external absolute positioning sources to operate over long periods. All perception sensors can perform either technique with varying performance and cost. For instance, cameras provide rich measurements at a relatively low cost, but they are prone to outages in dark environments and in scenarios of heavy rain, snow, or fog  \cite{agostinho2022practical}. Like cameras, lidars can provide rich point cloud measurements and are heavily affected by weather conditions. Yet, in contrast to cameras, they can operate in dark conditions, they are bulky and costly \cite{li2020lidar}. Finally, radars are low-cost and can operate in all weather and lighting conditions, but have limited accuracy and resolution compared to cameras and lidars \cite{dawson2023merits}.

\subsubsection{Motion Sensors} Typically, modern vehicles are equipped with an odometer and an \gls{IMU} which houses three orthogonal accelerometers and three orthogonal gyroscopes. As the name suggests, accelerometers measure the 3D acceleration of the vehicle, including the gravity of the Earth, and gyroscopes measure the 3D angular rotation of the vehicle, which is also affected by the rotation of the Earth. Finally, the odometer measures the forward velocity of the vehicle by counting the number of wheel turns relative to time and the diameter of the wheel. Motion sensors' main advantage is that they can operate in all environments, lightning conditions, and weather conditions. Hence, they do not experience outages due to external factors. Also, they can typically provide a positioning solution at a much higher rate compared to perception- and wireless-based positioning technologies \cite{noureldin2012fundamentals}. However, to estimate the position of the vehicle, \gls{IMU} measurements are integrated twice, and odometer measurements are integrated once. Such integration will lead to high position errors if the \glspl{IMU} are biased, which is typically the case. Hence, \glspl{IMU} require external corrections to estimate their biases to perform properly. Odometers, on the other hand, can sustain position errors due to skidding, wheel diameter calibration errors, or operation at low velocities \cite{noureldin2012fundamentals}. The pros and cons of motion sensors are complementary to wireless technologies, and hence, the integration between the two garnered the attention of many researchers.

\subsection{Sensor Fusion Methods}
Over the years, sensor fusion research has matured and defined staple sensor fusion techniques and architectures. In this section, we categorize and review these sensor fusion methods from a vehicular positioning perspective.

\subsubsection{Techniques}
Sensor fusion techniques can be categorized in various ways. In this paper, we categorize them into Bayesian (recursive) filtering techniques, batch processing techniques, and data-driven techniques. Each category holds advantages and disadvantages when it comes to vehicular applications. Since the sensor fusion literature is rich with methodologies proposed under each category, we will briefly cover the basics of each category along with their benefits, limitations, and applicability in vehicular applications.

\paragraph{Bayesian Filtering}
Bayesian filtering methods are characterized by the recursive state prediction and correction cycle. They propagate one or more hypotheses about the state of the user (i.e., its position and possibly its velocity and orientation).
Chief among the single hypothesis methods is the family of Kalman filtering methods, which form the majority of the works proposed in this field. The classical \gls{KF} has very tight assumptions about the linearity of the transition and measurement models as well as the Gaussian distribution associated with the process and measurement noises to achieve optimality. These assumptions are usually not met in vehicular applications, which leads to sub-optimal performance.\footnote{For instance, mobility/transition models for vehicles are usually far from linear, and the measurement models for most positioning sensors are highly non-linear.} To deal with these issues, research works might use other flavors of the classical \gls{KF}, like the \gls{EKF} and the \gls{UKF}, which are better at handling non-linearities. 
On the other hand, \glspl{PF} are the most noteworthy example of multi-hypothesis filtering, which propagates multiple ``\textit{particles}" that carry the individual hypotheses. Unlike \gls{KF}, \gls{PF} does not have any constraints on the linearity of the models or the distribution of the noise. For that reason, their performance cannot be guaranteed to be optimal. On the other hand, \gls{PF} is considered to be more computationally heavier than its \gls{KF} counterparts. 
In terms of applicability to vehicular applications, Bayesian filters are usually favored due to their relatively low complexity and real-time response, as opposed to batch-processing approaches, and because they do not require prior training to operate, as opposed to data-driven approaches.

\paragraph{Batch-processing}
Batch-processing methods are characterized by their approach of processing all available data after the entire observation period has concluded, rather than relying on recursive online updates. This enables them to more effectively handle large amounts of noisy, non-linear data by using global optimization techniques, such as least squares, maximum likelihood estimation, and factor graphs to name a few, to minimize the error over the full dataset. This approach often leads to more accurate estimates than those from real-time filtering methods, particularly in complex environments like vehicular applications. 
However, this comes at the cost of higher computational complexity and the inability to provide real-time estimates, which makes them less suitable for applications that require continuous updates, such as real-time vehicular tracking. Additionally, batch-processing methods often require prior knowledge about the entire data set before processing can begin, which can limit their flexibility in dynamic environments.
In terms of applicability to vehicular applications, batch-processing methods are typically used in post-processing scenarios or in applications where the data is collected in bursts and can be processed in batches offline.

\paragraph{Data-driven} Data-Driven Methods rely on learning patterns directly from the available data, without relying on explicit system models. These methods use \gls{ML} or \gls{DL} techniques to identify complex, non-linear relationships between sensor measurements and the system state. Unlike Bayesian filtering and batch-processing methods, data-driven approaches adapt to the data itself, improving as more data is provided.
Common examples include supervised learning methods like regression models, support vector machines, and deep \glspl{NN}, as well as unsupervised learning techniques such as clustering and semi-supervised methods like reinforcement learning. These methods can handle non-linear, noisy data, making them useful for vehicular systems where dynamics are difficult to model.
The main advantage of data-driven methods is their ability to learn from the data and adapt to complex patterns. However, they often require large datasets for training and can be computationally intensive, especially deep learning models. Their reliance on large-scale data and heavy computational resources can limit their real-time applicability, especially in time-sensitive vehicular tracking scenarios. However, since the number of vehicles equipped with onboard positioning sensors is increasing by the day, data-driven models stand to gain a lot of value soon and in the long run.

\subsubsection{Fusion Schemes} 
\begin{figure}[t]
    \centering

    \begin{subfigure}[b]{0.4\textwidth}
        \centering
        \includegraphics[width=\textwidth]{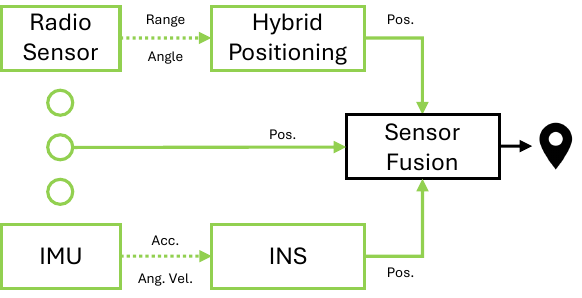}
        \caption{\rev{Loosely-coupled integration}}
        \label{fig:first_sub}
    \end{subfigure}
    
    \vspace{1cm}

    \begin{subfigure}[b]{0.35\textwidth}
        \centering
        \includegraphics[width=\textwidth]{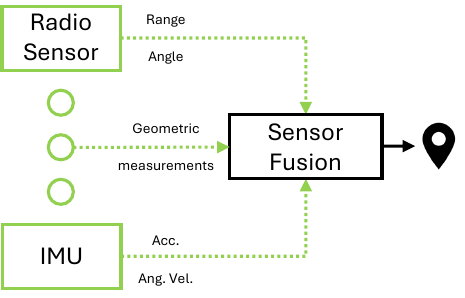}
        \caption{\rev{Tightly-coupled integration}}
        \label{fig:second_sub}
    \end{subfigure}
    
    \vspace{1cm}

    \begin{subfigure}[b]{0.4\textwidth}
        \centering
        \includegraphics[width=\textwidth]{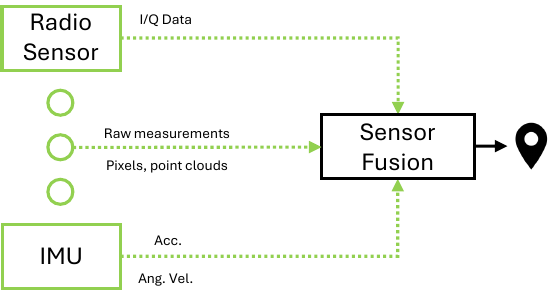}
        \caption{\rev{Ultra-tightly-coupled integration}}
        \label{fig:third_sub}
    \end{subfigure}

    \caption{\rev{Illustration of the integration of radio sensors with other onboard sensors using (a) LC, (b) TC, and (c) UTC integration schemes. In LC, sensor fusion is conducted on the position level. On the other hand, TC integrates the geometric measurements of the sensors while UTC integrates the raw measurements of the sensors.}}
    \label{fig:Integration_schemes}
\end{figure}

Fusion of measurements from multiple technologies (or from multiple nodes of the same technology) can be done in three main ways, regardless of the filter used. These schemes are known as \gls{LC}, \gls{TC}, and \gls{UTC} integration schemes\rev{, shown in Fig.~\ref{fig:Integration_schemes}}. Each has its advantages and disadvantages based on the type of measurements used. In the following, we detail the applicability of each scheme from a vehicular positioning lens.

\paragraph{Loosely-coupled Integration} \gls{LC} integration, also known as high-level or late integration, is the simplest of the three schemes and the easiest to implement, hence its popularity among research works. \gls{LC} architectures integrate the technologies on the positional level, meaning that each individual technology is expected to provide a standalone positioning solution prior to participating in the fusion process. This method is effective when using Kalman filtering while the measurement models are highly non-linear, as it forces the relationship between the states and the measurement to be linear. On the other hand, the requirement for a standalone solution before integration can be restrictive, as in the case when only three \glspl{GNSS} satellites are in view, i.e., trilateration cannot be performed. Hence, from a vehicular perspective, \gls{LC} integration is suitable in cases where the vehicle does not have enough processing power while having access to multiple standalone positioning solutions.

\paragraph{Tightly-coupled Integration} \gls{TC} schemes, which are also known as low-layer or early integration schemes, hold similar popularity compared to \gls{LC} schemes. \gls{TC} schemes directly integrate the geometrical measurements of the technologies without the need to compute a standalone positioning solution for each technology. The implementation of such schemes is more complex compared to \gls{LC} schemes but benefits technologies that cannot provide a standalone solution. Moreover, since integration occurs on the measurement level, non-linear measurement models will be used, which can hamper the efficacy of linear estimators like the \gls{KF}. However, with the aid of other non-linear fusion algorithms, \gls{TC} schemes can provide higher positioning accuracy, compared to \gls{LC}, at the cost of computational complexity. Hence, \gls{TC} schemes are more suitable in scenarios when the vehicle has access to high computational power without tight energy restriction or when standalone positioning solutions are not available (i.e., access to less than four \gls{GNSS} satellites).

\paragraph{Ultra-tightly-coupled Integration} \gls{UTC} integration is the most complex among the three schemes and is only realized in the literature for \gls{GNSS}–\gls{IMU} integration. \gls{UTC} schemes integrate the technologies on a much deeper level as the integration takes place at the signal level, i.e., prior to the computation of geometrical measurements like ranges and angles. Hence, such integration requires access to the I/Q samples of the received signal, which might not be provided by cellular operators to vehicles for security and privacy reasons.

\subsection{Radio Sensor Fusion Works}
In this section, we first divide the sensor fusion works into two categories: (i) fusion with onboard sensors only and (ii) fusion with other wireless technologies. We then divide the works according to the wireless technology used (i.e., \gls{GNSS}, cellular, and IEEE-based). In each sub-category, we further divide works based on the experimental setup used for validation (i.e., real-world and simulation setups), the level of integration (i.e., \gls{LC}, \gls{TC}, and \gls{UTC} schemes), and finally the sensor fusion filter used (e.g., Bayesian, batch, or \gls{AI}-based). We report the accuracy statistic achieved by each work if available, providing insights into the trends and trade-offs observed across different approaches.

\subsubsection{Fusion with Onboard Sensors Only}
This category focuses on enhancing vehicular positioning by integrating a primary wireless technology (\gls{GNSS}, cellular, or IEEE-based) with sensors already present on the vehicle, such as \gls{IMU}s, odometers, cameras, and lidar. \rev{Table~\ref{tab:radio_sensor_fusion_onboard} summarizes the works presented in this section.}

\begin{table*}
\centering
\caption{\rev{Summary of radio sensor fusion works with onboard sensors.}}
\label{tab:radio_sensor_fusion_onboard}
\vspace{2mm}
\footnotesize
\renewcommand{\arraystretch}{1.2} 
\begin{tabular}{|m{0.09\textwidth}|m{0.08\textwidth}|m{0.05\textwidth}|m{0.35\textwidth}|m{0.19\textwidth}|m{0.09\textwidth}|}
\hline
\textbf{Technology} & \textbf{Environment} & \textbf{Scheme} & \textbf{Measurements \& Techniques} & \textbf{Accuracy} & \textbf{Validation} \\ \hline
\rowcolor{pastel2} GNSS \cite{kaczmarek2022experimental} & Urban & LC & GNSS (RTK) + IMU + odometers, EKF & Centimeter-level & Experimental \\ \hline
\rowcolor{pastel2} GNSS \cite{zhang2024robust} & Outdoor & LC & GNSS + multiple IMUs, KF & Several meters & Experimental \\ \hline
\rowcolor{pastel2} GNSS \cite{he2021integrated} & Outdoor & LC & GNSS (RTK) + lidar, SLAM & Decimeter-level & Experimental \\ \hline
\rowcolor{pastel2} GNSS \cite{zhu2023tightly} & Urban & TC & GNSS (PPP-RTK) + INS, Bayesian & Decimeter-level & Experimental \\ \hline
\rowcolor{pastel2} GNSS \cite{li2022continuous} & Urban & TC & GNSS (PPP-RTK) + INS + vision, EKF & Decimeter-level & Experimental \\ \hline
\rowcolor{pastel2} GNSS \cite{xu2023high} & Urban & TC & GNSS (SF-RTK) + IMU + camera, MSC-KF & Sub-meter-level & Experimental \\ \hline
\rowcolor{pastel2} GNSS \cite{zhu2023improved} & Rural/Urban & TC-LC & GNSS (RTK) + IMU + vision + lidar, factor graph & Sub-meter-level & Experimental \\ \hline
\rowcolor{pastel2} GNSS \cite{zhumu2023multisensor} & Urban & TC-LC & GPS + IMU + camera, adaptive EKF & Decimeter-level & Experimental \\ \hline
\rowcolor{pastel2} GNSS \cite{zhou2020cascaded} & Outdoor & LC & GPS + IMU, cascaded KFs & Meter-level & Simulation \\ \hline
\rowcolor{pastel2} GNSS \cite{yin2023sensor} & Urban & TC & GNSS + IMU, error-state KF + RTS & Sub-meter-level & Simulation \\ \hline
\rowcolor{pastel2} GNSS \cite{tao2022multi} & Urban & TC & GNSS (RTK) + IMU + VIO, pose graph optimization & Sub-meter-level & Simulation \\ \hline \hline
\rowcolor{pastel3} Cellular \cite{saleh_integrated_2023} & Urban & LC & 5G (range + AoD) + IMU, EKF & Decimeter-level & Semi-experimental \\ \hline
\rowcolor{pastel3} Cellular \cite{saleh2024integration} & Urban & LC & 5G (range + AoD) + IMU, EKF & Decimeter-level & Semi-experimental \\ \hline
\rowcolor{pastel3} Cellular \cite{bader2024leveraging} & Urban & LC & 5G (AoA + AoD + ToA + RSS) + IMU, EKF & Decimeter-level & Semi-experimental \\ \hline
\rowcolor{pastel3} Cellular \cite{mostafavi2020vehicular} & Highway & TC & 5G (range + AoA) + IMU, EKF & Decimeter-level & Simulation \\ \hline
\rowcolor{pastel3} Cellular \cite{hammarberg2022architecture} & Highway & TC & 5G (range + AoA) + accelerometer, EKF & Decimeter-level & Simulation \\ \hline \hline
\rowcolor{pastel1} IEEE \cite{li2024novel} & Outdoor & TC & UWB (range) + IMU, MC-RUKF & Centimeter to Decimeter-level & Experimental \\ \hline
\rowcolor{pastel1} IEEE \cite{sanmartin2020precise} & Outdoor & TC & UWB (range) + IMU + odometer, EKF & Decimeter-level & Experimental \\ \hline
\rowcolor{pastel1} IEEE \cite{sun2020research} & Indoor & LC & Wi-Fi (RSS, kNN fingerprint) + PDR (IMU + magnetometer), EKF& Few meters & Experimental \\ \hline
\rowcolor{pastel1} IEEE \cite{wang2022fusion} & Indoor & LC & Wi-Fi (RSS, K-means fingerprint) + IMU, EKF & Meter-level & Experimental \\ \hline
\rowcolor{pastel1} IEEE \cite{wang2022wifipdr} & Indoor & LC & Wi-Fi (CSI, kNN fingerprint)  + PDR (IMU + magnetometer), EKF & Meter-level & Experimental \\ \hline
\rowcolor{pastel1} IEEE \cite{lu2022ONavi} & Indoor & LC & Wi-Fi (RSS-SSID, fingerprint)  +  PDR (IMU, CNN-transformer), optimization & Decimeter-level & Experimental \\ \hline
\rowcolor{pastel1} IEEE \cite{zhang2023from} & Indoor & LC & Wi-Fi (RSS) + PDR (IMU) + vision, matching algorithm & Centimeter-level & Experimental \\ \hline
\rowcolor{pastel1} IEEE \cite{tang2024sequential} & Indoor & LC & Wi-Fi (RSS, GPR fingerprint) + vision, whale optimization & Meter-level & Experimental \\ \hline
\rowcolor{pastel1} IEEE \cite{tang2023novel} & Indoor & LC & Wi-Fi (RSS, kNN fingerprint) + vision, Lagrangian fusion & Few meter-level & Experimental \\ \hline
\rowcolor{pastel1} IEEE \cite{lin2022pedestrian} & Indoor & TC-LC & Wi-Fi (RSS, TC) + IMU, PF + chaos particle swarm & Millimeter-level & Simulation \\ \hline
\rowcolor{pastel1} IEEE \cite{kang2022lidar} & Outdoor & LC & DSRC (position, velocity, heading) + lidar, EKF + PF & Decimeter-level & Simulation \\ \hline
\end{tabular}
\end{table*}

\paragraph{GNSS-based}
Fusion with \gls{GNSS} is a cornerstone of vehicular positioning, leveraging its global coverage while compensating for its vulnerabilities using onboard sensors. The choice of experimental setup, integration scheme, and filtering method significantly impacts performance.

\textbf{Real-world Experimental Setups:} These studies provide validation under practical conditions, crucial for assessing real-world viability. Authors in \cite{kaczmarek2022experimental} employed an \gls{EKF} for a loosely coupled integration of \gls{GNSS} with \gls{IMU} and odometers, augmented with \gls{RTK}, achieving accuracies of less than 5 cm in urban experiments. Likewise, \cite{zhang2024robust} utilized a \gls{KF} to fuse \gls{GNSS} with multiple \glspl{IMU} in an \gls{LC} approach termed eNav-Fusion, demonstrating few-meter accuracy in outdoor experiments, emphasizing the importance of redundancy to achieve robustness. For partially \gls{GNSS}-denied outdoor environments, \cite{he2021integrated} combined \gls{RTK} \gls{GNSS} with lidar-\gls{SLAM} in an \gls{LC} fashion to build a 3D-map of the surroundings. They achieved a decimeter-level of accuracy experimentally, which shows the potency of lidar-\gls{SLAM} when aided by absolute \gls{GNSS} data. It is worth noting here that the \gls{LC} approach in the above-mentioned works means that the standalone \gls{GNSS} solution was first computed using a \gls{TC} method, then it was \gls{LC} integrated with the other sensors on the position level.

Compared to \gls{LC}, \gls{TC} integrations are more prevalent in experimental works seeking higher robustness and accuracy, especially when dealing with challenging \gls{GNSS} conditions. Like its \gls{LC} counterpart, Bayesian filters dominate this sub-category as well. For instance, \cite{zhu2023tightly} presented a tightly coupled multi-frequency \gls{PPP}-\gls{RTK} \gls{GNSS} and \gls{INS} integration, achieving 10 cm accuracy in urban experimental settings. The tight coupling allows the \gls{INS} to bridge \gls{GNSS} outages and helps in resolving carrier phase ambiguities. Similarly, \cite{li2022continuous} developed a tightly coupled \gls{PPP}-\gls{RTK} \gls{GNSS}, \gls{INS}, and vision system using an \gls{EKF}, yielding decimeter-level accuracy in urban canyons. This work underscores the benefit of incorporating visual information in a \gls{TC} framework to mitigate outlier measurements originating from multipath and \gls{NLoS} effects. Another experimental study by \cite{xu2023high} in urban areas utilized a multi-state constraint (MSC)-\gls{KF} for a tightly coupled fusion of single-frequency (SF)-\gls{RTK} \gls{GNSS}, \gls{IMU}, and a monocular camera, resulting in submeter-level accuracy. Batch processing methods, while computationally more intensive, can offer superior accuracy by processing all data collectively. In an experimental context, \cite{zhu2023improved} explored multi-sensor fusion involving \gls{RTK} \gls{GNSS}, \gls{IMU}, vision, and lidar. They employed semi-tight coupling (raw measurements from all sensors except \gls{GNSS}, which provided direct position measurements) and factor graph optimization to achieve an \gls{RMSE} of 0.695m in rural and urban canyon experiments. Authors in \cite{zhumu2023multisensor} presented a hybrid \gls{LC}-\gls{TC} framework that utilizes adaptive \gls{EKF} for the fusion of \gls{GPS}, \gls{IMU}, and a monocular camera in urban settings, achieving decimeter-level accuracy. The fusion scheme fuses the outputs of a \gls{LC} filter and a \gls{TC} filter. Adaptive filtering is key to handling varying noise characteristics in dynamic environments.

\textbf{Simulation Setups:} Simulations allow for controlled evaluation of algorithms and exploration of scenarios that may be difficult or costly to replicate in the real world. For \gls{LC} integrations in simulation, \cite{zhou2020cascaded} proposed cascaded \gls{KF}s for \gls{GPS} and \gls{IMU} data fusion in outdoor simulations, achieving meter-level accuracy. Cascaded architectures can simplify filter design but may not be as optimal as fully \gls{TC} systems. For instance, \cite{yin2023sensor} developed a \gls{TC} error-state \gls{KF} with Rauch–Tung–Striebel (RTS) smoothing to integrate \gls{GNSS} and \gls{IMU} in urban simulation environments, reporting sub-meter accuracy. The RTS smoother is a classic batch technique to improve state estimates from a forward \gls{KF}. Additionally, the paper showed the efficacy of error-state \gls{KF} in circumventing the linearization errors caused by traditional \gls{EKF} when estimating and correcting quaternion states (used to represent the orientation of the vehicle). A batch localization method was demonstrated by \cite{tao2022multi}, who investigated a \gls{TC} global pose graph optimization strategy for \gls{RTK} \gls{GNSS}, \gls{IMU}, and \gls{VIO} in an urban setting (hybrid real/simulation data), reporting absolute submeter-level and relative centimeter-level accuracy.

\paragraph{Cellular-based}
Cellular networks, especially 5G, are increasingly explored for integration with onboard sensors, mostly \glspl{IMU}, for vehicular positioning. These studies, to date, predominantly rely on simulation for validation, as \gls{3GPP}-based positioning signals are not widely implemented by infrastructure vendors worldwide, which limits real-world experimentation. However, some works rely on quasi-real datasets that consist of real sensor measurements and simulation-based 5G measurements.

\textbf{Quasi-real Setups:} The following works utilize real \gls{IMU} datasets and fuse them with simulation-based 5G measurements from ray-tracing tools \cite{saleh_integrated_2023, saleh2024integration, bader2024leveraging}. The ray-tracing tools operate on digital-twin replicas of the cities from which the real-world \gls{IMU} data was taken, with artificial 5G \glspl{BS} deployed in them. To emulate the vehicle's real trajectory in the digital twin, they utilize the positioning output of a high-end \gls{GNSS}-\gls{IMU} solution (co-mounted with the other onboard sensors on the real-world vehicle), which also acts as the ground-truth. In \cite{saleh_integrated_2023,saleh2024integration}, authors propose a \gls{LC} \gls{EKF} approach to fuse 5G's range and \gls{AoD} measurements from multiple \glspl{BS} with \gls{IMU} data, achieving 14 cm level of accuracy for 95\% of the time. In \cite{bader2024leveraging}, authors utilized a \gls{LC} \gls{EKF} approach to fuse 5G \gls{AoA}, \gls{AoD}, \gls{ToA}, and \gls{RSS} measurements from \gls{LoS} and single-bounce reflections with \gls{IMU} measurements in an urban setting, resulting in 30 cm accuracy. They showed that the inclusion of first-order reflections, though challenging, can significantly enhance availability and accuracy in urban canyons.

\textbf{Simulation Setups:} In \cite{mostafavi2020vehicular}, authors explored the use of \gls{TC} \gls{EKF} to fuse 5G range and \gls{AoA} measurements from a single \gls{BS} with an \gls{IMU}, achieving several decimeters of accuracy in a 2D highway scenario. A similar highway simulation by \cite{hammarberg2022architecture} also used a \gls{TC} \gls{EKF} to fuse range and \gls{AoA} measurements from multiple 5G \glspl{BS} and an accelerometer, reporting decimeter-level accuracy. This suggests that increasing the number of \glspl{BS} can improve positioning geometry and thus accuracy.

\paragraph{IEEE-based}
Works that integrate IEEE wireless technologies with onboard motion sensors usually focus on indoor applications. Although this survey focuses on vehicular applications, we will cover such indoor works as their findings can be easily extended to vehicular positioning in parking lots and underground garages.

\textbf{Real-world Outdoor Experimental Setups:} Robust filtering techniques like the maximum correntropy robust \gls{UKF} (MC-RUKF), a Bayesian method resilient to outliers, have been applied by \cite{li2024novel} to \gls{TC} fuse \gls{UWB} and \gls{IMU} in outdoor \gls{LoS}/\gls{NLoS} conditions, yielding centimeter to decimeter-level accuracy. In \cite{sanmartin2020precise}, authors fused \gls{UWB} range measurements (with \gls{NLoS} detection and exclusion capabilities) with \gls{IMU} and odometer measurements using an \gls{EKF} in a \gls{TC} fashion, leading to decimeter-level accuracy in both experimental and simulation setups.

\textbf{Real-world Indoor Experimental Setups:} Most of the Bayesian-based works proposed in this area rely on \gls{LC} integration of inertial sensors and fingerprinting-based Wi-Fi solutions via an \gls{EKF}. For instance, authors in \cite{sun2020research} combined \gls{kNN}-based Wi-Fi fingerprinting with \gls{PDR} using an accelerometer, gyroscope, and magnetometer via an \gls{EKF} in an \gls{LC} scheme, achieving 2-meter accuracy indoors. 
Another experimental indoor system by \cite{wang2022fusion} fused a K-means clustering-based Wi-Fi fingerprinting solution with \gls{IMU} data using an \gls{EKF} in an \gls{LC} fashion, reporting 1.76 m of average accuracy. Similarly, \cite{wang2022wifipdr} combined \gls{CSI}-based Wi-Fi fingerprints (using weighted \gls{kNN}) and \gls{IMU}-magnetometer-based \gls{PDR} with an \gls{EKF} in an \gls{LC} format, achieving 1.23-meter accuracy indoors. In \cite{lu2022ONavi}, the authors presented a \gls{LC} optimization-based framework to fuse \gls{RSS}-\gls{SSID}-based fingerprint Wi-Fi solution and an \gls{IMU}-based \gls{PDR} solution,  reporting several decimeters of relative accuracy in indoor settings. The \gls{IMU}-based \gls{PDR} solution was developed using a convolutional \gls{NN}-transformer-based algorithm. In \cite{zhang2023from}, authors fused Wi-Fi, \gls{PDR}, and surveillance vision cameras to locate a user. The Wi-Fi's \gls{RSS}-based fingerprints are used to coarsely locate the user and to perform data association with the pedestrians detected with the surveillance camera. Consequently, the image-based position is fused with the user's \gls{IMU}-based \gls{PDR} solution in a \gls{LC} fashion to have a finer estimate of the user's position, achieving an average error of 4.61 cm in simulations. It is worth noting that all the above-mentioned \gls{PDR} methods can be replaced with vehicular-based transition models and be used for vehicular applications in indoor environments. In \cite{tang2024sequential}, authors used a hybrid whale optimization algorithm to \gls{LC} fuse a Gaussian process regression (GPR) fingerprinting-based Wi-Fi solution and a vision solution indoors, resulting in a position \gls{RMSE} of 2 m. \cite{tang2023novel} fused \gls{kNN}-based Wi-Fi fingerprints with vision-based measurements via an unsupervised Lagrangian fusion algorithm, achieving 1.51 m \gls{RMSE} accuracy in indoor experiments. 

\textbf{Simulation Setups:} Authors in \cite{lin2022pedestrian} proposed \gls{LC} fusion of a Wi-Fi-based positioning solution and an \gls{IMU} using a \gls{PF} that utilizes a chaos particle swarm algorithm, achieving millimeter-level accuracy, which is optimistic. The Wi-Fi-based solution is constructed by \gls{TC} fusion of \gls{RSS}-based range measurements from multiple Wi-Fi access points. In the context of \gls{V2V} communication, \cite{kang2022lidar} fused \gls{DSRC} with lidar for multi-object tracking using \gls{EKF} and \gls{PF} in \gls{GNSS}-denied simulations, achieving several decimeters accuracy. The \gls{DSRC} signals are used to communicate the absolute position, velocity, and heading states of other vehicles to the user, while the lidar measurements are used to estimate their relative states to the user. The data are fused in a \gls{LC} fashion using an \gls{EKF} to detect, associate, and track the states of the surrounding vehicles. The output of the \gls{EKF} is then fed to a \gls{PF} to track the user's ego states.

\subsubsection{Fusion with other Wireless Technologies}
This category explores the synergistic combination of different wireless technologies, often pairing a global system like \gls{GNSS} with a local one like 5G or an IEEE-based technology, to achieve more ubiquitous and reliable positioning. \rev{Table~\ref{tab:radio_sensor_fusion_other} summarizes the works presented in this section.}

\begin{table*}
\centering
\caption{\rev{Summary of radio sensor fusion works with other wireless technologies.}}
\label{tab:radio_sensor_fusion_other}
\vspace{2mm}
\footnotesize
\renewcommand{\arraystretch}{1.2} 
\begin{tabular}{|m{0.12\textwidth}|m{0.09\textwidth}|m{0.05\textwidth}|m{0.32\textwidth}|m{0.16\textwidth}|m{0.09\textwidth}|}
\hline
\textbf{Technology} & \textbf{Environment} & \textbf{Scheme} & \textbf{Measurements \& Techniques} & \textbf{Accuracy} & \textbf{Validation} \\ \hline
\rowcolor{pastel2} GNSS-5G \cite{ge2024experimental} & Urban & TC-LC & 5G (RTT + AoD/AoA) + GNSS (RTK), EKF & Meter to several meters & Experimental \\ \hline
\rowcolor{pastel2} GNSS-5G \cite{liu2024hybrid} & Urban & TC & 5G (ToA + DL-AoA) + GNSS (pseudoranges), weighted LS & Sub-meter to meter-level & Experimental \\ \hline
\rowcolor{pastel2} GNSS-5G \cite{bai2022gnss} & Suburban & TC & GNSS (RTK) + 5G (ToA + UL-AoA), adaptive KF & Sub-meter-level & Simulation \\ \hline
\rowcolor{pastel2} GNSS-5G \cite{peralrosado2020exploitation} & Urban & TC & GNSS (pseudoranges) + 5G (RTT) + city maps, weighted LS & Several meters & Simulation \\ \hline
\rowcolor{pastel2} GNSS-5G \cite{campolo20245GNSS} & Outdoor & TC & GNSS (pseudoranges) + 5G (ToA), weighted LS + batch processing & Decimeter-level & Simulation \\ \hline
\rowcolor{pastel2} GNSS-5G \cite{klus2021neural} & Urban & LC & 5G (RSS, fingerprint) + GNSS, neural networks & Meter-level & Simulation \\ \hline \hline
\rowcolor{pastel3} GNSS-IEEE \cite{shin2022uwbgps} & Outdoor & LC & GPS + UWB + IMU, KF & Decimeter-level & Experimental \\ \hline
\rowcolor{pastel3} GNSS-IEEE \cite{jiang2023research} & Outdoor & TC & GNSS (RTK, range) + UWB (range) + IMU, EKF & Decimeter-level & Experimental \\ \hline
\rowcolor{pastel3} GNSS-IEEE \cite{li2023indoor} & Outdoor & TC & GNSS (PPP, range) + UWB (range) + INS, EKF & Decimeter-level & Experimental \\ \hline
\rowcolor{pastel3} GNSS-IEEE \cite{chang2024advanced} & Outdoor & LC & GNSS (RTK) + UWB + INS + camera (stero, SLAM), EKF/batch & Decimeter-level & Experimental \\ \hline
\rowcolor{pastel3} GNSS-IEEE \cite{lv2022seamless} & Indoor/Outdoor & LC & GNSS + UWB, weighted average + EKF & Centimeter-level & Experimental \\ \hline
\rowcolor{pastel3} GNSS-IEEE \cite{luo2023accurate} & Indoor/Outdoor & TC & GPS (pseudorange) + UWB (range), LS & Centimeter-level & Experimental \\ \hline
\rowcolor{pastel3} GNSS-IEEE \cite{ibrahim2020wigo} & Urban & TC & GPS (pseudoranges) + Wi-Fi (FTM RTT) + odometer, PF-based SLAM & Meter to several meters & Experimental \\ \hline
\rowcolor{pastel3} GNSS-IEEE \cite{gao2023performance} & Outdoor & TC & GPS (pseudoranges) + UWB (range) + V2V DSRC (position), robust KF & Decimeter to meter-level & Experimental \\ \hline
\rowcolor{pastel3} GNSS-IEEE \cite{yan2022tightly} & Outdoor & TC & GNSS (pseudoranges + Doppler) + RSU DSRC (range + Doppler) + V2V DSRC (position + range) + IMU + maps (map-matching), Rao-Blackwellized PF & Sub-meter to meter-level & Experimental \\ \hline
\rowcolor{pastel3} GNSS-IEEE \cite{xiong2020carrier} & Outdoor & TC & GNSS (pseudoranges + carrier phase) + V2V UWB (range) + V2V DSRC (Doppler) + IMU, EKF,  & Centimeter to meter-level & Simulation \\ \hline \hline
\rowcolor{pastel1} 5G-IEEE \cite{yang2024dynamic} & Indoor & LC & 5G (RSS,  KF + neural network fingerprint) + WiF (RSS, KF + neural network fingerprint), PF & Meter to several meters & Experimental \\ \hline \hline
\rowcolor{pastel5} IEEE-IEEE \cite{zhu2021indoor} & Indoor & LC & Wi-Fi (RSS, kNN fingerprint) + Bluetooth (RSS, kNN fingerprint) + PDR (IMU + magnetometer), UKF & Meter-level & Experimental \\ \hline
\end{tabular}
\end{table*}

\paragraph{GNSS-5G}
The fusion of \gls{GNSS} and 5G aims to leverage the global coverage of \gls{GNSS} with the potentially high-accuracy and low-latency measurements from 5G networks, especially in urban areas where \gls{GNSS} can be challenged.

\textbf{Real-world Experimental Setups:} Experimental validation is in its early stages in this area due to the lack of off-the-shelf 5G mmWave testing equipent that has positioning signals/functionalities. Authors in \cite{ge2024experimental} demonstrated a hybrid \gls{TC}-\gls{LC} \gls{EKF}-based fusion of 5G's \gls{RTT} and \gls{AoD}/\gls{AoA} measurements and \gls{RTK}-\gls{GNSS} position measurements, showing 1.71 m accuracy in \gls{LoS} and several meters in \gls{NLoS} urban conditions. This highlights the current accuracy achievable and the persistent challenge of \gls{NLoS} for 5G signals. Authors in \cite{liu2024hybrid} fused 5G's \gls{ToA} and \gls{DL}-\gls{AoA} with \gls{GNSS}'s pseudoranges via a \gls{TC} weighted \gls{LS} estimator, yielding meter to sub-meter accuracy in both simulation and experimental validation. 

\textbf{Simulation Setups:} Simulations are widely used to explore the potential of \gls{GNSS}-5G fusion. In \cite{bai2022gnss}, authors proposed a \gls{TC} multiple-rate adaptive \gls{KF} to fuse \gls{RTK}-\gls{GNSS} range measurements with 5G's \gls{ToA} and \gls{UL}-\gls{AoA} measurements (at higher rate) in suburban areas, achieving sub-meter accuracy through simulations and semi-physical experiments (real \gls{GNSS} data and simulate 5G measurements). The adaptive approach alters the measurement covariance matrix of the 5G measurements (originally computed using a reference \gls{CRB} computation) based on the estimated range between the given \gls{BS} and the \gls{UE} (i.e., the greater the range, the higher the covariance entree). The exploitation of 3D city maps to aid \gls{GNSS}-5G fusion was investigated by \cite{peralrosado2020exploitation} using a ray-tracing approach. A weighted \gls{LS} estimator was used to \gls{TC} fuse \gls{GNSS} pseudorange measurements with 5G's \gls{LoS} \gls{RTT} measurements in deep urban canyon simulations, resulting in 10-meter accuracy. Another simulation study by \cite{campolo20245GNSS} focusing on urban, suburban, and rural scenarios fused multiple 5G \glspl{BS}' and \gls{GNSS}'s range measurements in a \gls{TC} fashion. Although 5G measurements were available at a much higher rate compared to their \gls{GNSS} counter parts, the sensor fusion (using weighted \gls{LS}) was conducted at the slower rate of the \gls{GNSS} measurements (i.e., batch processing/smoothing was utilized), achieving decimeter-level of accuracy. \gls{AI}/\gls{ML} methods are also being investigated as shown in \cite{klus2021neural}, where they employed neural networks for \gls{LC} fusion of 5G \gls{RSS}-based fingerprints and \gls{GNSS} data in urban environments, achieving meter-level accuracy.

\paragraph{GNSS-IEEE}
Fusing \gls{GNSS} with IEEE-standard technologies is a well-established strategy to enhance positioning continuity and accuracy, especially where \gls{GNSS} is weak. Since \gls{GNSS} is present in these works, they all operate in outdoor urban scenarios, unlike works that focus on integrating IEEE-based technologies with other onboard sensors or with 5G systems, which focus on indoor localization.

\textbf{Real-world Experimental Setups:} This area boasts a significant number of experimental validations, indicating technological maturity. Several works have employed Kalman filters to fuse \gls{GPS}/\gls{RTK}/\gls{PPP}-\gls{GNSS}, \gls{UWB}, and \gls{IMU} integration in outdoor experimental settings, consistently achieving decimeter-level accuracy \cite{shin2022uwbgps, jiang2023research, li2023indoor}. In \cite{shin2022uwbgps}, authors utilized a linear \gls{KF} to fuse the technologies in a \gls{LC} fashion. On the other hand, \cite{jiang2023research} and \cite{li2023indoor} utilized an \gls{EKF} to fuse the range measurements in a \gls{TC} fashion. In \cite{chang2024advanced}, authurs \gls{LC} fused \gls{RTK}-\gls{GNSS}, \gls{UWB}, \gls{INS}, stero-camera-based \gls{SLAM} position estimates, also achieving decimeter-level accuracy in outdoor experiments. Similarly, \cite{lv2022seamless} achieved 8 cm accuracy by fusing \gls{RTK}-\gls{GNSS}, \gls{UWB}, and \glspl{IMU} using a two stage approach for seamless indoor/outdoor vehicular positioning. The first stage fuses the \gls{GNSS} and \gls{UWB} measurements in a \gls{LC} fashion using a weighted average approach. In the second stage, the fused solution is integrated with \gls{IMU} measurements using an \gls{EKF} in a \gls{LC} approach. Authors in \cite{luo2023accurate} fused \gls{GPS} and \gls{UWB} range measurements using a \gls{TC} \gls{LS} estimator in seamless indoor/outdoor vehicular experiments, achieving centimeter-level accuracy. Authors in \cite{ibrahim2020wigo} fused Wi-Fi-based \gls{FTM} \gls{RTT} measurements with \gls{GPS} pseudoranges and vehicular odometers in a \gls{TC} fashion to localize vehicles driving in urban neighborhoods. Since the exact location of the Wi-Fi anchors is unknown, the authors utilized a \gls{PF}-based \gls{SLAM} technique (to simultaneously localize the vehicle and map the positions of the Wi-Fi anchors), achieving achieving meter to several meters of accuracy in experimental and simulation setups. In \cite{gao2023performance}, authors investigate the integration of traditional \gls{GPS} and \gls{UWB} measurements (from satellites and roadside units) with cooperative positioning measurements (provided through \gls{DSRC} with neighboring vehicles). The authors utilized a \gls{TC} robust \gls{KF} to better handle measurements outliers while positioning in urban canyons, achieving meter- to decimeter-level accuracy. Similarly, authors in \cite{yan2022tightly} fused \gls{GNSS} pseudo ranges and Doppler measurements, roadside unit's range and Doppler measurements (through \gls{DSRC}), the estimated position and range measurements of surrounding vehicles (through \gls{DSRC}), \gls{IMU} measurements, and map matching using a \gls{TC} Rao–Blackwellized \gls{PF}. The method achieved meter to sub-meter accuracy in experimental and simulation setups.

\textbf{Simulation Setups:} Authors in \cite{xiong2020carrier} proposed a novel framework to \gls{TC} integrate \gls{GNSS}'s pseudoranges and carrier phase measurements, \gls{V2V} \gls{UWB}-based range measurements, \gls{V2V} \gls{DSRC}-based Doppler measurements, and \gls{IMU} measurements. The framework follows a cascaded filtering approach to resolve carrier phase ambiguities. The method utilizes \gls{EKF} as the filter of choice,  demonstrating meter to centimeter-level accuracy in realistic outdoor simulations.

\paragraph{5G-IEEE/IEEE-IEEE}
The integration of 5G with other IEEE wireless technologies (and fusion among multiple IEEE technologies) is an emerging field, primarily targeting robust indoor positioning where \gls{GNSS} is unavailable and 5G deployment might be sparse. We will cover those indoor methods as they can be easily extended to deep-urban and parking environments where vehicles can operate.

\textbf{Real-world Experimental Setups:} Experimental work by \cite{yang2024dynamic} combined 5G and Wi-Fi \gls{RSS} measurements in a \gls{LC} fashion. The \gls{RSS} measurements from both sources are first filtered via an \gls{EKF} before passing them into individual neural network-based fingerprinting processes. The individual positioning estimates of all the sources are then fused via a \gls{PF}, achieving several meters to meter-level of accuracy in indoor environments. Another experimental study by \cite{zhu2021indoor} focused on a multi-stage \gls{LC} fusion of multiple Wi-Fi, Bluetooth, and \gls{IMU}-magnetometer-based \gls{PDR} using an \gls{UKF} for indoor environments, achieving meter-level accuracy. In the first stage, the proposed method first computes a separate Wi-Fi and Bluetooth positioning estimate through a \gls{kNN}-based fingerprinting method using their respective \gls{RSS} signals and then conducts a weighted average to fuse the two solutions. The resulting positioning output is then fused with the \gls{PDR} solution in a \gls{LC} fashion.

\subsection{Challenges and Open Problems}
Despite significant advancements in sensor fusion techniques for vehicular localization, several pressing challenges and unresolved issues remain. These obstacles must be addressed to achieve robust, real-time, and highly accurate positioning systems suitable for safety-critical applications such as autonomous driving. One of the most fundamental limitations lies in the current inability to accurately assess and adapt to the quality of incoming sensor data. Most fusion systems assume fixed or heuristic-based confidence levels for each sensor, yet the reliability of data from these sensors can vary drastically depending on environmental conditions. This is even exacerbated when considering other dimensions of sensor quality, like security and trustworthiness. For instance, spoofing attacks on \gls{GNSS}, falsified \gls{V2X} messages, or compromised sensors can also undermine the entire localization pipeline. Fusion architectures must incorporate mechanisms to detect and mitigate such threats, whether through redundancy, anomaly detection, or secure message verification. Building trust in both individual sensors and shared data is critical for system resilience. Such systems need to be adaptive, where certain sensors are used to validate or cross-check the reliability of others. For example, visual tracking can detect \gls{GNSS} anomalies in urban canyons, or when the positioning message is being spoofed. This calls for deeper research into uncertainty quantification and sensor quality monitoring, allowing for context-aware and dynamically weighted fusion.

Another core issue is the need for better sensor calibration, alignment, and synchronization. Multi-sensor systems require tight spatial alignment across different reference frames, precise temporal synchronization, and compensation for individual sensor biases and drifts. Although initial calibration can be performed offline, real-world operation demands continual recalibration, especially in the presence of mechanical wear, environmental variations, or hardware changes. Yet, continuous online calibration remains an underexplored topic, particularly for large-scale deployments across heterogeneous vehicular fleets.

Equally important is the current gap in ultra-tight sensor coupling. Most fusion systems operate on processed outputs, such as position estimates (\gls{LC} approaches) or feature tracks (\gls{TC} approaches), rather than directly integrating raw measurements of fused sensors. Ultra-tight fusion at the signal level holds the potential to significantly improve accuracy and robustness, but it remains largely unexplored due to the complexity of joint signal modeling and the high computational demands it introduces. Nonetheless, this direction is essential for pushing the limits of localization performance. However, as fusion becomes tighter and more complex, computational efficiency emerges as a major concern. High-fidelity fusion algorithms are often computationally intensive, introducing delays that are unacceptable for real-time control. Future systems must balance accuracy with latency, requiring innovation in algorithm design, model simplification, and deployment strategies such as edge computing or distributed inference. Even with such advanced fusion strategies, achieving centimeter-level accuracy with 3-sigma reliability (i.e., 99\% of the time) is still far from realization in dynamic and cluttered environments. Multipath, signal blockage, sensor occlusion, and other real-world impairments remain major barriers. To meet the stringent accuracy and reliability requirements of future autonomous systems, improvements in both sensing modalities and their integration are needed.

Furthermore, more emphasis is needed on cooperative localization and live mapping. Sharing observations between vehicles, such as relative ranges or landmarks, can enhance individual positioning accuracy and robustness. However, implementing such systems at scale introduces challenges in data association, bandwidth management, and privacy. Moreover, the integration of real-time map updates into the localization process remains limited. Future systems should aim to jointly localize and map their environment in a dynamic, distributed, and cooperative manner.

Finally, the lack of comprehensive datasets also hampers the development of advanced fusion architectures. Most available datasets primarily focus on \gls{GNSS}, vision, lidar, and inertial data, with minimal or no inclusion of cellular 5G and IEEE-based technologies (Wi-Fi, \gls{UWB}, Bluetooth, etc.). This fragmentation hinders the development and benchmarking of fully integrated systems. To enable progress, there is a pressing need for datasets that combine all key sensing/positioning modalities—wireless, inertial, and perception—captured under realistic conditions with proper synchronization and ground-truth alignment.

In conclusion, while sensor fusion has already profoundly improved vehicular localization, substantial work remains to address uncertainty handling, resilience to threats, collaborative calibration, ultra-tight integration, real-time performance, cooperative localization, live mapping, and comprehensive dataset completeness. Tackling these challenges is essential for the next generation of localization systems that are reliable, scalable, and ready for full autonomy.

\section{\rev{Lessons Learned} and \rev{Open Problems}}\label{Sec: outlook}
\rev{This survey has navigated the extensive landscape of wireless vehicular positioning, starting from the various vehicular use cases, \glspl{KPI}, and requirements, diving into the history, standards, theoretical fundamentals, and contemporary research works of satellite-, 3GPP-, and IEEE-positioning, and ending with a survey of sensor fusion with onboard sensors.} Throughout the survey, we encountered a handful of overarching themes and trends within the individual wireless research communities. These themes range from the prominent \glspl{KPI} focused on by these communities to the varying technology readiness/maturity levels, and the complementary aspects between them. \rev{In the following, we first unpack those themes and comment on what we think could be learned from the past decades of research on these topics. Then, we discuss the open problems that are yet to be solved by the research community.}

\subsection{\rev{Lessons Learned}}
\subsubsection{KPIs} The survey showed a clear historical focus from the research community on positioning accuracy as the main \gls{KPI} of interest. Although the pursuit of sub-meter and, eventually, centimeter-level accuracy is still paramount, it should no longer be the sole objective of this research community. The future of this field will be defined by how robustly we address the holistic performance requirements of safety, reliability, and scalability demanded by fully autonomous systems. Hence, future work must broaden its scope to embrace the other critical \glspl{KPI} \rev{discussed in Section~\ref{Sec: requirements}. Most important among these are reliable position uncertainty estimates and position integrity.} For safety-critical applications, knowing the confidence of a position estimate is as important as the estimate itself. While integrity has a long history in aviation \gls{GNSS}, defining and validating robust protection levels and alert limits for cellular, IEEE-based, and fused positioning systems is a critical and largely open problem that requires immediate attention. Likewise, metrics such as availability, continuity, latency, and scalability must be pushed to the forefront of system design and evaluation.

\subsubsection{\rev{Technology Readiness Level}} Another prominent theme throughout this review is the varying maturity and validation levels across different technologies. \gls{GNSS}, enhanced by correction services like NRTK and \gls{PPP}-{RTK}, stands as the established workhorse, backed by decades of real-world deployment and extensive experimental validation, achieving centimeter-level accuracy under favorable conditions. In contrast, 5G (and beyond 5G) cellular positioning is an emerging powerful challenger. Its capabilities are evolving rapidly through 3GPP standardization, with carrier phase and sidelink functionalities showing theoretical promise. However, its validation remains largely in the simulation and quasi-real experimental domains—widespread infrastructure deployment and the availability of public datasets are essential next steps to bridge the gap to real-world performance. IEEE-based technologies like \gls{UWB} and the latest generations of Wi-Fi are proving their mettle, particularly for cooperative V2V and niche applications like parking garages. Recent standards (e.g., IEEE 802.11bk) are significantly boosting their raw capabilities, bringing their potential closer to that of cellular systems and creating exciting opportunities for outdoor vehicular use.

\subsubsection{\rev{Sensor Fusion}}
The path forward is unequivocally centered on sensor fusion. \rev{As the comparative analysis in Table~\ref{tab:localization_comparison_survey_revised} demonstrates,} no single technology is a silver bullet; their strengths and weaknesses are complementary. Future research must move beyond the prevalent loosely-coupled and tightly-coupled architectures towards more sophisticated integration schemes. Ultra-tightly-coupled fusion, which integrates raw sensor measurements at the signal level, holds the potential to unlock significant gains in robustness, especially in environments where one or more technologies are degraded. Furthermore, fusion frameworks must become more intelligent and adaptive, capable of assessing the quality and uncertainty of each data stream in real-time to dynamically adjust their weight and influence within the overall solution.

\vspace{-5pt}
\subsection{\rev{Open Problems}}
Several fundamental challenges and emerging frontiers must be addressed to realize the full potential of vehicular positioning:

\begin{itemize}[leftmargin=*]
    \item Security and trust: As vehicles become increasingly connected and cooperative, they also become more vulnerable. Future systems must incorporate robust mechanisms to defend against threats like \gls{GNSS} spoofing, malicious \gls{V2X} messages, and compromised sensor data. Building trust in shared cooperative data is paramount.

    \item Datasets and openness: A significant bottleneck hindering progress is the lack of comprehensive, large-scale, multi-modal public datasets. There is a pressing need for datasets that include synchronized measurements from 5G, Wi-Fi, and UWB alongside traditional \gls{GNSS}, \gls{IMU}, and perception sensors, captured under diverse and challenging real-world driving conditions. \rev{There are published experimental datasets for most of the above-mentioned technologies except for 5G. This is mainly due to that 5G positioning has not been deployed widely at the market yet, and experimental datasets are limited to internal testbeds and innovation project experiments, not made publicly available. Deploying vehicular specific cellular positioning capabilities and enablers will further accelerate vehicular-specific applications.}

    \item \rev{NLoS positioning: Accurate and reliable positioning in areas where all radio-based technologies are operating under \gls{NLoS} conditions, e.g., in deep urban areas like Manhattan, is still an open problem. Although many works attempted solving the problem using multipath signals, most of them were validated in theoretical/simulation settings. Rigorous real-world experimentation and validation of these methods is yet to be seen.}

    \item Continuous calibration and synchronization: The practical difficulty of maintaining precise spatial and temporal calibration between a heterogeneous suite of sensors on a dynamic platform remains a significant hurdle. Developing robust, online auto-calibration methods is essential.

    \item \rev{Information interfaces and provisioning: For scalability and efficient deployment, standardized interfaces for information exchange, exposure, and provisioning mechanisms are vital, including functional partitioning between vehicles, positioning servers, and vehicle cloud.}

    \item Emerging technologies: The integration of nascent technologies will continue to push boundaries. \gls{LEO} satellite constellations offer a new layer of global signals, \glspl{RIS} promise to reshape the radio environment for better positioning, and the full realization of \gls{ISAC} will create unprecedented opportunities for both communication and sensing tasks.
\end{itemize}

In conclusion, the future of vehicular positioning is not about finding a single superior technology, but about mastering the art of intelligent, secure, and resilient fusion. Success will be measured not just by raw accuracy, but by the ability to deliver a holistic and trustworthy positioning solution that can reliably meet the stringent demands of the next generation of autonomous vehicles.

\vspace{-5pt}
\section*{\rev{Acknowledgment}}
\rev{The authors wish to thank Professor Sujit Kumar Chakrabarti (Department of Computer Science, IIIT Bangalore, India) for his valuable contribution of the illustrative sketch presented in Fig.~\ref{scenario_pic}.}

\vspace{-10pt}
\appendices
\section{\rev{Generalized Wireless Channel Modeling}}
\label{appendix}
\rev{The baseband received signals in most radio technologies can be generalized to have the following form
\begin{equation}
\begin{split}
   \boldsymbol{y}(t) = \sum_{\ell=0}^{L-1} \rho_\ell s(t-\tau_\ell+\nu_\ell t) e^{-\jmath2\pi f_c(\tau_\ell-\nu_\ell t)} \\
    \mathbf{a}_{\text{rx}}(\boldsymbol{\phi}_{\ell}) \mathbf{a}_{\text{tx}}^T(\boldsymbol{\theta}_{\ell})\boldsymbol{f}(t) +\omega(t)\,,
\end{split}\label{eq_rec}
\end{equation}
where $\boldsymbol{y}(t)\in\mathbb{C}^{\Ntx}$ is the received baseband signal at time $t$ after demodulation, with $\Ntx$ being the number of antennas at the receiver, $s(t)$ is the transmitted baseband signal, $\omega(t)$ is the complex additive white Gaussian noise, $L$ is the total number of paths the received signal traversed (including the \gls{LoS} one), $\ell$ is the path index, $\rho_\ell$, $\tau_\ell$, $\nu_\ell$, $\phib_\ell= [\phiazl, \phiell]$, and $\thetab_\ell= [\thetaazl, \thetaell]$, are the geometric channel parameters of the $\ell$-th path relating to the channel gain, delay/\gls{ToA}, Doppler, \gls{AoA}, and \gls{AoD}, respectively, while $\arx(\cdot)\in\mathbb{C}^{\Ntx}$ and $\atx(\cdot)\in\mathbb{C}^{\Nrx}$ denote the receiver's \gls{AoA} and transmitter's \gls{AoD} steering vectors, respectively, where $\Nrx$ denotes the number of antennas at the transmitter, and $\boldsymbol{f}(t)\in\mathbb{C}^{\Ntx}$ is the transmitter's precoder vector. Variations in $s(t)$ and hardware environments---such as the high attenuation and Doppler shifts of satellite links versus terrestrial networks---dictate the scale of the channel parameters. These technological differences determine whether certain effects are negligible or significant, directly defining the overall complexity of channel estimation.}


\vspace{10pt}

\bibliographystyle{IEEEtran} 
\bibliography{ref_vehicular_survey}
\end{document}